\documentclass{pasa}%

\title[Galactic Bulge Composition]{The Chemical Composition of the Galactic Bulge and Implications for its Evolution}
\author[Andrew McWilliam]{Andrew McWilliam \\
\affil{Carnegie Observatories, Pasadena, CA} }%
\jid{PASA}
\doi{10.1017/pas.\the\year.xxx}
\jyear{\the\year}

\begin{document}%
\begin{abstract}

The bulge appears to be a chemically distinct component of 
the Galaxy; at {\it b}$=$-$4^{\circ}$ the average [Fe/H] and [Mg/H]
values are $+$0.06 and $+$0.17 dex respectively, roughly 0.2 dex
higher than the solar neighborhood thin disk, and $\sim$0.7 dex
greater than the local thick disk.  This high average metallicity suggests 
a larger {\it effective yield} for the bulge compared to the solar neighborhood, 
perhaps due to more efficient retention of supernova ejecta.


A vertical metallicity gradient in the bulge, at $\sim$0.5 dex/kpc, is
attributed to the changing mixture of metal-rich and metal-poor sub-populations
(at [Fe/H] $+$0.3 and $-$0.3 dex from Hill et al. 2011;
but $+$0.15, $-$0.25 and $-$0.7 dex from Ness et al. 2013), where the metal-poor
sub-populations have a larger scale height than the metal-rich population.

Abundances of O, Mg, Si, Ca, Ti, and Al are enhanced in the bulge compared to 
solar composition, with [$\alpha$/Fe]=$+$0.15 dex at solar [Fe/H]; 
below [Fe/H]$\sim$$-$0.5 dex, the bulge and local thick disk [$\alpha$/Fe] ratios
are very similar.
Small enhancements in [Mg/Fe] and possibly [$<$SiCaTi$>$/Fe] relative to the
thick disk trends are apparent, suggesting slightly higher SFR in the bulge.
This is supported by low [s-/r-] process ratios, as measured by [La/Eu],
and dramatically enhanced [Cu/Fe] ratios compared to the thick disk.
However, the differences between thick disk and bulge composition trends 
could, conceivably, be due to measurement errors and non-LTE effects.
Unfortunately, the comparison of bulge with solar neighborhood thick disk 
composition may be confused by uncertainties in the identification 
of local thick disk stars; in particular, the local thick disk [$\alpha$/Fe] 
trend is not well defined above [Fe/H]$\sim$$-$0.3 dex.

The unusual zig-zag abundance trends of [Cu/Fe] and [Na/Fe] are qualitatively
consistent with the Type~Ia supernova time-delay scenario of Tinsley (1979) 
and Matteucci \& Brocato (1990) for elements made principally by 
core-collapse supernovae, 
but with metallicity-dependent yields.

The metallicities, [$\alpha$/Fe] ratios and kinematics of the metal-poor and metal-rich
bulge sub-populations
resemble the solar neighborhood thick and thin disks, respectively, but with 
higher [Fe/H] 
than at the solar circle.  If these sub-populations really represent the inner
thin and thick disks, but at higher [Fe/H], then
both the thin and thick disks
possess a radial [Fe/H] gradient, within the solar circle, near $\sim$$-$0.04 to 
$-$0.05 dex/kpc.

In the secular bulge scenario, the bulge was built from entrained inner disk
stars driven by a stellar bar. Thus, it appears that the inner thin and thick disk 
stars retained vertical scale heights characteristic of their kinematic origin, 
resulting in the vertical [Fe/H] gradient seen today.

\end{abstract}
\maketitle%

\section{Introduction and Motivation}

The Galactic bulge is major component of the Milky Way (MW) Galaxy, 
morphologically distinct from the disk and halo, composed of mostly old
stars with an embedded bar.  It is the closest bulge and bar 
to us and we can study it in greater detail than for any other galaxy, 
down to individual stars.  Not only does the MW bulge provide a way to
understand bulges and bars in extra-galactic spiral galaxies, but its 
population is similar to distant giant elliptical galaxies.

We would, naturally, like to know how the bulge came to be: how did it 
evolve?  Because the chemical element abundance patterns contain a 
fossil record of past star formation, much could be learned from a 
study of the bulge chemical composition.  However, an impediment to 
reading this fossil record is that we do not fully understand the 
nucleogenesis of all the elements.  Thus, we must try to simultaneously
understand both the mechanisms and astrophysical sites of element 
synthesis as well as the star formation history of the bulge.

Because the bulge is situated in a deep gravitational well, compared to
the MW disk and halo, and because its stars seem to be mostly old, 
chemical evolution occurred under different environmental conditions in
the bulge.  Thus, a comparison of the bulge chemical properties, to 
those in other locations, offers a way to understand how environment 
can affect chemical evolution.  This should inform us about the sites of 
nucleosynthesis and clues to how the bulge evolved.  Certainly, the 
chemical evolution models, developed to explain the composition of stars 
near the sun, should work everywhere.

To address these questions and issues we must first measure the bulge's
chemical properties; good measurements are the basis for understanding.
Once we have good measurements we need to compare them to something.
It would be nice to compare with the output of chemical evolution models,
but at the present time it is more informative to compare to other
chemically evolving systems.  Here, we compare the bulge chemical
composition with the Milky Way thin and thick disks, and then ask how
the evolution of these systems could have produced the measured composition
differences.

\section{A Few Ideas in Chemical Evolution}

The idea that the chemical composition of the Galaxy has evolved over time
sprang from the identification of metal-poor stars by
Chamberlain \& Aller (1951) and the theoretical predictions of
Hoyle (1946), who proposed that element synthesis occurred in stars
and supernova explosions.

The Simple model of chemical enrichment (e.g., Schmidt 1959;
Searle \& Sargent 1972) assumed zero metal starting point, gas in a closed box,
consumed by multiple generations of star formation;
each generation locked-up some gas in the form of low-mass stars and
returned gas enriched in metals from massive stars.  The return of metals was
assumed to occur instantaneously and the interstellar gas immediately homogenized.

Upon complete consumption of the gas in this model, the metallicity distribution
function (henceforth MDF) has a predictable mean and standard deviation.  In
particular, the mean metallicity of the final MDF is equal to the ratio
of the mass of metals produced to mass of gas locked-up in dwarf stars per generation;
this parameter is called the {\em yield} (Searle \& Sargent 1972).
For systems that lose gas or metal-rich ejecta the yield
is lowered; for systems that over-produce low-mass stars the yield is also
lowered, while for systems that under-produce low-mass stars or over-produce
high mass stars, the yield is increased.  In such situations we often refer
to the {\it effective yield}.
Early comparison of the MDF predicted from the Simple model
with metallicity measurements of G-dwarf stars 
(e.g., Schmidt 1963) showed that the Milky Way disk has fewer metal-poor stars
than expected; this lacuna was called ``The G-dwarf Problem''.  
The {\em G-dwarf problem} is thought to be due to infall of fresh material
(e.g. Larson 1972; Pagel 1989) during the chemical evolution of the MW disk.

The Simple model provides a tight age-metallicity relation, linear for 
a constant star formation rate, which occurs when not much of the gas 
has been consumed.

The factor of two enhancement in [X/Fe] for even-numbered light elements in 
metal-poor stars of the Galactic halo, has been known for over 50 years
(e.g., Wallerstein et al. 1962, 1963; Conti et al. 1967).  Initially,
it was thought that these elements (e.g., O, Mg, Si, S, Ca, Ti) were produced
in the $\alpha$-process, suggested by Burbidge et al. (1957),
by successive addition of $\alpha$ particles.  Although alpha-capture 
in massive stars accounts for $^{16}$O and some $^{24}$Mg, the remaining
$^{24}$Mg is produced by carbon burning, while Si, S, and Ca are thought
to be produced during explosive oxygen burning in core-collapse SNII events
(e.g. Woosley \& Weaver 1995, abbreviated WW95).  Thus, ``alpha element'' is not a very appropriate
name.

Tinsley (1979) suggested that the observed decline of the [O/Fe] ratio with increasing
[Fe/H] (from halo to disk) was due to the delayed addition of iron from
Type~Ia supernovae (henceforth
SNIa), whose progenitors are $\leq$8M$_{\odot}$, to an oxygen-rich composition
produced by more massive stars.  These massive stars end as core-collapse supernovae
(henceforth SNII) with O/Fe yields that increase with progenitor mass (e.g. WW95).
A small, but useful, idea is that the frequency of SNIa declines with
delay time roughly 1/$\tau_{delay}$ ; most SNIa occur promptly, in less than 2Gyr, with
a long tail out to 10Gyr (see Maoz et al. 2010; Greggio et al. 2008).

\begin{center}
\begin{figure}[ht!]
\includegraphics[width=3.0in]{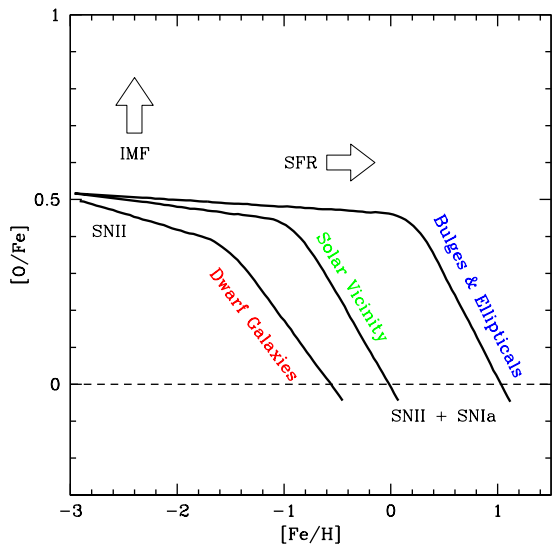}
\caption{Matteucci \& Brocato (1990) predicted that the knee in the trend of
[O/Fe] with [Fe/H] depends on the SFR: high SFR systems, like bulges 
and elliptical galaxies, should show enhanced [O/Fe] to high [Fe/H], while 
the low SFR dwarf galaxies show reduced [O/Fe] relative to the Solar vicinity.  
Also shown is the direction of the [O/Fe] plateau with an enhanced
fraction of massive stars, marked as IMF, which over-produce oxygen.  }
\label{fig-mb90}
\end{figure}
\end{center}

Matteucci \& Brocato (1990, henceforth MB90) produced a marvellous sketch of the expected
[O/Fe] trend with [Fe/H] for stellar systems with different star formation 
rates (henceforth SFR).  Since infall time scales with 1/$\sqrt{\rho}$, where $\rho$ is
the mass density, dense systems collapse more quickly and have a higher SFR than
low-density or loose systems.  Thus at high SFR, expected for bulges and giant elliptical
galaxies, MB90 predicted that [O/Fe] remains high to much higher [Fe/H] than
than low SFR systems like the solar vicinity or dwarf galaxies.  This idea of 
MB90 provides a useful prediction for the composition of the Galactic bulge, 
that can constrain the bulge SFR and formation timescale.  
Figure~\ref{fig-mb90} shows a representation of the MB90 [O/Fe] prediction,
and also the sense of the effect due to increased fraction of massive stars.
Remarkably, the low [O/Fe] ratios predicted by MB90 for dwarf galaxies is
verified (e.g., Shetrone et al. 2001, 2003), although
some dwarf galaxies show low [O/Fe] due to IMF modification (e.g. McWilliam,
Wallerstein \& Mottini 2013).

These simple ideas provide a framework for understanding evolution of the 
Galactic bulge, compared to the solar neighborhood, by the study of the
detailed chemical composition.  We may hope to learn how much
time it took chemical evolution to build the bulge and what the star formation
rate was.  We may also learn whether, or not, the initial mass function 
(henceforth IMF; Salpeter 1955) of stars in the bulge 
was like that in the solar vicinity.  While such detailed abundance ratios may
elucidate the chemical evolution history of the bulge, and  provide a test for 
these ideas, abundance ratios in the bulge can be used to constrain the sites of
nucleosynthesis of various elements, whose origin is uncertain.

\begin{center}
\begin{figure}[ht!]
\includegraphics[width=3.0in]{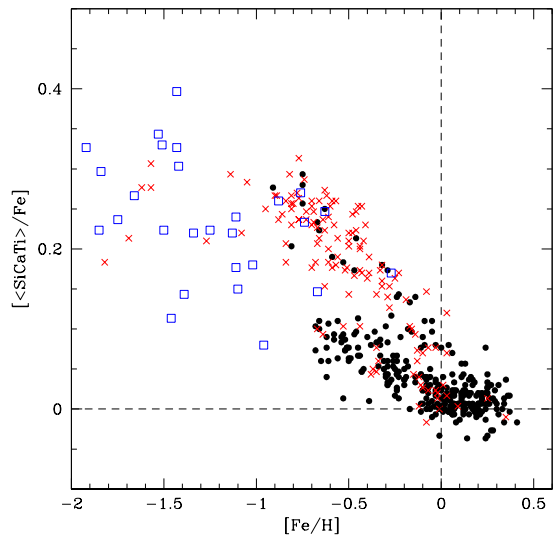}
\caption{[$<$SiCaTi$>$/Fe] for halo (open blue boxes), thick disk (red crosses)
and thin disk (filled black circles) stars from Bensby et al. (2014).  Differences
and confusion between thick and thin disks are apparent.  Kinematically identified
thick disk stars plummet in [$<$SiCaTi$>$/Fe] near [Fe/H]=$-$0.2 dex and merge
with the thin disk trend.  However, an extrapolation of the alpha-enhanced main thick
disk trend of [$<$SiCaTi$>$/Fe] with [Fe/H] seems continue beyond solar [Fe/H],
dominated by a sub-sample of thin disk stars.
}
\label{fig-b14sicati90}
\end{figure}
\end{center}

The current observational situation for alpha-elements in the solar 
neighborhood disk, from analysis of 714 nearby dwarf stars by 
Bensby et al. (2014), is summarized in Figure~\ref{fig-b14sicati90},
revealing a situation much more complex than initially considered.
Most of the results were previously seen by Reddy et al. (2003, 2006)
and Bensby et al. (2003, 2005).

Figure~\ref{fig-b14sicati90} shows kinematic identifications, at P$>$90\%,
for the halo, thick disk and thin disk.  There is a large
dispersion in [$\alpha$/Fe] for halo stars at a given [Fe/H],
probably due to different SNII/SNIa ratios of accreted dwarf galaxies.
The stars identified as
thick disk have larger [$\alpha$/Fe] than the thin disk, declining with
increasing [Fe/H], but near [Fe/H]=$-$0.2 dex the thick disk [$\alpha$/Fe]
declines sharply and merges with the thin disk.  These low [$\alpha$/Fe] thick 
disk stars could well be mis-identified thin disk stars, or simply
thin disk stars that suffered kinematic heating to thick disk velocities.


Figure~\ref{fig-b14sicati90} shows a number of intriguing sub-populations.  For example,
the $\alpha$-enhanced metal-poor stars with thin disk kinematics could
be thick disk stars mis-identified as thin disk.  However, they may simply be old,
$\alpha$-enhanced, stars but with thin disk kinematics.
If this is true, a more accurate name for the thick disk might be
the ``early disk'' or ``old disk'.

Another unusual sub-population in Figure~\ref{fig-b14sicati90} includes stars
identified with thick disk kinematics, but showing low [$\alpha$/Fe] ratios,
similar to most thin disk stars.  While these stars may
simply be mis-identified thin disk stars, taken at face value it is possible that
they reflect the composition of the late thick disk; this might occur in a
prolonged evolution of the thick disk, where there was sufficient time
to permit late SNIa ejecta to be included in the chemical enrichment.  Such
a slow-down of thick disk chemical enrichment might be expected after the main
MDF peak, by which time a significant fraction of gas had been lost.
Another possibility is that these stars formed in the thin disk, but
later experienced gravitational interactions that resulted in increased vertical
scale heights and kinematics similar to the thick disk.

Finally, the most interesting sub-population in Figure~\ref{fig-b14sicati90} include
the kinematically identified thin disk stars with slightly sub-solar [Fe/H], which 
seem to extrapolate the alpha-enhanced thick disk trend, from
lower [Fe/H].  These stars cannot be explained by mis-identification and
composition measurement error seems highly unlikely.  
The chemical similarity to the bulge suggests that these stars might be due
to radial migration of low-vertical scale height stars from the inner thick
disk or bulge.

If these stars have not migrated from inner regions, but are related to the local
thick disk, their thin disk kinematics suggests low vertical scale heights and, thus,
a vertical metallicity gradient in the thick disk; this may have occurred
shortly after the time when thick disk molecular clouds relaxed into thin disk
kinematics.  Other scenarios, some heretical, may also explain this unusual sub-population 
of stars.

%
%
%

It appears as if there are two modes of chemical enrichment in the MW disk that
are not strictly confined to kinematic sub-populations: the metal-poor
$\alpha$-rich mostly thick disk trend and the metal-rich, low-$\alpha$ 
mostly thin disk trend.  The two populations both show [$\alpha$/Fe]
decline with increasing [Fe/H] suggesting the delayed addition of SNIa iron.

Another inconsistency with simplistic ideas of chemical enrichment, 
seen in Bensby et al. (2014) and Edvardsson et al. (1993), is the relatively large
dispersion in the age--metallicity relation.  While metal-rich disk 
stars are mostly young, older metal-poor stars cover a large full range of 
[Fe/H] in both MW disks; the tight age-metallicity relations predicted by
the Simple model is not evident.

For more detailed discussions of chemical evolution see Pagel (1997) and
Matteucci (2012).

\section{A Brief History of the Bulge Metallicity}

The presence of both RR Lyrae stars and M giant stars in the Galactic 
bulge indicate a large metallicity range, from halo metallicities up
to at least the solar value (e.g., Baade 1946; Nassau \& Blanco 1958).  
Strictly, the word ``metallicity'' 
refers to the mass fraction of metals (elements heavier than He), 
represented as {\it Z}.  However, in modern times the abundance of iron
relative to the sun, [Fe/H], has been used, interchangeably, with the 
word ``metallicity''.  
The connection is that in order to compute the metallicity, {\it Z}, one
must scale the solar abundance distribution by a reference element,
typically iron, which is easily measured.  Here, I will use 
an imprecise meaning of the word, generally correlated with overall Z
or [Fe/H].

The seminal work of Rich (1988) provided the first metallicity estimate
for the bulge from low-resolution stellar spectra of 88 bulge K giants 
in Baade's Window, at {\it b}=$-$3.9$^{\circ}$; he found a mean [Fe/H] 
of $+$0.3 dex; 1$\sigma$ measurement uncertainties were 0.20 dex.  This
was not a model atmosphere abundance analysis, but rather a correlation
of the equivalent widths (henceforth EWs) of prominent Fe, Mg and Na 
optical features versus (J$-$K) color, calibrated against high-resolution
model atmosphere abundance results from solar neighborhood stars.

Terndrup (1988) used BVI photometry to estimate the
bulge metallicity, at latitudes ranging from $-$4$^{\circ}$ 
to $-$10$^{\circ}$, that relied upon color-metallicity calibrations
from detailed abundance analyses of solar neighborhood stars.
Terndrup claimed mean solar [Fe/H] in Baade's Window, 
and a decrease in [Fe/H] by 0.5 dex for the highest latitude field, i.e., a
vertical metallicity gradient in the bulge.  

Rich (1990) showed that the metallicity distribution of his 88 bulge K giants
compared well with the predicted metallicity distribution function (MDF) of the 
Simple Closed-Box model of Searle \& Sargent (1972).  This was unlike the MDF
of the solar neighborhood, which shows a distinct deficit of metal-poor stars,
known as ``The G-Dwarf Problem'' (van den Bergh 1962; Schmidt 1963).  Thus, it
appeared that the bulge did not suffer from the {\it G-Dwarf Problem} seen in the MW
disk.  Presumably, chemical evolution in the bulge was not 
affected by significant, prolonged, infall of metal-free gas, unlike the
MW disk.

Geisler \& Friel (1992) used the Washington photometric system to estimate 
the metallicity of 314 G and K giants in Baade's Window, giving a mean [Fe/H]
of $+$0.17$\pm$0.15 dex; they also confirmed a very good fit of their MDF to
a closed-box model of chemical evolution.

McWilliam \& Rich (1994, henceforth MR94) were the first to attempt model
atmosphere chemical abundance analysis of bulge stars, for 11 K 
giants in Baade's Window, with R=17,000 echelle spectra.  From only 11 
[Fe/H] values of the brightest bulge K giants a true iron distribution 
function (IDF) could not be measured, but a calibration against the 
Rich (1988) metallicities suggested a low mean [Fe/H] of $-$0.25 dex.

Minniti et al. (1995) obtained R$\sim$2,000 spectra of bulge giants at 1.5-1.7 kpc 
from the Galactic center and determined metallicities using a similar method to 
Rich (1988).  By combining previous photometric and spectroscopic metallicity
estimates, Minniti et al. (1995) showed convincing support for the vertical
metallicity gradient claimed by Terndrup (1988).

Sadler, Rich \& Terndrup (1996) measured line-strength indices, from 
R$\sim$1,000 spectra, of 400 K and M giant stars in Baade's Window.
A correlation of the indices against [Fe/H] from solar neighborhood giants
resulted in an average [Fe/H] of $-$0.11$\pm$0.04 dex for the bulge.

High resolution (R=45,000 and 67,000) echelle spectra of 25 bulge and 2
non-bulge K giants in Baade's Window, by Fulbright, McWilliam \& Rich 
(2006; 2007 henceforth FMR06, FMR07) performed model atmosphere abundance
analysis, improving upon MR94.  
The average difference of MR94 minus FMR07 [Fe/H] values was 0.02 dex,
but for the 6 stars above [Fe/H]=$-$0.34 dex, FMR07 were higher, on average,
by 0.05 dex.  Since 25 [Fe/H] measurements of bright bulge K giants is 
still too few (and too biased) to measure the bulge IDF, FMR06 correlated their
[Fe/H] measurements against the Sadler et al. (1996) and R88 metallicities, 
and found a mean Baade's Window [Fe/H] of $-$0.10$\pm$0.04  dex.

Zoccali et al. (2008) performed model atmosphere abundance analysis on 
high-resolution spectra of 800 bulge K giants at three latitudes.  They
found a large range of [Fe/H], from $-$1.5 to $+$0.5 dex, a mean near
the solar [Fe/H], and confirmed the vertical metallicity gradient, at 
0.6 dex per kpc, previously found by Minniti (1995) and Terndrup (1988).
Although Gonzalez et al. (2011) focused on alpha-element trends, their
[Fe/H] values agreed with this vertical metallicity gradient (see also
Uttenthaler et al 2012).

The preceding methods all measured, or estimated, the compositions of 
mostly bright bulge K giants, but resulted in a selection bias against
the most metal-rich stars.  Metal-rich giants form TiO more readily in 
their atmospheres than metal-poor giants, particularly for the coolest,
most evolved, giants at high luminosity.  Thus, while bright K giants
are more amenable for high resolution abundance analysis, they systematically
sample the low side of the MDF.  This problem is exacerbated by increased
TiO blanketing, due to the enhancement of Ti, in solar metallicity bulge stars.
The solution to this bias is to study the chemical composition of Red Clump
(henceforth RC) giant stars, since they are considerably warmer than the
cool K and M giants, yet the RC samples all metallicities.  The small scatter
in RC luminosities reduces the chance of foreground contamination from the disk,
which is a problem for bright K giants.

The excellent study by Hill et al. (2011), based on high-resolution 
spectra of 219 red clump (RC) stars in Baade's Window included [Fe/H] 
and [Mg/H] measurements.
The RC stars allowed improved bulge identification and more fully 
encompassed the bulge metallicity range, compared to luminous K giants 
of some previous studies.  They found an asymmetric [Fe/H] distribution,
with mean and median values of $+$0.05 and $+$0.16 dex respectively.   This 
asymmetric [Fe/H] distribution was de-composed into two Gaussian 
components, with average [Fe/H] centered at $-$0.30 and $+$0.32 dex.  
A Simple closed-box model consisting of two components fit the results.
Hill et al. (2011) also speculated that the bulge vertical metallicity
gradient could be explained by the changing dominance of one
sub-population to the other with increasing latitude.

\begin{center}
\begin{figure}[ht!]
\includegraphics[width=3.0in]{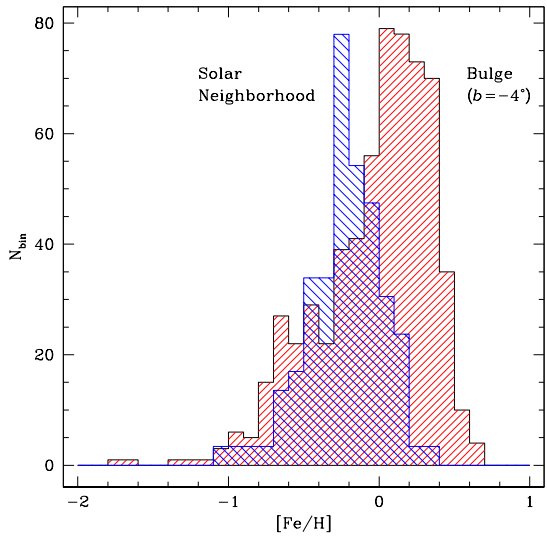}
\caption{
A comparison of the [Fe/H] distribution function in the solar 
neighborhood (blue-filled histogram) for stars within 15pc
from Allende Preito et al. (2004, scaled to the peak number of bulge
stars) compared to RC stars in the bulge,
at {\it b}=$-$4$^{\circ}$ (red-filled histogram), from Gonzalez et al. (2015)
and Hill et al. (2011).  The average and median separations are 0.19 dex
and 0.28 dex, respectively.  }
\label{fig-mdf}
\end{figure}
\end{center}

Gonzalez et al. (2015) measured [Fe/H] from high-resolution spectra of
400 bulge RC stars, near {\it b}=$-$4$^{\circ}$, at four positions
in longitude between {\it l}=$-$10$^{\circ}$ to $+$10$^{\circ}$.  They
found no variation in the IDF with longitude.   The global IDF was
well fit by two Gaussians, with average [Fe/H]=$-$0.31 and $+$0.26 dex, 
similar to the Hill et al. (2011).  The combined IDF from the 
Hill et al. (2011) and Gonzalez et al. (2015) RC results are presented 
in Figure~\ref{fig-mdf}.

The [Fe/H] distribution for RC stars over multiple {\it b}=$-$4$^{\circ}$
fields, shown in Figure~\ref{fig-mdf}, indicate a mean [Fe/H] of $+$0.06 dex, but
a median [Fe/H] of $+$0.15 dex; the difference is due to the asymmetry
towards the metal-poor sub-population.  The Solar Neighborhood histogram
in Figure~\ref{fig-mdf} is from the study of 118 stars within 15pc of the sun
(nearly all dwarfs) by Allende Prieto et al. (2004)
and dominated by the thin disk; both the mean and median [Fe/H] are 
$-$0.13 dex.  In Figure~\ref{fig-mdf} the number of stars in the Solar
Neighborhood sample has been scaled up by a factor of 3.4 so that the peak
bin matches the bulge sample.

Casagrande et al. (2010) revised the photometric temperature scale for
dwarf stars up by 100K, from the Alonso et al. (1999) temperature scale employed
by in the Solar Neighborhood study of Allende Prieto et al. (2004).
However, Allende Prieto et al. (2004) found good agreement of
their photometric T$_{\rm eff}$ values with those computed from
H$_{\beta}$ line profiles.  An increase in T$_{\rm eff}$ of 100K
typically results in an increase in [Fe/H] by $\sim$0.1 dex.

From a recalibration of the Str\"omgren photometric system, based
on their higher temperature scale, Casagrande et al. (2011) found
the median [Fe/H] of Solar Neighborhood stars at $-$0.05 dex, some
0.08 dex higher than the Allende Prieto et al. (2004) result.  If this
correction is applied, then the {\it b}=$-$4$^{\circ}$ bulge
average and median [Fe/H] values
are higher than the nearby thin disk by $+$0.11 dex and $+$0.20 dex,
respectively.

The [Fe/H] abundance ratios for the 58 lensed bulge dwarf stars, studied by
Bensby et al. (2013), reveal remarkably similar IDF envelope to the RC giants 
in Hill et al.  (2011).  A larger sample of lensed dwarf stars would enable
more detailed comparison, particularly for the position of the [Fe/H]
peaks of sub-populations.

From R=11,000 spectra of 28,000 mostly RC stars in various bulge fields,
Ness et al. (2013) found [Fe/H] ranging from $-$2.8 to $+$0.6 dex.  
Multiple [Fe/H] sub-populations were identified, with the three most 
prominent at [Fe/H] = $+$0.15, $-$0.25 and $-$0.71 dex for a latitude 
of {\it b}=$-$5$^{\circ}$; at this latitude the median [Fe/H] is 
$-$0.12 dex.  While the peaks of these individual sub-populations vary 
slowly with latitude, the overall vertical metallicity gradient is 
obtained due the changing proportions with latitude, similar to the
suggestion of Hill et al. (2011).  Based on the change in median
[Fe/H] of their {\it b}=$-$5$^{\circ}$ field, at$-$0.12 dex, to
$-$0.46 dex for their {\it b}=$-$10$^{\circ}$ field, and a Galactic
center distance of 8.0 kpc, a vertical [Fe/H] gradient of 0.48 dex/kpc
is obtained, quite similar to Zoccali et al. (2008), who obtained 0.6 dex/kpc.
See Ness \& Freeman (2016, this volume)
for a further discussion of their bulge MDF and APOGEE results.

The MDF as measured by the [Mg/H] distribution offers the advantage
that Mg is dominated by SNII progenitors, whose lifetimes are relatively 
short.  Therefore, with Mg there is no need to account for production on
long timescales, as occurs with the delayed production of Fe from SNIa;
thus, Mg chemical enrichment is closer to the instantaneous recycling
approximation than Fe.

\begin{center}
\begin{figure}[ht!]
\includegraphics[width=3.0in]{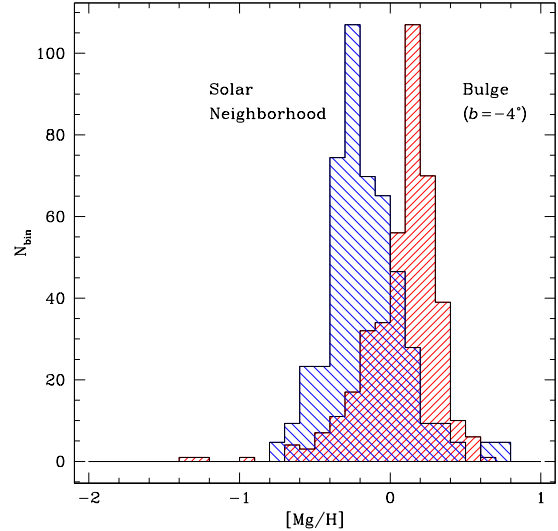}
\caption{
A comparison of the [Mg/H] distribution function in the solar 
neighborhood (blue-filled histogram) for stars within 15pc
from Allende Preito et al. (2004, scaled to the peak number of bulge stars)
compared to RC stars in the bulge,
at {\it b}=$-$4$^{\circ}$ (red-filled histogram), from Gonzalez et al. (2015)
and Hill et al. (2011).  The average and median separations are 0.24 dex
and 0.33 dex, respectively.  }
\label{fig-mgmdf}
\end{figure}
\end{center}

In Figure~\ref{fig-mgmdf} the separation in [Mg/H] between the Solar
Neighborhood and bulge at {\it b}=$-$4$^{\circ}$ is even greater than 
for [Fe/H]; the average and median [Mg/H] of the bulge are 0.24 dex
and 0.33 dex higher, respectively, than Solar Neighborhood.  These
differences are much larger than the possible 0.08 dex correction
expected if the hotter temperature scale of Casagrande et al. (2010, 2011)
had been used in the Allende Prieto et al. (2004) analysis of nearby 
thin disk stars.

These plots show that the bulge yield of Fe and Mg are significantly
higher than the in MW thin disk.  Clearly, with the MW thick disk mean 
[Fe/H] at $-$0.7 dex (Gilmore et al 1995) the bulge yield also exceeds
the thick disk's.

\subsection{Bulge Alphas Overview}

\begin{center}
\begin{figure}
\includegraphics[width=3.5in]{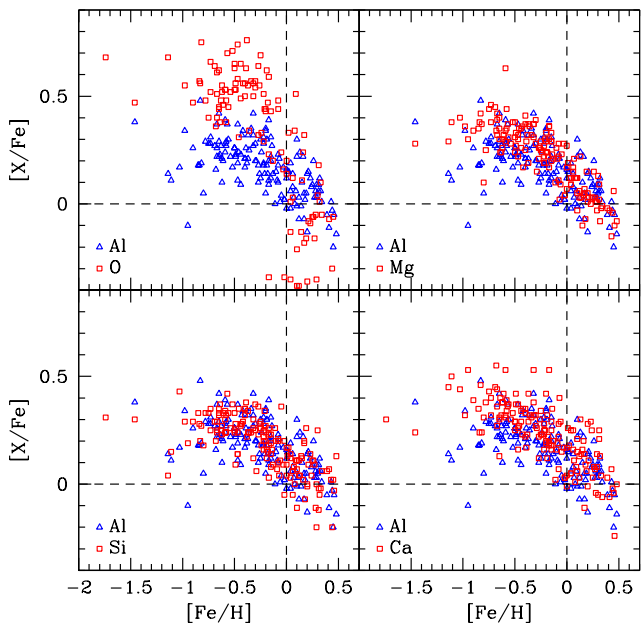}
\caption{ 
A plot inspired by Johnson et al. (2014), showing [$\alpha$/Fe] versus
[Fe/H] in the bulge for O, Mg, Al, Si and Ca.  All 5 elements are enhanced
by $\sim$0.15 dex at [Fe/H]=0.0 dex, and all trends show the classic decline
in [$\alpha$/Fe] with increasing metallicity, but the [O/Fe] range is greater
than the other alphas.  Aluminum shows a clear alpha-like trend, consistent
with its production in massive stars.  }
\label{fig-j14alphas}
\end{figure}
\end{center}

An overview of some of the features of the bulge [$\alpha$/Fe] trends 
are displayed in Figure~\ref{fig-j14alphas}, inspired by a plot from
Johnson et al. (2014).
This figure shows three elements (O, Mg and Al) principally made in the
hydrostatic phase of massive stars, and two elements (Si and Ca) mostly
produced during explosive nucleosynthesis of SNII events, although
some small production also occurs in SNIa events.  Thus, O and Mg
are hydrostatic alpha-elements while Si, S, Ca, Ti are explosive 
alpha-elements.  Significant production of Al, Na and Cu occurs in
massive star progenitors to SNII events, synthesized in hydrostatic
phases, so they are hydrostatic elements; notably, their production
is thought to be sensitive to metallicity (e.g., WW95; Nomtoto et al 2006).

Obvious in Figure~\ref{fig-j14alphas} is the fact that all 5 elements are
over-abundant in [X/Fe]
at lower metallicity and decline roughly linearly with increasing [Fe/H],
similar to the alpha-element trends seen from MW halo to thick and thin disks.
At [Fe/H]=0.0 all 5 elements are overabundant, by $\sim$0.15 dex, suggesting
a higher SFR in the bulge than the solar neighborhood.  

The use of [Al/Fe] as a reference in Figure~\ref{fig-j14alphas} shows
that Mg, Si and Ca have roughly the same trends in the bulge, and also
that [Al/Fe] displays an alpha-like trend.  While Al is not 
often appreciated to be an alpha-element, its alpha-like trend 
reflects the fact that it is mostly produced by massive stars.  The
alpha-like trend for Al is also seen in the MW thick and thick disk
abundance results of Bensby et al. (2005) and Reddy et al. (2006).
Notably, the predicted metal-dependent Al yields (e.g., Arnett 1971; WW95)
 are not apparent in the [Al/Fe] trend seen in the bulge.

Finally, the low-metallicity [O/Fe] ratio in the bulge is higher than
the other 4 elements, and declines more steeply, in Figure~\ref{fig-j14alphas}.
Although the Johnson et al. (2014) [O/Fe] are slightly higher than other
studies, and their measurement uncertainties larger, all show this effect,
indicating that O behaves differently and
must have evolved differently.

\subsection{Oxygen}

The first bulge oxygen abundances, measured by MR94 for 6 bulge K 
giants, were quite noisy due to the low resolution (R$\sim$17,000),
low S/N, spectra, blending with an Sc~II line and lack of identifiable 
continuum.  The [O/Fe] ratios also suffered from a 0.20 dex zero-point 
error, due to the high solar oxygen abundance reference (Anders \& 
Grevesse 1989).  The average [O/Fe] from MR94, corrected for the 
zero-point systematic error, near $+$0.3 dex, is roughly consistent 
with present-day values.

McWilliam \& Rich (2004, henceforth MR04) performed a high-resolution 
(R$\sim$35,000--60,000), model atmosphere, abundance analysis of 10 bulge 
RGB stars.  The [O/Fe] results were arguably consistent with the MW 
thick disk for metallicities below solar. Above solar [Fe/H] the [O/Fe]
values appeared to plummet, as if no further oxygen production occurred
above [Fe/H]=$-$0.5 dex.  The data used by MR04 was a subset of 
FMR07, ultimately 
re-analyzed by McWilliam, Fulbright \& Rich (2010; henceforth MFR10). 
However, the re-analysis by MFR10 gave higher [O/Fe] than MR04 for the 
super-metal-rich (henceforth SMR) stars, and the steep, zero-oxygen 
production, slope claimed by MR04 was not evident.

\begin{center}
\begin{figure}
\includegraphics[width=3.0in]{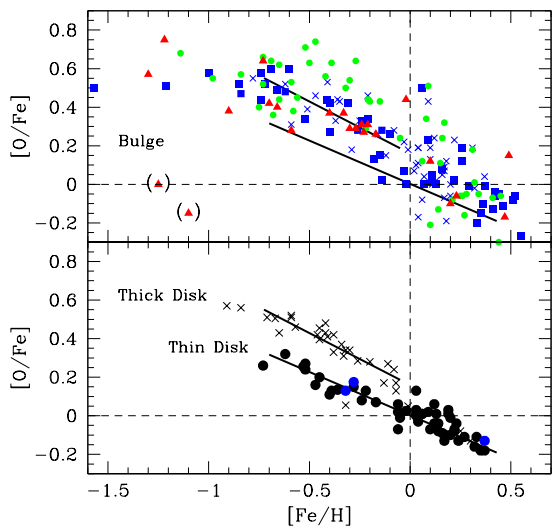}
\caption{{\it Top panel:} [O/Fe] versus [Fe/H] in the Galactic bulge. 
Green filled circles: Johnson et al. (2014) RC stars at R$\sim$22,000; 
blue crosses: giants from Lecureur et al. (2007), at R$\sim$47,000; 
filled red triangles: McWilliam, Fulbright, \& Rich (2010) re-analysis
of Fulbright et al. (2007) EWs for RGB stars at R$>$45,000;
filled blue squares: Bensby et al. (2013) lensed dwarfs at R$>$40,000.\quad
{\it Bottom panel:} [O/Fe] versus [Fe/H] in the thick disk (crosses) 
and thin disk (filled circles), from Bensby et al. (2005).  The 
thick/thin disk lines, reproduced in the top panel, are estimated, not 
fit.}
\label{fig-ofe}
\end{figure}
\end{center}

Rich \& Origlia (2005) performed the first detailed abundance study of 14 bulge
M giant stars in Baade's Window, using R=25,000 Keck-NIRSPEC spectra at 
1.5--1.8$\mu$m.  The M giants covered a relatively narrow range in [Fe/H], 
with the mean [Fe/H] at $-$0.19 dex, and no stars above
[Fe/H]=$-$0.03 dex.  Their [O/Fe] ratios, with oxygen abundances derived from
OH lines, were uniformly high, with a 1$\sigma$ scatter of 0.07 dex, and a mean
value of $<$[O/Fe]$>$=$+$0.32$\pm$0.02 dex.  The location of the mean
fits very well with the present-day results shown in Figure~\ref{fig-ofe}.
Similar enhancements of [O/Fe] and [Ti/Fe] were also found using near-IR
echelle spectra of a handful of Baade's Window RGB stars by Cunha \& Smith (2006).

Zoccali et al. (2006) performed an high-resolution detailed abundance study
of [O/Fe] versus [Fe/H] for 50 bulge K giants using S/N=50, R=45,000, VLT UVES 
spectra.  Like all optical studies, they employed the [O~I] line at 6300\AA , a
line that is very robust against deviations from LTE.  Zoccali et al. (2006)
found [O/Fe] declining with increasing [Fe/H], parallel to the MW thick disk 
trend, but enhanced by $\sim$0.10 dex.

FMR07 and Lecureur et al. (2007) measured [O/Fe] for bulge giant stars
in Baade's Window.  Both studies employed high resolving power
(R$\sim$45,000 to 67,000) spectra, and both reached similar conclusions: that
[O/Fe] is over-abundant in Baade's Window bulge stars relative to both the MW thin
and thick disks;  the [O/Fe] trend decreases with increasing [Fe/H], parallel to,
but enhanced by $\sim$0.1 dex compared to the MW thick disk trend, similar to
the results of Zoccali et al. (2006). 

An important point is that the FMR07 study employed a line-by-line
differential abundance technique, relative to the nearby red giant
$\alpha$~Boo.

Melendez et al. (2008), based on high resolving power near-IR
spectra, R=50,000, at 1.55$\mu$, combined with published EWs from the
optical spectra of FMR07.  Melendez et al. (2008) also acquired
spectra of MW thick and thin disk RGB stars for comparison, using the same
spectrograph.  While Melendez et al. (2008) found a mean difference
with the [O/Fe] of FMR07 of $+$0.03$\pm$0.13 dex, their [O/Fe] trend
was consistent with the MW thick disk.  Inspection of the results
shows that the FMR07 bulge stars near [Fe/H]=$-$0.5 dex are near
[O/Fe]$\sim$$+$0.5 dex and $\sim$0.10 dex higher than the thick disk
comparison, whereas for Melendez et al. (2008) the bulge stars are
only marginally higher than the thick disk.  At solar [Fe/H],
Melendez et al. (2008) and FMR07 found [O/Fe]$\sim$$+$0.2 dex,
while Lecureur et al. (2007) and MFR10 found [O/Fe]$\sim$0.10 dex.

Melendez et al. (2008) found that the choice of model atmosphere
[$\alpha$/Fe] ratio has a significant effect on the derived [O/Fe]
ratio, with a difference in [$\alpha$/Fe]=$+$0.2 dex leading to an
increase in [O/Fe] of $\sim$0.1 dex.

A re-analysis of the FMR07 EWs by MFR10 found that the [O/Fe] trend 
with [Fe/H] in bulge RGB stars, is indeed, consistent with the MW thick
disk, as claimed by Melendez et al. (2008).

The chemical composition of 58 microlensed bulge dwarf and subgiant 
stars were measured by Bensby et al. (2010, 2013; henceforth B13), 
using R=40,000, S/N$\sim$50, UVES VLT spectra.  Notably, B13 employed 
published non-LTE corrections to Fe and O abundances (e.g. Lind, 
Bergemann \& Asplund 2012).  The oxygen abundance measurements
employed the allowed O~I triplet lines at 7772, 7774, and 7775\AA , 
with excitation potential of the lower level of 9.14 eV.  These lines 
are sensitive to the adopted temperature, arising from a small fraction
of the total oxygen population; thus, the correction for non-LTE effects 
is important.  Because the solar lines are used in the calibration, the
differential non-LTE corrections, relative to the sun, are more 
important than the absolute non-LTE corrections.

The 6300\AA\ [O~I] lines in the B13 dwarf stars are too weak 
for reliable abundance measurement, at the S/N of
the spectra; furthermore, the Ni~I blend has a significant impact on 
the derived abundance.  This is unlike the situation for RGB stars, 
where the [O~I] line is stronger (30--60m\AA ) and the Ni~I blend is 
negligible, by comparison.

B13 claimed that their bulge alpha-element abundance ratios
show trends that are similar to the nearby thick disk, but 
the knee in the alpha-trends (other than oxygen) occur at a
higher [Fe/H], by $\sim$0.10 dex, than the local thick disk (near
[Fe/H]=$-$0.3 to $-$0.2 dex).  The B13 results indicate that the
[O/Fe] knee occurs near [Fe/H]=$-$0.6 dex, followed by a steep, linear,
decline with [Fe/H].  Although the decline in [O/Fe] versus [Fe/H] is 
steep, the slope is shallower than would result if no oxygen had been 
produced.

Johnson et al. (2014; henceforth J14) analyzed S/N=70, R=20,000, GIRAFFE fiber spectra of
156 RGB and RC stars at $(l,b)$=($+$5.25,$-$3.02) and (0,$-$12), employing 
both spectrum synthesis and EW abundance analysis techniques.  Like other
abundance studies of evolved bulge stars, oxygen abundances were derived 
from the [O~I] line at 6300\AA .  Given the relatively low spectral
resolution, which results in line blending and poor continuum definition,
it is not surprising that the dispersion in element abundances is greater
than other studies at higher resolving power; in particular, the scatter
is relatively large above [Fe/H]=0.00.  Still, the work was expertly
done and the results remarkably similar to other studies.  The [O/Fe]
values form a plateau near [O/Fe]=$+$0.55 dex below [Fe/H]$\sim$$-$0.25
dex, followed by a decline in [O/Fe] with increasing [Fe/H].  The knee in the
plateau is located at a higher [Fe/H] than the thick disk plateau,
indicating a high SFR and rapid chemical enrichment timescale.  
This high-metallicity knee is similar to the general alpha-element 
trends found by B13, but at higher [Fe/H] than Bensby et al's oxygen
plateau.  It is notable that both fields in J14 
show the same [X/Fe] trends with [Fe/H], although the stars in the 
$b$=$-$12$^{\circ}$ field have lower overall [Fe/H] than the $b$=$-$3$^{\circ}$ 
field, consistent with the vertical metallicity gradient.  

The results of J14 indicate [O/Fe]=$\sim$$+$0.2 dex at solar metallicity; however,
there is considerable scatter
above [Fe/H]=0.00, which confuses this value.  The [O/Fe] ratio appears to decline
from $+$0.55 to roughly 0.00 dex between [Fe/H]=$-$0.25 and $+$0.20; although, 
an intercept with solar composition at [Fe/H]=$+$0.30 dex is not unreasonable.
If the latter is true, then there appears to be no production of oxygen
above [Fe/H]=$-$0.25 dex.  If the steeper [O/Fe] slope is correct, then
the composition at [Fe/H]=$-$0.25 is not related to that at [Fe/H]=$+$0.20
dex, or there is a systematic measurement error.  

Taken at face value, the simplest interpretation of the J14 
[O/Fe] trend is that oxygen production in the bulge ceased after
[Fe/H]=$-$0.25 dex, similar to the suggestion of MR04. 
However, the scatter in J14 and the revision of the MR04 scale by MFR10
suggests that such excessively steep [O/Fe] slopes are due to measurement
error about the already steep slope.


A visual summary of the observational situation for the bulge [O/Fe] 
trend, and comparison 
with the MW thick and thin disks, is shown in Figure~\ref{fig-ofe}.  
The top panel shows measured [O/Fe] ratios in bulge RGB stars from J14,
FMR10 and Lecurer et al. (2007) and the lensed bulge dwarfs of B13.  The 
lower panel shows [O/Fe] for dwarf stars in the MW thick and thin disks
measured by Bensby et al. (2005).  Lines in the lower panel of 
Figure~\ref{fig-ofe}, showing the approximate trends for thick and thin
disks, are reproduced in the upper panel for comparison with the bulge 
trend.  Obviously, as first found by Melendez et al. (2008), the
bulge [O/Fe] trend follows that of the thick disk, and lies roughly 0.15
dex above the thin disk trend; indeed, the trend line shows that, at
solar [Fe/H], the thick disk [O/Fe] is $+$0.16 dex.
Note that the thick disk [O/Fe] trend 
above solar [Fe/H] is not clear: a few putative thick disk stars lie 
co-incident with the thin disk line at [Fe/H]$\sim$$+$0.2 dex; but, are
these really from thick disk, or simply mis-identified thin disk members?.
On the other hand, the bulge stars seem to continue a linear decline of
[O/Fe] versus [Fe/H] up to [Fe/H]$\sim$$+$0.5 dex, passing through solar
[O/Fe] near [Fe/H]=$+$0.25 dex.  The SMR bulge stars lie above the thin
disk trend, but the separation from the thin disk trend here is smaller
than at sub-solar [Fe/H].  Future, improved measurements of [O/Fe]
versus [Fe/H] up to [Fe/H]$\sim$$+$0.5 dex would be useful for
understanding whether the bulge approaches the thin disk trend at high
metallicity.

Inspection of Figure~\ref{fig-ofe} suggests that a metal-poor plateau of
[O/Fe] exists near $+$0.52 dex; but, the data are consistent with a
mean ranging from $+$0.47 to $+$0.57 dex.  Notwithstanding, a handful
of the B13 lensed dwarfs show [O/Fe] near $+$0.60 dex.  It is notable
that the range of the [O/Fe] plateau in the bulge is similar to the
two most metal-poor stars in the Bensby et al. (2005) thick disk sample.
However, [O/Fe] measurement of thick disk stars by Nissen et al. (2002)
and Ramirez et al. (2013) find [O/Fe] of $+$0.47 and $+$0.45 dex, 
respectively, for the plateau between [Fe/H]=$-$1 to $-$2 dex.  To avoid
a confusing figure we do not show these results in Figure~\ref{fig-ofe}.
Thus, while the bulge and thick disk [O/Fe] plateaus agree to within 
reasonable systematic measurement uncertainties, it appears that the
average [O/Fe] of bulge plateau stars is 0.05 dex higher than for the
thick disk, but the bulge could be 0.10 dex higher, mostly due to the
high values of J14. Clearly, improved
[O/Fe] measurements of the plateau is desirable.

In the context of the  SNIa time-delay scenario of Tinsley (1979) and 
MB90: if the [O/Fe] plateau in bulge and thick disk are the same, then
the IMF of the bulge and thick disk are equal.  However, the plateau 
difference of 0.05 dex suggests that, in fact, the 
bulge IMF included slightly more massive stars than the 
thick disk, but the [O/Fe] difference is within measurement
uncertainties.  The plateau difference is not enough to explain the observed 
bulge/thick disk [O/Fe]=$+$0.16 dex at solar [Fe/H], and why the bulge
does not reach solar [O/Fe] until [Fe/H] near $+$0.25 dex.  From these 
measurements, the SNIa time-delay idea (see Figure~\ref{fig-mb90}) suggests 
that the SFR during bulge and thick disk evolution must have been higher 
than the thin disk, and the formation timescale correspondingly shorter.
In this way, the bulge [O/Fe] ratios indicate that the bulge 
chemical enrichment timescale was faster than for the thin disk, but
similar to the thick disk (although the thick disk stopped forming 
stars near solar [Fe/H]).

%
%

As discussed later, an important feature of the [O/Fe] trend in the 
bulge, especially evident in the B13 results, and indeed, the
MW thin and thick disks, is that oxygen has a larger [X/Fe] decline with
metallicity than other alpha elements.  This cannot occur in the SNIa 
time-delay scenario if alpha-element [X/Fe] ratios decline only by the
addition of Fe from SNIa; an additional mechanism must be at work.
Two possible methods for this extra [O/Fe] decline are: 1. reduction
of the O yield via stellar wind mass-loss, and 2. a steepening of the
upper-end of the IMF with increasing metallicity.

\subsection{Magnesium}


Like oxygen, it is thought that most magnesium production is associated with 
SNII events,
during the hydrostatic evolution of the progenitor star.  Thus, the evolution
of both Mg and O should follow the expectations of the SNIa time-delay scenario
of Tinsley (1979) and MB90.  For this reason, the
trend of [Mg/Fe] with [Fe/H] should depend on the bulge SFR.  

Beginning with MR94, Lecureur et al. (2007), FMR07 and through to today's
excellent bulge Mg abundance studies by Hill et al. (2011), J14 and Gonzalez et al.
(2015, henceforth G15), [Mg/Fe] has been found to be enhanced
relative to the thick and thin disks, defined by dwarf stars.  The earlier
studies compared to the Edvardsson et al. (1993) MW disk dwarf star results,
rather than the MW disk dwarfs of Bensby et al. (2005, 2014); notably,
MR94, Lecureur et al. (2007) and FMR07
also over-estimated the size of the bulge Mg enhancements.

The [Mg/Fe] ratios measured by all MW bulge studies show enhancements at low
metallicity, and a decline with increasing [Fe/H].  However, there is an
unsatisfactory range in the reported [Mg/Fe] ratios; in particular, the
results of FMR07 are higher than more recent studies (e.g. G15, J14).

In the FMR07 abundance tables\footnote{The FMR07 tables show different
abundance ratios than the FMR07 figures, for some unknown reason.  I favor the
numbers presented in the tables.} [Mg/Fe] for
stars below [Fe/H]=$-$1 have a mean near $+$0.46 dex; subsequently, there is a
very shallow decline in [Mg/Fe] as [Fe/H] increases to $\sim$$-$0.3 dex.  From
[Fe/H] = $-$0.3 to $+$0.1 dex, [Mg/Fe] drops sharply, slightly above the MW thick
disk trend. Finally, for the 5 SMR stars in FMR07 with [Fe/H]$\sim$$+$0.2 to
$+$0.5 dex the mean reported [Mg/Fe] values is $+$0.28 dex.

\begin{center}
\begin{figure}
\includegraphics[width=3.0in]{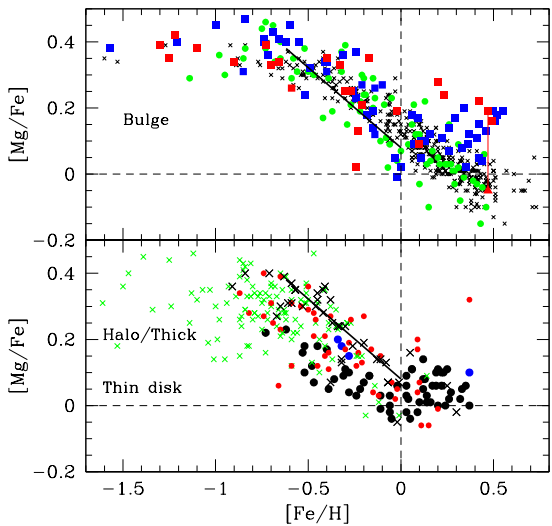}
\caption{{\it Top Panel:} [Mg/Fe] versus [Fe/H] in the Galactic bulge. 
Small black crosses: RC star results from
Gonzalez et al. (2015), at R$\sim$22,000; 
filled red squares: Fulbright et al. (2007) RGB stars at R$>$45,000,
including a $-$0.07 dex zero-point correction (this work);  filled red triangle: re-analysis
(this work) of I-025 RGB star from data from Fulbright et al. (2007); 
other symbols are the same as Figure~\ref{fig-ofe}.
{\it Bottom Panel:} [Mg/Fe] versus [Fe/H] for MW thin disk (black filled circles) and
thick disk (large black crosses) from Bensby et al. (2005). Halo and thick disk stars 
(green crosses) and thin disk stars (filled red circles) are from Reddy et al. (2006).
The black line is the estimated linear trend for the thick disk from the 
Bensby et al. (2005) results, reproduced in the top panel.
}
\label{fig-mgfe}
\end{figure}
\end{center}

In contrast to FMR07, the bulge [Mg/Fe] trend 
defined by J14 and G15 (see Figure~\ref{fig-mgfe}) shows 
[Mg/Fe]$\sim$$+$0.37 dex for the metal-poor stars, flat up to roughly 
[Fe/H]=$-$0.6 dex, followed by a linear decline toward increasing 
metallicity.  At solar metallicity the mean G15 and J14 [Mg/Fe] ratio 
is $+$0.15 dex.  
Thus, the J14 and G15 [Mg/Fe] ratios are lower than 
FMR07 at low [Fe/H] by 0.09 dex; but, at the metal-rich end,
for stars near [Fe/H]$\sim$$+$0.5 dex, the J14 and G15 points
are lower than the original FMR07 values by $\sim$0.20 dex.

In order to understand the difference between FMR07 and J14/G15 
[Mg/Fe] trends, for this review, I have investigated the FMR07 results,
using original spectra.  In particular, I re-consider the analyses for the
Arcturus standard star and for bulge star I-025, with [Fe/H] values 
near $-$0.5 and $+$0.5 dex, respectively.

In the re-analysis, I find [Mg/Fe]=$+$0.32 dex for Arcturus, which is lower
than FMR07 by 0.07 dex and higher than the J14 result by 0.02 dex.  Since
FMR07 employed Arcturus as a standard, this removes a zero-point difference
of 0.09 dex between FMR07 and J14 [Mg/Fe] ratios.  This change
is mostly due to the strongly damped wings of the Mg~I lines in the sun, 
used in the line-by-line differential abundance analysis of Arcturus relative 
to the sun.
Because
of the absence of damping constants for their Mg~I lines, the Uns\"old
Van der Waals damping constants were default in FMR07; but the Uns\"old damping
constants for the red Mg~I lines used by FMR07 under-predict the wings
in the solar spectrum.

An important Mg~I abundance indicator, employed by several bulge 
studies, is the 6319\AA\ triplet, which is, unfortunately, blended with a 
very broad, shallow, Ca~I auto-ionizing line.  
For giant star spectra, the Ca~I auto-ionization line profile can be reasonably well
treated as a pseudo-continuum, in which the Mg~I lines sit, and an EW abundance
analysis, as performed by FMR07, give good results.  This works well for the bulge giants 
and the FMR07 standard star, Arcturus.  However, in the solar spectrum the
local pseudo-continuum merges the wide Mg~I line wings into the Ca~I auto-ionizing 
line profile, resulting in a degeneracy between Mg and Ca profiles.  In a
pseudo-continuum analysis the depth of the Mg~I lines is under-estimated in 
the sun, leading to an 
over-estimate of the Mg abundance of Arcturus relative to the sun.  Unfortunately,
the Ca~I auto-ionizing line wavelengths and damping constants are only 
approximately known (e.g.  Newsom 1968), resulting in additional covariances.  

In the re-analysis performed here, the Ca~I auto-ionization line is 
treated properly in the LTE line radiative transfer code (i.e., no pseudo-continuum).
I find a reasonable fit to the solar Ca~I auto-ionization line profile 
using a wavelength of 6318.209\AA , a radiative damping constant 
$\Gamma$=1.095$\times$10$^{12}$ s$^{-1}$, and log\,gf$\epsilon$=6.52.  Notably,
the Mg~I 6319\AA\ triplet Van der Waals line damping constants are not
known and even the log\,gf values are highly uncertain.  I employed 
log\,gf$\epsilon$=5.567, 5.346, and 4.867 dex for the solar Mg~I lines at 
6318.72, 6319.24, and 6319.50\AA\ respectively, with Barklem damping 
constants copied from the Mg~I line at 9255.78\AA , which has a similar
upper-level energy.

Strongly damped wings are very obvious for most Mg~I lines in the red
region of the solar spectrum.  For these lines, the upper level of the transition
is close to the ionization potential; thus, the electron orbits
are large, leading to an enhanced collision cross section.  This results in a high
collisional rate and strong damping at the 
density of the solar atmosphere, but not for the relatively thin RGB atmospheres.  
For such lines, the strong solar line wings can depress the local continuum
and/or be mistaken for blending features.
In these cases, the EW of the solar Mg~I lines may easily be under-estimated, while 
in Arcturus and bulge giant spectra these difficulties do not arise, due to the absence of
significantly damped wings.  This effect would result in a high estimate for the
Arcturus Mg abundance relative to the sun.
%
%
%
%

The re-analysis of the super-metal-rich (SMR) bulge RGB star I-025, from FMR07 spectra,
resulted in a decrease from [Mg/Fe]=$+$0.24 reported by FMR07, to $-$0.05 dex, 
found here.  The analysis performed here relies heavily on synthesis of the 
6319\AA\ Mg~I triplet, because these lines are relatively unsaturated.  Other red 
lines of Mg are so strong, and the continuum so blended, in I-025, that Mg 
abundance changes of $\pm$0.1 dex occur with quite small differences in the assumed 
continuum level.

The result obtained here for I-025 is in excellent agreement with the J14, G15 and
even the MR94 [Mg/Fe] ratios seen near [Fe/H]=$+$0.5 dex, and fits the consensus
values better than the [Mg/Fe] value from FMR07.  
As a result of the lower [Mg/Fe] and [O/Fe] found here (and in MFR10) for star
I-025, a scaled-solar composition model atmosphere is appropriate,
instead of the alpha-enhanced atmosphere used by FMR07.  In 
that case, [Fe/H]=$+$0.47 according to FMR07.  Note that, for the differential
line-by-line analysis, relative to Arcturus and ultimately the sun, the
FMR07 and corrected FMR07 results are on a solar scale and independent 
of log\,gf values.

\begin{table*}[!btp]
\caption{Summary of Galactic Bulge [Mg/Fe] Measurements}
\begin{tabular}{lcccl}
\hline\hline       
Reference   &  [Mg/Fe]    &  [Fe/H]   &  [Mg/Fe] at   &   [Mg/Fe] at  \\
            &  Plateau           &  Knee          &  [Fe/H]=0.0   &   [Fe/H]=$+$0.5 \\
\hline                  
McWilliam \& Rich 1994 &  $+$0.41   &   $-$0.3            & $+$0.28    &   $+$0.1   \\
Lecureur et al. 2007   &  $+$0.46   &   $-$0.3            & $+$0.29    &   $+$0.1:  \\
Fulbright et al. 2007$^{a}$ & $+$0.36 &  $-$0.4:          & $+$0.18\rlap{$^{a}$}    &   $-$0.05\rlap{$^{a}$}  \\ 
Alves-Brito et al. 2010$^{b}$ & $+$0.37   &   $-$0.4 to $-$0.5  & $+$0.03    &   $+$0.04  \\
Hill et al. 2011       &  $+$0.36   &   $-$0.4 to $-$0.5  & $+$0.20    &   $+$0.06  \\
Bensby et al. 2013$^{c}$ &  $+$0.41   &   $-$0.5 to $-$0.6  & $+$0.15    &   $+$0.18  \\
Johnson et al. 2014    &  $+$0.37   &   $-$0.6 to $-$0.7  & $+$0.15    &   $-$0.08  \\
Gonzalez et al. 2015   &  $+$0.35   &   $-$0.7\rlap{$^{d}$}           & $+$0.15    &   $-$0.03  \\
\hline                  
\end{tabular}

\tabnote{{$a$ } Fulbright et al. (2007) entries were corrected in a partial re-analysis
performed for this review; see text for details.}
\tabnote{{$b$ } Alves-Brito et al. (2010) re-analyzed EWs of Fulbright et al. (2007).}
\tabnote{{$c$ } Bensby et al. (2013) abundances for lensed bulge dwarf stars, other studies were bulge
red giants.}
\tabnote{{$d$ } Gonzalez et al. (2015) measured the knee at [Fe/H]=$-$0.44 dex, based on the same data.}
\label{tab-mgfe}
\end{table*}

Table~\ref{tab-mgfe} shows my estimate of the parameters defining the [Mg/Fe] trends
of various studies, based on the published results.
Except for the dwarfs of B13, all the studies listed in Table~\ref{tab-mgfe} provide 
[Mg/Fe] and [Fe/H] values for Arcturus.  In this way, we may attempt to evaluate zero-point 
differences between these studies.  For example, FMR07 originally obtained 
[Mg/Fe]=$+$0.39 dex for Arcturus, but the value found in this review is $+$0.32 dex; thus, 
a $-$0.07 dex correction should be applied to the published [Mg/Fe] values of FMR07.  
Similarly, G15 found [Mg/Fe] for Arcturus of $+$0.22 dex;
Johnson et al. (2014) obtained $+$0.30 dex; for Hill et al. (2011) it was $+$0.38 dex, while
Lecureur et al. (2007) found $+$0.32 dex, and MR94 $+$0.30 dex.  The implied correction for
the Gonzalez et al. (2015) [Mg/Fe] values, of $+$0.10 dex, seems too large in 
Figure~\ref{fig-mgfe}, and suggests random error.

As for the knee in the [Mg/Fe] versus [Fe/H] trend: a point near [Fe/H]=$-$0.6 dex would
reasonably agree with the most reliable of the published results.

Considering these issues, my best estimate of the [Mg/Fe] plateau and [Mg/Fe] at solar [Fe/H]
are $+$0.39$\pm$0.02 dex and $+$0.17$\pm$0.02 dex respectively.  
For [Fe/H]=$+$0.50 dex my best estimate of the bulge [Mg/Fe] is $-$0.05 dex; however, scatter
between different studies suggests that further investigation of these fiducial points
would be useful.

Figure~\ref{fig-mgfe} shows the uncorrected [Mg/Fe] versus [Fe/H] 
results from the studies of Bensby et al. (2013), Johnson et al. (2014),
and Gonzalez et al. (2015); small zero-point shifts to these studies 
are probably required (as discussed earlier) to place these onto the 
same scale and reproduce my best estimate of the [Mg/Fe] at the 
plateau, knee, at [Fe/H]=0.0 and $+$0.5 dex.  Also included in 
Figure~\ref{fig-mgfe} are results from Fulbright et al. (2007), with 
the $-$0.07 dex zero-point correction applied, as well as the re-analysis 
of star I-025 performed here.

The lower panel of Figure~\ref{fig-mgfe} shows the MW thin disk, thick disk, 
and some halo results for solar neighborhood dwarf stars analyzed by Bensby et al. (2005)
and Reddy et al. (2006).  while there is good agreement for solar metallicity
thin disk stars, the Reddy et al. (2006) thick disk points appear to suffer from
more contamination by thin disk stars, plus a larger dispersion than the Bensby et al. (2005)
results.  Reddy et al. (2006) had a much larger sample of metal-poor thick disk and 
halo stars than Bensby et al. (2005).  Notwithstanding, it appears that the Bensby et al. (2005)
metal-poor (below [Fe/H]$\sim$$-$0.6 dex) plateau has higher [Mg/Fe] than the 
Reddy et al. (2006) value.  A high estimate of the plateau from Reddy et al. (2006) 
is [Mg/Fe]=$+$0.34 dex, whereas the Bensby et al. (2005) thick disk data are consistent
with the bulge plateau, at [Mg/F]=$+$0.39$\pm$0.02 dex.  In this way, difficulties
in measuring the [Mg/Fe] ratios for solar neighborhood thick disk stars interferes
with our interpretation of the relative IMF of the bulge.

The line in the lower panel of Figure~\ref{fig-mgfe} indicates the thick disk
trend from the Bensby et al. (2005) results, which is slightly higher, by 
$\sim$0.05 dex, than would be obtained from the Reddy et al. (2006) values.
This line is reproduced in the upper panel of Figure~\ref{fig-mgfe}.

The most noticeable features of the bulge [Mg/Fe] trend shown in Figure~\ref{fig-mgfe}
is that, past the knee, the bulge [Mg/Fe] ratios lie above the Bensby et al. (2005)
thick disk trend (as noted by B13), by
about $+$0.05 dex, and at solar [Fe/H] the bulge has [Mg/Fe] of $+$0.17 dex, well
above the solar composition and about 0.09 dex higher than the Bensby et al. (2005)
thick disk.  The
solar [Mg/Fe] ratio is not reached in the bulge until [Fe/H]$\sim$$+$0.3 dex.

Perhaps, the simplest interpretation of the data is that the bulge 
[Mg/Fe] trend appears shifted to higher [Fe/H] compared to the solar 
neighborhood thick disk trend of Bensby et al. (2005), by approximately $+$0.16 dex
in [Fe/H].  This indicates a higher SFR in the bulge than the thick disk.
 Furthermore, the Bensby et al. (2005) metal-poor [Mg/Fe] plateau is similar
to the B13 bulge results, conceivably identical, suggesting a similar IMF.

However, if the Reddy et al. (2006) thick disk results are
correct, then the metal-poor plateau of the bulge has higher [Mg/Fe]
than the thick disk, by at least $+$0.05 dex and possibly $+$0.10 dex; 
but, the bulge [Mg/Fe] at solar [Fe/H] is at least $+$0.10 dex higher 
than the Reddy et al. (2006) thick disk.  Thus, it is possible that the
bulge is both shifted to higher [Mg/Fe] and higher [Fe/H] than the 
Reddy et al. (2006) thick disk, suggesting both an IMF weighted to 
higher mass stars and a higher SFR in the bulge.

Interestingly, at the most metal-rich end both the bulge and disk
[Mg/Fe] trends appear flat with [Fe/H], suggesting a convergence in Mg 
and Fe production.
However, at the metal-rich end the bulge points are sub-solar,
near [Mg/Fe]$\sim$$-$0.05 dex, whereas the Bensby et al. (2005) thin disk points lie 
above solar, near $+$0.05 dex.  Thus, the bulge shows a larger range of
[Mg/Fe] than the Bensby et al. (2005) thick plus thin disks.  

This difference in SMR [Mg/Fe] levels, between disk and bulge, may be due to systematic
measurement error, perhaps resulting from different Mg~I line treatment for dwarfs and
giants.  Indeed, the uncharacteristic increase in [Mg/Fe] with [Fe/H] for the B13 SMR
lensed dwarfs, compared to the RC and RGB bulge star studies, suggests that such an
error is likely.  However, if the difference in SMR [Mg/Fe] ratios is due to a genuine
abundance effect, then the sun is a relatively Mg deficient thin disk star and the SMR
bulge may have had a different IMF than the SMR thin disk.

Compelling similarities between bulge and thick disk abundance ratios
were first pointed out for [O/Fe] by Melendez et al. (2008), and extended to Na, 
Mg, Al, Si, Ca and Ti by Alves~Brito et al. (2010); although, this was
based on rather few data points.  More recently, Johnson et al. (2014) 
found bulge [O/Fe], [Mg/Fe], [Si/Fe] and [Ca/Fe] ratios,
overlapping but slightly above the thick disk trends, based on their large 
sample.  For Mg, at least, the bulge trend is measurably different
than the thick disk trend.  
 

\subsection{[O/Mg]: Stellar Winds, Element Yields, and WR Stars}

The production of both oxygen and magnesium is thought to be dominated by SNII progenitors
(e.g., WW95, Nomoto et al. 1997, Kobayashi et al. 2006),
during hydrostatic nuclear burning phases.  

Inspection of the bulge and disk points in Figures~\ref{fig-ofe} and \ref{fig-mgfe}
show that the oxygen plateau has higher [O/Fe], at $+$0.52 dex, than the magnesium
plateau [Mg/Fe] at $+$0.39 dex.
It is clear that the range of [O/Fe] is larger than the range of [Mg/Fe] for the galactic
bulge and disks; consequently, the trend of [O/Mg] must show a decline with
increasing [Fe/H].  

\begin{center}
\begin{figure}
\includegraphics[width=3.0in]{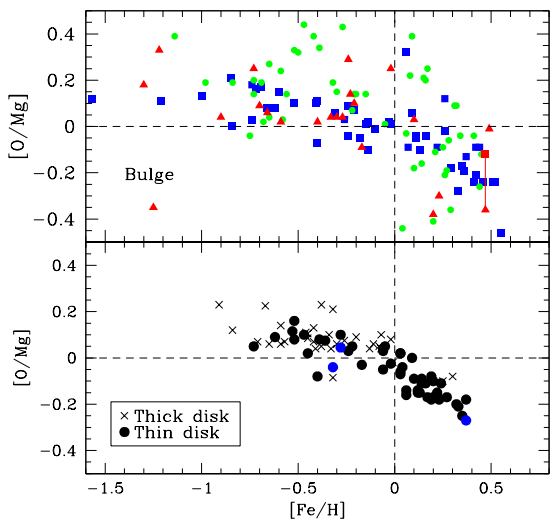}
\caption{{\it Top Panel:} [O/Mg] versus [Fe/H] in the Galactic bulge. Key to symbols is the same
as in Figure~\ref{fig-ofe}.  Note the $\sim$0.2 dex decline in [O/Mg] with increasing
[Fe/H]. {\it Bottom Panel:} The thick disk (crosses) and thin disk stars (filled circles)
show the same metal-dependent decline.  The trend is unchanged for the larger sample of
disk stars in Bensby et al. (2014).}
\label{fig-omg}
\end{figure}
\end{center}

Figure~\ref{fig-omg} shows the [O/Mg] decline with increasing [Fe/H] from
the galactic bulge chemical abundance studies compared to the thin and thick disk results
of Bensby et al. (2005).  Interestingly, the [O/Mg] trends with [Fe/H] 
for MW bulge and both thick and thin disks look remarkably similar, although 
the bulge data has larger scatter, presumably due to 
measurement error.  I note that if the upturn in the B13 bulge [Mg/Fe] ratios
(see Figure~\ref{fig-mgfe}),
for their subset of lensed dwarf stars above [Fe/H]=$+$0.2 dex, is 
assumed to be spurious, and if these are corrected to the median giant
star [Mg/Fe] trend, then the bulge [O/Mg] ratios would move closer to 
the disk [O/Mg] trend.

Clearly, Figure~\ref{fig-omg} shows that there is an additional reduction of [O/Fe] 
that is not seen in [Mg/Fe] (and other alpha elements).

Bensby et al. (2004) was the first to find this decline in [O/Mg] 
with [Mg/H] for MW thick and thin disk stars; the same result holds
for the 714 disk stars studied in Bensby et al. (2014).
While FMR07 noted different bulge trends for O compared to other
$\alpha$ elements, Lecureur et al. (2007) and McWilliam et al. (2008; henceforth M08)
emphasized the metal-dependent decline in [O/Mg] versus [Mg/H], [O/H], or [Fe/H] ratios,
seemingly the same in the galactic bulge and MW thick and thin disks.  

Because the [O/Fe] decline is greater than the decline in [Mg/Fe]
(or other alphas), it is not possible to explain the [O/Fe] and other [$\alpha$/Fe]
trends simultaneously with only the delayed addition of SNIa iron, as in the
SNIa time-delay scenario of Tinsley (1979) and MB90.

This steep decline in [O/Fe], in excess of other alpha-elements, indicates
a significant decline in the production of oxygen with increasing metallicity.
Although other alpha elements (e.g. Mg, Si, Ca, Ti) show
a smaller range of [X/Fe] than O, it could be argued that for Si, Ca and Ti
this is due to some small synthesis in SNIa.  However, Mg is not 
produced in significant qualities by SNIa (e.g., Nomoto et al. 1984; 
Fink et al. 2014) and has no detectable effect on the [Mg/Fe] ratio.  
Therefore, the decline
in [O/Mg] with increasing metallicity indicates a relative decrease in
the production of oxygen by SNII progenitors.

Based on the relatively steeper decline of [O/Fe] with [Fe/H] in the 
bulge, compared to other alpha-elements, McWilliam \& Rich (2004) 
suggested that metal-dependent winds from massive stars (that 
produce the Wolf-Reyet phenomenon, henceforth WR) may have played an
important role in
bulge chemical evolution.  McWilliam et al. (2008)
employed the chemical evolution model of the Galactic bulge constructed
by Ballero et al. (2007), but used the Maeder (1992) oxygen yields for
massive stars that included the effects of mass-loss.  Previous work
(e.g. Ballero et al. 2007) had used the WW95 SNII yields, where mass-loss
was not considered.  Although the Maeder (1992) calculations did not include
the SNII explosion this had no effect on the oxygen yields, because the
oxygen is produced in the hydrostatic phase of massive star evolution. 
The M08 calculations resulted in a slope of [O/Mg] versus [Mg/H]
that matched the slope from the bulge abundances of FMR07 and 
Lecureur et al. (2007), as well as the MW thin and thick disks.  However,
a zero-point shift in the computed [O/Mg] ratios, compared to the observations,
was evident, which may be accounted for with a slightly different bulge IMF.

A difficulty with the M08 calculations is that the Maeder (1992) mass-loss
rates are a factor of 2 to 3 higher than subsequent observations of
massive stars indicated.
In this regard, episodic mass-loss from single massive stars may be involved
and the time-averaged mass-loss higher than from steady-state winds.
However, it is also likely that mass-transfer in massive binary systems
plays a significant role.  Notwithstanding these issues, the winds from
WR stars are radiatively driven and do depend 
on metallicity. In that case, it seems possible that oxygen yields are
reduced at high metallicity due to shorter WR lifetimes.  I refer the reader
to the review by Smith (2014) for a detailed discussion of stellar wind mass-loss.
Also relevant, the downward revision in the FMR07 [Mg/Fe] ratios for
SMR bulge stars, discussed here, would make the [O/Mg] slope shallower
at high [Fe/H] than 
considered in MB08; thus, the metal-dependent oxygen yield reduction in
the bulge is less than previously thought, and less severe mass-loss 
rates from massive stars would be required to match the observed [O/Mg]
slope with [Mg/H] or [Fe/H] than in MB08.

Relevant to the issue of the effects of stellar winds and WR stars on 
the chemical evolution of the bulge, Cescutti et al. (2009) employed 
the oxygen abundances from MRF10 and carbon and nitrogen abundances for
the same bulge stars by Melendez et al. (2008) to estimate the original
[C/O] ratios of bulge stars, considering that C$+$N is roughly 
conserved through dredge-up onto the RGB.  A plot of original [C/O]
versus [O/H] for the bulge was compared to four chemical evolution models;
notably, the comparison to [O/H] minimized consideration of SNIa contributions
in the chemical evolution models.  Three of the models included the effects
of mass-loss and rotation on the yields of C and O from massive stars,
from Maeder (1992) and Meynet \& Maeder (2002).  An additional model was
computed, using nucleosynthesis yields from massive stars taken from WW95,
that did not include mass-loss.
While the observational data showed a strong increase in [C/O] above 
[O/H]$\sim$$-$0.2 dex, only the models including yields affected by 
mass-loss from massive stars could reproduce this behavior, and
indicated mass-loss rates somewhere between the Maeder (1992) and
Meynet \& Maeder (2002) predictions.  On the other hand, the model 
including WW95 yields, which did not include mass-loss, completely 
failed to match the observations.
Thus, the bulge [C/O] ratios indicate an important role for 
metallicity-dependent mass-loss from massive stars on the chemical 
evolution of the bulge, as suggested by M08 based on the observed 
[O/Mg] trend with metallicity.  On the other hand, Ryde et al. (2010)
suggest that carbon enhancements should result from this WR mass-loss
scenario, which they do not find in their bulge [C+N/Fe] ratios.

Measurements of the fluorine abundance in bulge stars have been made by
Cunha et al. (2008) and J\"onsson et al. (2014), using high-resolution
K-band spectra of HF lines in bulge K and M giants.  Both studies found a
metal-dependent increase in [F/O] and [F/Fe] with [Fe/H] that cannot be 
explained by neutrino spallation in SNII, nor AGB s-processing alone; 
instead, they conclude that WR stars are an important source of fluorine 
in the galactic bulge.

\subsection{Explosive Alpha-Elements: Si, Ca and Ti}

I now discuss trends of the average of Si, Ca and Ti: three alpha elements,
mainly produced during the the explosive nucleosynthesis of SNII events (WW95).
Small amounts of these explosive alpha-elements are also thought to be 
made during SNIa events; however, the predicted SNIa yields, relative 
to iron, are quite low (e.g. Nomoto et al. 1984; Fink et al. 2014).  
We average these three explosive alpha-elements because FMR07 showed 
that their trends follow each other closely.  The average explosive 
alpha-element abundances, $<$SiCaTi$>$, discussed here are based only 
on results from neutral lines (Si~I, Ca~I and Ti~I).

FMR07 measured [$<$SiCaTi$>$/Fe] with [Fe/H] in their sample of bulge 
and MW disk stars; however, they did not identify the thick/thin disk 
sub-populations.  The disk giant abundances were compared to the 
Bensby et al. (2005) thin and thick disk dwarf star results; while Ca 
and Ti from the disk giants followed the Bensby et al. (2005) abundances,
FMR07 found that their [Si/Fe] ratios were higher than the 
Bensby et al. (2005) disk giants by 0.09 dex.  Thus, a correction of 
$-$0.09 dex must be added to the FMR07 [Si/Fe] ratios to put them on 
the MW disk dwarf scale.

Following the finding, by Melendez et al. (2008) that the bulge [O/Fe] 
ratios follow the MW thick disk trend (confirmed in the re-analysis of 
FMR07 data by MFR10), Alves-Brito et al. (2010) used the EWs of FMR07, 
plus a sample of MW thick disk giants, and found that Mg, Si, Ca and Ti
abundances also followed the MW thick disk trend.

Although it is easy to compare bulge abundances with thick disk
stars near [Fe/H]=$-$0.6 dex, and thin disk stars near solar [Fe/H],
a comparison with thick disk stars near solar [Fe/H] is uncertain, 
because of the paucity of solar metallicity thick disk stars.  Although the 
Bensby et al. (2005) and Reddy et al. (2006) studies do include putative 
MW thick disk stars near solar [Fe/H], there are only a handful of such
relatively metal-rich thick disk stars, and their [$\alpha$/Fe] ratios
more closely resemble thin disk than the metal-poor thick disk stars.
 Thus, the solar-metallicity thick disk stars are either mis-identified thin disk 
members, or the [$\alpha$/Fe] ratios of the thick disk merges with the 
thin disk by [Fe/H]=0.00 dex.  To resolve this uncertainty requires a 
sample of reliable solar-metallicity thick disk stars.

\begin{center}
\begin{figure}
\includegraphics[width=3.0in]{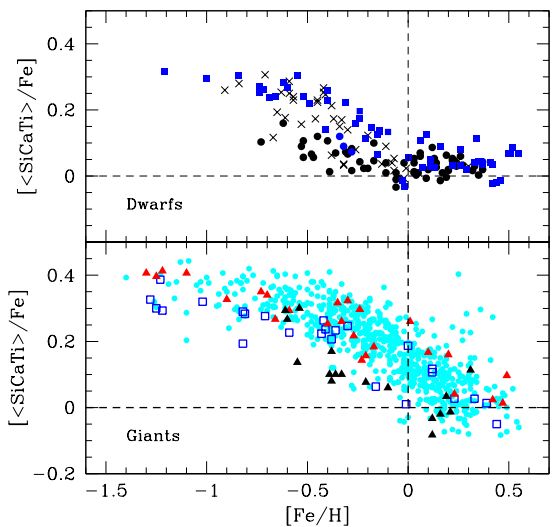}
\caption{{\it Top panel:} [$<$SiCaTi$>$/Fe] versus [Fe/H] for the
lensed turnoff bulge stars of B13 (filled blue squares), compared
to the MW thin and thick disk dwarf stars from Bensby et al. (2005),
filled black circles and black crosses, respectively.
{\it Bottom panel:} [$<$SiCaTi$>$/Fe] versus [Fe/H] for bulge giants
from Gonzalez et al. (2011; filled cyan circles), FMR07 (filled
red triangles), and Alves-Brito et al. (2010; open blue squares).
Filled black triangles are the disk giants analyzed by FMR07.}
\label{fig-sicati}
\end{figure}
\end{center}

As evident from Figure~\ref{fig-sicati}, the bulge [$<$SiCaTi$>$/Fe] trend 
is slightly different when derived from RGB stars than the lensed dwarf
bulge stars of B13.

The top panel of Figure~\ref{fig-sicati} shows the B13 lensed dwarf bulge stars,
with [$<$SiCaTi$>$/Fe]=$+$0.06 dex at solar [Fe/H], but constant for stars above solar
metallicity.  The bulge trend is certainly higher than the MW thin disk stars.  Near
[Fe/H]$\sim$$-$0.6 dex the bulge dwarfs show similar [$<$SiCaTi$>$/Fe] to the top
of the range for Bensby et al. (2005) MW thick disk stars, near $+$0.28 dex, above the 
thin disk by $\sim$0.15 dex.

The lower panel of Figure~\ref{fig-sicati} shows the Gonzalez et al. (2011)
bulge giant [$<$SiCaTi$>$/Fe] ratios with filled cyan circles; these are in quite
good agreement with the FMR07 bulge RGB stars, represented by filled 
red triangles.  At solar metallicity the Gonzalez et al. (2011) results suggests
[$<$SiCaTi$>$/Fe]=$+$0.13 dex,  but the FMR07 results suggest a slightly higher
intercept near between $+$0.15 and $+$0.17 dex.  At the metal-rich end 
[$<$SiCaTi$>$/Fe]=0.00 at [Fe/H]$\sim$$+$0.3 dex; the metal-poor 
plateau occurs below [Fe/H]$\sim$$-$0.6 to $-$0.8 dex at a value of 
[$<$SiCaTi$>$/Fe]$\sim$$+$0.39 dex.  These fiducial points indicate that, to
within the measurement uncertainties, the trend of [$<$SiCaTi$>$/Fe] 
with [Fe/H] is the same as the [Mg/Fe] trend for bulge giant stars.

The FMR07 disk giants, indicated by filled black triangles in the lower
panel of
Figure~\ref{fig-sicati}, fall well below the bulge trend, but mostly 
similar to the thin disk points (filled black circles) in the upper 
panel.  However, the cluster of three FMR07 disk giants with 
[Fe/H]$\sim$$-$0.6 dex have thick disk ratios; thus, they could 
reasonably be thick disk stars.   It appears that at [Fe/H]=$-$0.6 the 
Gonzalez et al. (2011) trend is slightly higher than the apparent trio 
of FMR07 thick disk giants, but they are in reasonable agreement with 
the FMR07 bulge giants.  Clearly, the FMR07 disk sample is too small to
define the thick disk near [Fe/H]=0.00 dex.

The Aves-Brito et al. (2010) points in the lower panel (represented by 
blue open squares) are$\sim$0.10 dex lower than the FMR07 and Gonzalez et al. (2011)
results.  Gonzalez et al. (2011) have traced this back to a difference in
the stars used to define the zero-point of the Alves-Brito et al. (2010)
scale.  Because the independent FMR07 and Gonzalez et al. (2011) 
abundance scales are both higher than Alves-Brito et al. (2010), by 
about 0.10 dex, it appears that Alves-Brito et al. (2010) [$<$SiCaTi$>$/Fe]
results may be erroneously low.  Furthermore, abundances for the FMR07 
zero-point standard star, Arcturus, are in excellent agreement with the
work of Ramirez \& Allende Prieto (2011); both give 
[$<$SiCaTi$>$/Fe]=$+$0.24 dex, whilst [Fe/H] for FMR07 and
Ramirez \& Allende Prieto (2011) give are close, at $-$0.50 and $-$0.52 dex, 
respectively.

It is important to note that, the Alves-Brito et al. (2010) finding 
that the bulge RGB stars overlap with their thick disk RGB star 
abundance trends, is not affected by potential zero-point errors.  Indeed,
the Gonzalez et al. (2011) re-analysis of the Alves-Brito et al. (2010)
thick disk giants found the same similarity to the bulge as 
Alves-Brito et al. (2010).  

The [$<$SiCaTi$>$/Fe] ratios of the bulge
dwarf stars in Figure~\ref{fig-sicati} lie slightly above the thick disk
trend in the range $-$0.3$<$[Fe/H]$<$0.0 dex; but, this is partly because
thick disk appears to merge with the thin disk trend. However, this merger
may simply be due to mis-identification of the thick disk stars.  The RC
stars in Figure~\ref{fig-sicati} lie above the trend established by the
thick disk dwarf stars, but this might  be due to an atmosphere effect.

It is probably best to give more weight to the dwarf comparison, and assume
that the bulge and thick disk alpha-element trends are probably coincident
in [$<$SiCaTi$>$/Fe]; at best the bulge is slightly higher.  Notably,
the bulge reaches much higher [Fe/H] than the thick disk.  However,
the Alves-Brito et al. (2010) bulge results for Ca and Ti
lie above their thick disk points, while Si follows the same trend.  The
differences could easily be due to T$_{\rm eff}$ errors, although they
should cancel in the $<$SiCaTi$>$ average, to some degree.  
While the Gonzalez et al. (2011) re-analysis of Alves-Brito et al. (2010)
thick disk giants agrees well with the Gonzalez et al. (2011) bulge 
results, there are very few thick disk stars at the critical 
metallicity in the comparison.  Although the similarity of
bulge and thick disk giant [$<$SiCaTi$>$/Fe] trends in Gonzalez et al. (2011)
very compelling, there are only four thick disk giant
comparison stars near solar metallicity. The fact that putative 
thick disk dwarfs show thin disk compositions, and are not alpha-enhanced,
unlike the small sample of thick disk giants, raises additional 
concerns regarding the comparison.

Non-LTE over-ionization of neutral Si, Ca and Ti atoms in dwarf
stars provides a plausible explanation for the $\sim$0.10 dex 
difference in [$<$SiCaTi$>$/Fe] between bulge dwarf and giant star 
trends in Figure~\ref{fig-sicati}.  The competition between collisional
and radiative excitation and ionization rates may result in departures
from LTE for the relatively hot dwarf stars, due to the presence of
ionizing radiation, and especially at lower metallicity, where 
stellar atmospheres become more transparent at UV wavelengths, due to
reduced line blanketing.  On the other hand, the atmospheres
of red giant stars are typically heavily blanketed in the UV, and cool
enough that the ionizing flux is low; for these reasons, conditions close
to LTE may exist, despite the reduced collisional rates in the low-density
giant atmospheres.

Non-LTE corrections have been computed for a variety of elements, but not all,
for restricted ranges of stellar parameters (but mostly for dwarf stars),
and clearly can affect alpha-element abundances (e.g. Gehren et al. 2006;
Lind, Bergemann \& Asplund 2012; Bergemann 2011)
at the level of interest here.  The review by Bergemann \& Nordlander (2014)
gives the recent status of such calculations; more needs to be done.

Finally, it is worthwhile noting that a dispersion in [Ca/Fe] in the
results of G15 suggests the possibility of a sub-population of Ca-rich
bulge stars.  These are may be due to errors in the microturbulent
velocity parameter for the rather strong Ca~I lines available for 
abundance analysis in near-solar metallicity bulge giants.  However,
it would be very significant if Ca-rich bulge stars were confirmed.

\section{Comments on Helium, Sodium and $^{12}$C/$^{13}$C}

  The identification of an X-shape morphology for the Galactic
  bulge, by McWilliam \& Zoccali (2010), was based on an adopted distance 
  calibration for RC stars.  However, the voracity of the X-shape was challenged
  by Lee, Joo, \& Chung (2015), who claimed that enhanced helium fractions near
  Y$\sim$0.4, similar to some sub-populations seen in a few globular
  clusters, could explain the RC magnitudes and colors without an X-shaped
  structure.  
  Unfortunately, due to its atomic structure it is very difficult to measure
  the photospheric abundance of helium in the relatively cool RC stars.

\begin{center}
\begin{figure}
\includegraphics[width=3.0in]{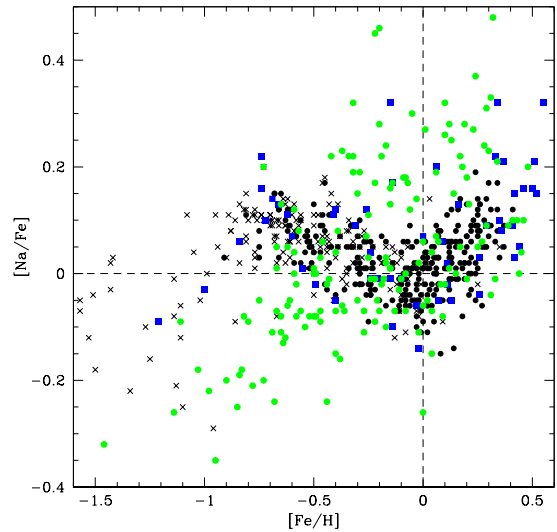}
\caption{ 
[Na/Fe] versus [Fe/H] for the thin and thick disks (filled black circles and
crosses respectively) from Bensby et al. (2014), compared with lensed bulge
dwarf stars (filled blue squares) from Bensby et al. (2013) and bulge RC
stars (filled green circles) of Johnson et al. (2014).  The zig-zag trend
line in the dwarfs is qualitatively consistent with Na produced by SNII
progenitors with a metallicity-dependent yield, in the SNIa time-delay scenario
of Tinsley (1979) and Matteucci \& Brocato (1990).  
}
\label{fig-bulgedisk-nafe}
\end{figure}
\end{center}

  However, it is well known that globular clusters with enhanced helium mass
  fractions, Y, show strong Na enhancements, typically $+$0.5 dex but up to
  $+$0.9 dex, and oxygen deficiencies; this results in and [Na/O] ratios up
  to $\sim$1.0 dex, and a steep Na--O anti-correlation.
  In this regard, Na abundances for lensed bulge dwarfs were measured by 
  Bensby et al. (2013), who found [Na/Fe]$\sim$0.1 dex for nearly all 
  metallicities (see Figure~\ref{fig-bulgedisk-nafe}), and no O deficiencies.
  Interestingly, the location of the Bensby et al. (2013) lensed bulge dwarf 
  sample was heavily weighted to the positive longitude side of the bulge, 
  where Lee et al. (2015) required high helium abundance.  Thus, if there is 
  an He enhancement in the bulge, it does not seem to be related to the
  mechanism seen in some MW globular clusters that also results in the
  strong Na--O anti-correlation.

Figure~\ref{fig-bulgedisk-nafe} compares measured [Na/Fe] ratios in
MW thin/thick disk dwarfs with the lensed dwarfs of B13 and the RC giants 
in J14.  Notably, the disk and bulge dwarf stars show identical trends, with
[Na/Fe] following a zig-zag pattern.  This unusual zig-zag trend can be
understood if Na is produced by massive stars (SNII or their progenitors) 
but with a yield that increases with increasing [Fe/H]. In the SNIa time-delay
scenario of Tinsley (1979) and Matteucci \& Brocato (1990), used to
explain the [O/Fe] trend with [Fe/H], SNIa iron is added above [Fe/H]$\sim$$-$1,
which produces declining ratios of both [O/Fe] and [Na/Fe]; however, as
[Fe/H] increases the production of Na increases strongly, leading to a
muted decline in [Na/Fe] compared to the decline of [O/Fe].  Finally, above
about solar metallicity, the metal-dependent Na yields have increased so 
much that [Na/Fe] increases with increasing [Fe/H], producing the observed
zig-zag locus.  I note, however, that for metal-rich stars saturation
effects and blends can cause systematic errors that increase the derived
abundances.

Figure~\ref{fig-bulgedisk-nafe} shows that the bulge RC stars have
lower [Na/Fe] at low [Fe/H] and a much larger scatter than the lensed bulge
dwarfs.
%
%
%
These differences
might be partly due to non-LTE effects in RC giants; indeed,
Lind et al. (2011) provide theoretical non-LTE corrections near $-$0.10 dex 
for Na~I lines in the J14 bulge RC stars.  On the other hand, the control
sample of MW disk RGB stars in FMR07 showed no difference in measured 
[Na/Fe] ratios compared to the Bensby et al. (2005) thin and thick disk 
dwarf stars.

  I note that measured $^{12}$C/$^{13}$C ratios in bulge red giant stars
  (e.g., Uttenthaler et al. 2015)  are similar to ratios seen in the Solar Neighborhood
  red giants (e.g., Smith \& Lambert 1990).  If the bulge helium abundance 
  had been strongly increased due to CNO cycle processing, the $^{12}$C/$^{13}$C ratios
  of all bulge stars would be expected to be lower than the disk; for the RGB stars,
  at least, there is no significant, obvious, difference between the bulge and disk.  
  Thus, it does not appear that proton burning products of the CNO cycle are enhanced
  in the bulge.  A useful test of this result could be obtained from the lensed
  bulge dwarf stars, whose $^{12}$C/$^{13}$C ratios should be near 90.
  Thus, there is currently no indirect chemical composition evidence favoring 
  enhanced bulge helium abundance; if Y is enhanced it likely resulted from the 
  p-p chain.

\section{Iron-Peak Elements}

\subsection{Chromium}

The abundance of Cr has been measured for bulge RGB stars by 
Johnson et al. (2014) and for lensed bulge dwarf stars by 
Bensby et al. (2013).  Both of these studies found that Cr scales
with Fe, i.e., [Cr/Fe]=0.0 dex, at all metallicities probed,
ranging from [Fe/H]$\sim$$-$1.5$-$1 to $+$0.5 dex.  In this regard,
the bulge [Cr/Fe] trend resembles that in both the thin and thick 
MW disks.

\subsection{Manganese}

The trend of [Mn/Fe] with [Fe/H] is either a probe of nucleosynthesis
and chemical evolution, or it reflects metal-dependent non-LTE effects
in stellar atmospheres.

\begin{center}
\begin{figure}
\includegraphics[width=3.0in]{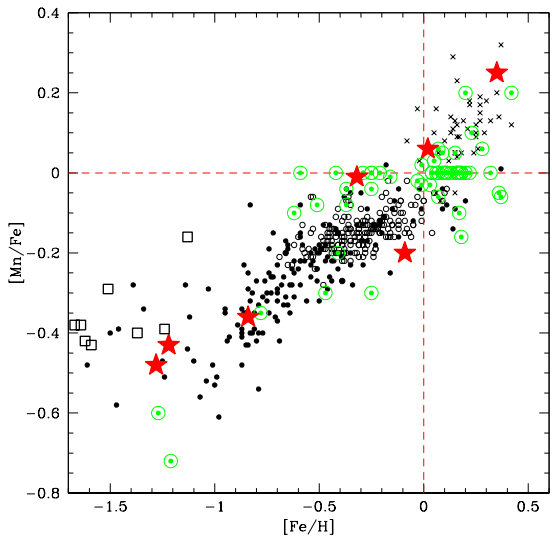}
\caption{ 
LTE [Mn/Fe] versus [Fe/H] abundance ratios in the bulge from McWilliam et al.
(2003; red stars) and Barbuy et al. (2013; green targets), compared to Reddy
et al. (2003, 2006) MW thin disk (open black circles), thick disk (filled
black circles), halo globular clusters from Sobeck et al. (2006; open black 
squares) and metal-rich thin disk stars from Feltzing \& Gustafsson 
(1998; black crosses).  The trend of [Mn/Fe] versus [Fe/H] in the bulge and
MW disks/halo appear identical to within the measurement uncertainties.
}
\label{fig-bulge-mnfe}
\end{figure}
\end{center}

Deficiencies of [Mn/Fe] in nearby G-dwarf stars were originally found
by Wallerstein (1962) and later also seen by Wallerstein et al. (1963) in
MW halo giants.  A significant finding came from a study of
MW halo and disk, giant and dwarf stars, by Gratton (1989).  Gratton (1989)
found a low [Mn/Fe] plateau, at $-$0.34 dex, for stars below
[Fe/H]~$\sim$$-$1, with a linear rise in [Mn/Fe] above this
metallicity, toward the solar composition.  This trend is the
inverse of the $\alpha$-elements (e.g., see Figure~\ref{fig-bulge-mnfe}).

Subsequent studies of [Mn/Fe] trends in the MW disks and halo
(e.g. Feltzing \& Gustafsson 1998; 
Reddy et al. 2003, 2006; Sobeck et al. 2006; Mishenina et al. 2015)
continued to show and extend the trend found by Gratton (1989), with 
the metal-poor plateau appearing closer to [Mn/Fe]=$-$0.40 dex.

Notable observational challenges for Mn abundance measurement include 
very strong hyperfine splitting of Mn~I lines; and, for absolute
abundances, a difference of 0.14 dex between the photospheric and
meteoritic Mn abundance in the sun (e.g. Grevesse \& Sauval 1998); 
although, the difference is now only 0.05 dex (e.g. Lodders et al. 2009; 
Asplund et al. 2009).

Based on the observed inverse-alpha trend, Gratton (1989) suggested 
that the [Mn/Fe] increase with [Fe/H] could be due to over-production
of Mn by SNIa.  

In this scenario, as the [O/Fe] ratio declines due to the late addition 
of iron from SNIa, the [Mn/Fe] ratio increases, from Mn over-production
by the same SNIa events. Another way to view it is that as the SNIa/SNII 
ratio increases at late times extra Fe and Mn from SNIa reduce [O/Fe]
and increase the [Mn/Fe] ratio.

If this idea is correct, then the [Mn/Fe] ratios provide a consistency check
for the SNIa time-delay scenario of chemical evolution, 
can be used as a probe of the SNIa/SNII ratio in different systems, and
to provide constraints on the progenitors and mechanism of Type~Ia supernovae.

Observational evidence for Mn over-production by SNIa
was recently identified, by means of  X-ray spectroscopy, for the 
supernova remnant 3C~397 (e.g., Yamaguchi et al. 2015).

On the theoretical side,
early nucleosynthesis calculations of core-collapse supernovae
indicated that SNII produce metallicity-dependent Mn yields that
largely explain the observed [Mn/Fe] trends in the MW
(Woosley \& Weaver 1995; Timmes, Woosley \& Weaver 1994), and
that completely overwhelm any Mn contributions from SNIa events.
However, more recent supernova nucleosynthesis calculations 
(e.g. Kobayashi et al. 2006; Tuguldur Sukhbold et al. 2015) 
indicate that SNII under-produce manganese at all metallicities, 
with [Mn/Fe] at roughly $-$0.3 to $-$0.6 dex.

Theoretical element yields for Chandrsekhar-mass deflagration SNIa
events (e.g. Nomoto 1984, 1997, Fink et al. 2014, Yamaguchi et al. 2015)
give enhanced [Mn/Fe] ratios near $+$0.3 dex, with Mn yields that are
insensitive to the progenitor metallicity.
On the other hand, sub-Chandrasekhar-mass SNIa (e.g. Woosley \& Kasen 2011;
Seitenzahl et al. 2013) result in deficient [Mn/Fe] ratios, but
with Mn yields that increase with metallicity 
(Yamaguchi et al. 2015; Kobayashi, Nomoto \& Hachisu 2015).  

From these theoretical results, it is clear that [Mn/Fe] ratios
are sensitive to the SNIa/SNII ratio, sensitive to the
SNIa mechanism and sub-type, and sensitive to metallicity in the case
of sub-Chandrasekhar SNIa.

Apparently, only Chandrasekhar-mass SNIa over-produce Mn, relative to Fe,
so the high [Mn/Fe] ratio in 3C~397, from Yamaguchi et al. (2015), 
indicates a Chandrasekhar-mass SNIa.

The above observational and theoretical results suggest
that the increasing [Mn/Fe] trend with [Fe/H] is consistent
with the delayed addition of Mn from Chandrasekhar-mass SNIa, 
as expected from the time-delay scenario of Tinsley (1979) and
Matteucci \& Brocato (1990) that explains the [O/Fe] trend by the
delayed addition of SNIa iron.


In stark contrast to the apparent observational consistency noted earlier,
Feltzing et al. (2007) found a flat, barely sub-solar, trend with [Fe/H].
Recently Battistini \& Bensby (2015) measured LTE abundances in agreement 
with the inverse-alpha trend seen by others, but the non-LTE abundance
corrections of Bergemann \& Gehren (2008), when applied, resulted in a 
flat [Mn/Fe] trend with [Fe/H], at roughly the solar value for all 
metallicities.  Interestingly, it seems possible that the non-LTE 
corrections could resolve the difference between solar photospheric 
and meteoritic Mn abundances.  On the other hand, smaller non-LTE
corrections, near $\sim$$+$0.1 dex, were found by Cunha et al. (2010)
for the red Mn~I lines in three Omega~Cen RGB stars, at
[Fe/H]~$\sim$$-$1 dex; these corrections would not erase the inverse
$\alpha$-trend of [Mn/Fe] with [Fe/H].

Somewhat contrary to the theoretical non-LTE corrections of Bergemann \& 
Gehren (2008), Sneden et al. (2016) measured 19 Mn~I and 10 Mn~II lines
in the metal-poor dwarf star, HD84927, and found 
the Mn~II abundance higher by only $+$0.04 dex, much smaller than the predicted
non-LTE correction, near $+$0.35 dex.  In particular,
Sneden et al. (2016) found [Mn II/Fe II]=$-$0.27 dex, consistent with
the trend derived from the neutral species; this is important, because the
singly ionized lines do not suffer severe non-LTE effects.

If the Mn~I non-LTE abundance corrections of Bergemann \& Gehren (2008) 
are correct, then the abundance of Mn scales with Fe in the MW halo, 
thick and thin disks, and there is no probe of Mn nucleosynthesis or 
chemical evolution.  Unfortunately, such a flat [Mn/Fe] trend leads to 
an inconsistency with the time-delay scenario explanation (Tinsley 1979; 
Matteucci \& Brocato 1990), for the decline of [$\alpha$/Fe] with increasing
[Fe/H], arising from
increased contributions of Fe from SNIa at late times.  The iron from
delayed SNIa reduce the [O/Fe] ratio, but should also increase [Mn/Fe]
from Chandrasekhar-mass SNIa.  

I conclude that, if the Mn~I non-LTE corrections Bergemann \& Gehren (2008) are
approximately correct, then either the time-delay chemical evolution
scenario, or the theoretical supernova nucleosynthesis yields are wrong 
and/or incomplete.

On the other hand, 
if there is an inverse-alpha trend of [Mn/Fe] with [Fe/H] in the MW halo, 
thick and thin disks, due to Mn over-production by SNIa, then we expect
enhanced [Mn/Fe] ratios in $\alpha$-poor systems, like dwarf galaxies,
where SNIa material is thought to dominate, and deficient [Mn/Fe] in 
$\alpha$-enhanced systems, like the MW bulge, where SNII material is thought
to be enhanced.  Thus, if Mn is over-produced in SNIa, the enhanced 
[Mg/Fe] and [O/Fe] ratios in the bulge should be accompanied by low
[Mn/Fe] ratios.

LTE abundance measurements of Mn in bulge stars have been performed by
McWilliam, Rich \& Smecker-Hane (2003) and Barbuy et al. (2013),
as shown in Figure~\ref{fig-bulge-mnfe} and compared 
with the [Mn/Fe] trend for MW disk stars, measured by
Reddy et al. (2003, 2006) and Feltzing \& Gustafsson (1998).
It is evident from Figure~\ref{fig-bulge-mnfe} that the [Mn/Fe]
trends for both bulge studies agree and are consistent with the MW disk trend,
to within the measurement uncertainties.  Thus, there is no evidence for a Mn
deficiency in the bulge, as expected from 
the observed enhanced [Mg/Fe] and [O/Fe] ratios.
For this reason, McWilliam et al. (2003) concluded that Mn
cannot simply be over-produced in all SNIa, but rather there must be a
metallicity-dependent yield of Mn from both SNII and SNIa, with increasing
[Mn/Fe] ratios to higher [Fe/H].  Such a strong metallicity effect 
is not supported by current supernova nucleosynthesis theory.

In addition to the unexpected [Mn/Fe] ratios, 
Barbuy et al. (2013) noted that the bulge [Mn/O] trend with [O/H]
is enhanced compared to the MW thick disk, and showed that, in this way, the
bulge is chemically distinct and likely evolved differently.

Of course, the alternate (and trivial) explanation for the similar [Mn/Fe]
trends in the bulge and MW disk stars is that the trend simply reflects the
metal-dependence of the non-LTE corrections to the LTE Mn~I abundances.

In summary, the observed LTE [Mn/Fe] trend with [Fe/H] in the bulge is similar
to the MW thin and thick disks; but, this is contrary to expectations of the
SNIa time-delay scenario of chemical evolution, given the enhanced
[$\alpha$/Fe] ratios for the same bulge stars.  The combination of non-LTE
corrections, theoretical supernova nucleosynthesis yields, and the
time-delay chemical evolution scenario are unable to explain the measured
[Mn/Fe] versus [Fe/H] trends in the MW disks or bulge; one of these inputs
must be wrong or incomplete.  


%
%

\subsection{Cobalt}

Bulge abundances for Co have only been measured for the 
sample of RGB stars studied by Johnson et al. (2014).  They found
roughly constant [Co/Fe] ratios with [Fe/H], albeit with a slight
overall enhancement, near $+$0.15 dex, and a 1$\sigma$ scatter of
about 0.1 dex, presumably due to measurement error.  

While this small [Co/Fe] enhancement lies above the MW thin disk values
(e.g., Reddy et al. 2003, 2006), it resembles the $\sim$$+$0.15 dex
enhancement of [Co/Fe] in the MW thick disk found by Reddy et al. (2006) and
the $\sim$$+$0.2 dex value for the thick disk stars in 
Battistini \& Bensby (2015).  These elevated [Co/Fe] ratios might be
related to the $\sim$$+$0.5 dex [Co/Fe] enhancements seen in extremely
metal-poor MW halo stars (e.g., McWilliam et al. 1995), thought to be due to
hypernovae (e.g., Nomoto et al. 2001).  However, hypernova nucleosynthesis
models also produce enhanced [Zn/Fe] and deficient [Cr/Fe] ratios.  
Johnson (1999) found enhancements of Zn in extreme metal-poor 
stars, later confirmed by Cayrel et al. (2004), that also show enhanced 
Co and low Cr, consistent with the hypernova 
origin at very low metallicity.  As mentioned below, 
although a small [Zn/Fe] enhancement is seen in somewhat metal-poor bulge 
stars, near [Fe/H]=$-$0.8 dex, Zn enhancements are not seen in the
solar-metallicity or metal-rich bulge stars where the Co enhancement
is found.  Furthermore, the solar [Cr/Fe] ratios measured in the bulge are
quite normal, and so inconsistent with the theoretical hypernova nucleosynthesis 
yields.  Thus, it is difficult to ascribe the small [Co/Fe] enhancement in bulge
stars to hypernova nucleosynthesis.  The enhancement is not likely due to
stellar atmosphere, non-LTE effects, since the non-LTE corrections of
Bergemann et al. (2010) would only increase the [Co/Fe] over-abundance.

\subsection{Nickel}

The abundance of nickel in the bulge has been measured by
Johnson et al. (2014) for red giant branch stars and by Bensby et al. (2013)
for lensed dwarf stars.

The Bensby et al. (2013) results show flat [Ni/Fe]=0.0 dex below solar metallicity;
above [Fe/H]=0.0 dex a small increase in [Ni/Fe] with increasing [Fe/H] is present,
reaching [Ni/Fe]$\sim$$+$0.1 dex at [Fe/H]=$+$0.5 dex.  Curiously, while the flat trend
is similar to [Ni/Fe] results for the MW thin and thick disk in both Reddy et al. 
(2003, 2006) and Bensby et al. (2003), the upward rise in [Ni/Fe]
above solar metallicity is seen only in the MW thin disk results of Bensby et al. (2003).
Notably, the analysis of SMR thin disk stars by Feltzing \& Gustafsson (1998)
found flat [Ni/Fe]=0.0 dex for all their stars: from solar metallicity to
[Fe/H]=$+$0.40 dex.  Thus, the up-turn in [Ni/Fe] for SMR thin disk and bulge stars,
found by Bensby et al. (2003, 2013), is not confirmed.

The Johnson et al. (2014) RGB bulge stars also show a flat trend of [Ni/Fe] below
solar metallicity, but slightly enhanced, at [Ni/Fe]$\sim$$+$0.1 dex.  Above
solar [Fe/H], and up to $+$0.5 dex, the same $\sim$0.1 dex in [Ni/Fe]
enhancement persists; although, a handful of points might be consistent with a
rise at the highest [Fe/H].  Given the disagreement with the bulge dwarf stars,
it is likely that the small [Ni/Fe] enhancement in Johnson et al. (2014) is due
to systematic error (for example, due to line blends, gf values, saturation effects).

Notwithstanding the inconclusive observational case for a small [Ni/Fe]
enhancement in the bulge, Ni deficiencies have been claimed for
some dwarf galaxies and $\alpha$-poor halo stars (e.g., Letarte 2007; Nissen \&
Schuster 2010).  Since these systems are deficient in SNII nucleosynthetic products, 
somewhat opposite to the bulge, it is possible that the small [Ni/Fe] enhancement
claimed by Johnson et al. (2014), may be due to a relative excess of SNII material.  
At present, however, I assume that there is no strong evidence for Ni enhancement
in the bulge and that the [Ni/Fe]$\sim$$+$0.1 dex value in Johnson et al. (2014)
is probably due to systematic error.

\subsection{Copper}

While deficiencies of Cu, relative to Fe, were found in studies of
small numbers of metal-poor MW halo stars (e.g., Cohen 1978, 1979; 
Peterson 1981; Luck \& Bond 1985; Sneden \& Crocker 1988), the 
large-scale trend of [Cu/Fe] with [Fe/H] in MW disk and halo stars was 
first clearly delineated by Sneden, Gratton \& Crocker (1991).  Subsequent 
studies (e.g., Mishenina et al. 2002; Simmerer et al. 2003; Reddy et al. 2003, 
2006) have improved the accuracy, metallicity coverage, number of stars,
and added halo globular clusters.

\begin{center}
\begin{figure}
\includegraphics[width=3.0in]{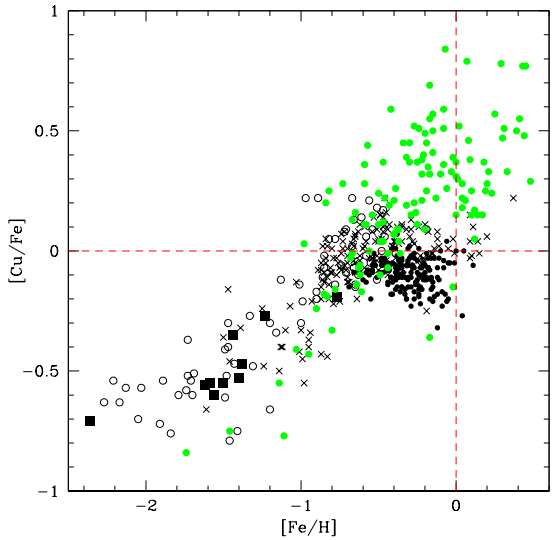}
\caption{ 
[Cu/Fe] versus [Fe/H] in the bulge from Johnson et al. (2014; filled green circles),
compared to the MW stars from Reddy et al. (2006; mostly thick disk and halo: black 
crosses), Reddy et al. (2003; thin disk: filled black circles), Mishenina et al.
(2002; halo and thick disk: open black circles), and Simmerer et al. (2003; halo 
globular clusters: filled black squares). The bulge [Cu/Fe] trend continues to high
values above [Fe/H]$\sim$$-$0.5 dex and exceeds the MW disk trend by up to
$\sim$0.5 dex.  Notice the large dispersion in [Cu/Fe] near solar metallicity.
}
\label{fig-bulge-cufe}
\end{figure}
\end{center}

As can be seen from Figure~\ref{fig-bulge-cufe}, the MW disk/halo trend is
roughly flat from solar [Fe/H] down to about [Fe/H]=$-$0.7 dex;
below that, [Cu/Fe] declines roughly linearly with decreasing [Fe/H], until
reaching a flat plateau of [Cu/Fe]=$-$0.6 to $-$0.7 dex at a metallicity
below [Fe/H]$\sim$$-$1.5.

A non-LTE analysis of copper in turnoff dwarfs, by Yan, Shi \& Zhao (2015),
indicated that the metal-poor plateau lay near [Cu/Fe]=$-$0.4 dex,
based on two stars (assuming collisions with hydrogen atoms scaled to
S$_H$=0.1).  For solar-metallicity stars, they found small non-LTE corrections
to the LTE Cu~I abundances, at most $+$0.05 dex, but this increased
to $+$0.15 dex by [Fe/H]=$-$1.5 dex.  Notably, non-LTE effects for
Fe~I lines were not computed in the analysis of Yan et al. (2015). 
Non-LTE corrections for Fe~I lines, using the on-line tool, INSPECT, by 
Lind et al.\footnote{inspect.coolstars19.com} gives values of 0.02--0.03 dex.
It is not known whether the non-LTE over-ionization effect for 
Cu~I levels is smaller or larger in the cool RGB globular cluster stars
of Simmerer et al. (2003).

Bihain et al. (2004) studied [Cu/Fe] in extreme metal-poor stars
and employed one of the Cu~I resonance lines at 3273\AA .  They found
[Cu/Fe] roughly solar above [Fe/H]=$-$1, but with a linear decline down
to [Fe/H]$\sim$$-$2.5 dex and a plateau at [Cu/Fe]=$-$1.0 dex.
A reasonable assumption is that the non-LTE corrections for the
most metal-poor stars in Bihain et al. (2004) exceed those found by
Yan et al. (2015) for more metal-rich stars; thus, the plateau
below [Fe/H]=$-$2.5 is probably higher than the Bihain et al. LTE 
value.  Notably, Bonifacio et al. (2010) concluded that the Cu~I
resonance lines were not reliable abundance indicators, likely due
to departures from LTE.

Although, many astrophysical sites of copper nucleosynthesis have been proposed,
it is now generally accepted that copper is mostly produced in the hydrostatic
He- and C-burning phases of massive stars, ($>$8M$_{\odot}$ which ultimately 
become SNe II), via weak s-process neutron-capture, driven by
$^{22}$Ne($\alpha$,n)$^{25}$Mg.  For extensive discussion of the evolution
and details of this idea see Prantzos et al. (1990), Raiteri et al. (1991, 1993),
The et al.  (2000), Bisterzo et al. (2004), Pignatari et al.  (2008, 2010),
and Pumo et al. (2010). 
Since the solar number ratio of iron to copper is $\sim$1,600, the addition of
only a few neutrons to the iron-peak elements can greatly increase the copper
abundance.  Obviously, the actual Cu enhancement depends on the details of the
s-process environment and the neutron-capture cross-sections.

In this scenario, the Cu yield increases with increasing metallicity (e.g.
Bisterzo et al. 2004; Kobayashi et al. 2006),
as expected from the metallicity dependence of the s-process,
but also increases with the mass of the massive star, presumably 
due to the size of the He- and C-burning regions. 

A minor component of the Cu is thought to be produced during
SNII explosive nucleosynthesis via the alpha-rich freeze-out
(e.g., Woosley \& Hoffman 1992; Woosley \& Weaver 1995).
On the other hand, negligible copper production ([Cu/Fe]$\sim$$-$3 dex)
resulted from the Chandrasekhar-mass deflagration model of SNIa by 
Fink et al. (2014).

Based on the above ideas, we can understand that [Cu/Fe] increases with
[Fe/H] because the s-process yield of Cu increases with metallicity.
The flattening of the [Cu/Fe] ratios above [Fe/H]$\sim$$-$0.8 can be
ascribed to the addition of Fe from SNIa events, which produce iron but
no copper.  Thus, although the yield of Cu from SNII increases with
metallicity the increasing addition of SNIa Fe, above [Fe/H]$\sim$$-$1 dex,
serves to maintain the [Cu/Fe] ratio at a roughly constant value.  In this
way, the abundance of copper provides a nice consistency check on the SNIa
time-delay effect used to explain the [$\alpha$/Fe] trends with [Fe/H].
In Figure~\ref{fig-bulge-cufe} the mostly thick disk stars of
Reddy et al. (2006) lie at slightly higher [Cu/Fe] than the mostly thin
disk stars of Reddy et al. (2003), suggesting that the SFR in the thick disk
was slightly higher/faster than the thin disk.

It seems possible that the [Cu/Fe] plateau below [Fe/H]$\sim$$-$1.5 dex
may be dominated by $\alpha$-rich freeze-out composition; thus, chemical
composition of stars in plateau may provide a useful nucleosynthesis 
diagnostic of the alpha-rich freeze-out.
Another nucleosynthetic clue is provided by the abnormally deficient 
[Cu/Fe] ratios observed in some dwarf galaxies, that also show 
greater deficiencies of hydrostatic alpha-elements (e.g., O, Mg) than the
explosive alphas (e.g. Ca, Ti):
for example, LMC (Pomp\'eia et al. 2008) and Sgr dSph (McWilliam \& Smecker-Hane 2005).
The low hydrostatic/explosive element abundance ratios in these 
two dwarf galaxies suggests a deficiency of the most massive stars 
(McWilliam et al. 2013).  Despite such low [Cu/Fe] ratios, the trend 
of [Cu/O] versus [Fe/H] is the same in Sgr and the MW disks, suggesting 
that Cu production is correlated with O production, but with a 
metal-dependent yield, as expected from a neutron-capture
origin in massive stars.  Similar Cu-deficiencies, combined with $\alpha$-element
deficiencies, have been seen in a sub-class of MW halo stars by
Nissen \& Schuster (1997; 2011), that may 
have originated from one, or more, dwarf galaxies accreted by the MW halo.
These chemical evolution clues support the idea that Cu is produced in SNII
progenitors, tracing massive stars, but with a metallicity-dependent yield.


To date, the only reported abundance measurements for Cu in bulge stars was performed
by Johnson et al. (2014). Figure~\ref{fig-bulge-cufe} shows that the bulge [Cu/Fe] 
ratio increases from low to high [Fe/H], similar to the MW disk trend below
[Fe/H]=$-$0.8 dex; however, instead of the flattening in the [Cu/Fe] slope above
this metallicity, the bulge [Cu/Fe] trend continues to rise, such that most points
above [Fe/H]=$-$0.6 dex lie well above the MW thin and thick disk values.  Clearly,
the bulge has a distinct [Cu/Fe] trend with [Fe/H], very different than the MW disks 
and halo.  A large
scatter in [Cu/Fe], for these more metal-rich stars, is also apparent.  An important
question is whether this scatter in [Cu/Fe] is real, or due to measurement errors.

\begin{center}
\begin{figure}
\includegraphics[width=3.0in]{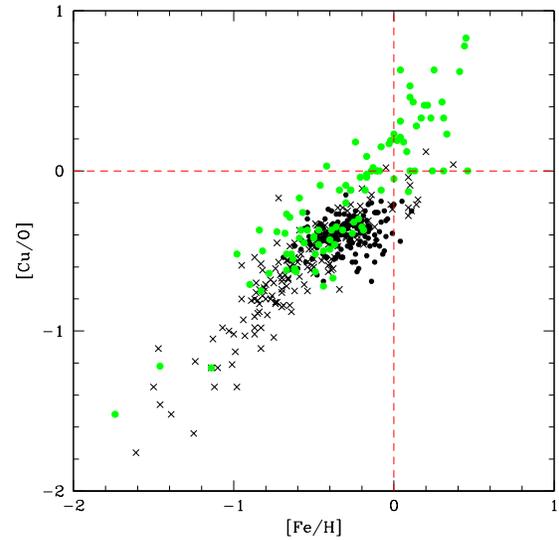}
\caption{ 
[Cu/O] versus [Fe/H] in the bulge and MW disks and halo; symbols have the same 
meaning as in Figure~\ref{fig-bulge-cufe}.  The bulge [Cu/O] trend is much tighter
than for [Cu/Fe], and continues to super-solar metallicity, supporting the idea
that synthesis of Cu and O are more closely related than that of Fe.
The Cu--O trend is consistent with copper production via the weak
s-process in SNII progenitors, as proposed by Raiteri et al. (1991, 1993) and 
Pignatari et al. (2008, 2010); in this case, the Cu yield depends on both the mass
and metallicity of the progenitor star.
}
\label{fig-bulge-cuo}
\end{figure}
\end{center}

Remarkably, the [Cu/O] versus [Fe/H] plot, shown in Figure~\ref{fig-bulge-cuo}, 
for the same bulge stars has much less scatter than the [Cu/Fe] versus [Fe/H]
diagram.  The small scatter in [Cu/O] versus [Fe/H], and the great similarity
to the MW halo/disk values, provides a qualitative check that Cu is produced
by massive stars (the oxygen producers) with a yield that increases with both
metallicity and mass.  I note that abundance ratios for stars in the LMC
(Pomp\'eia et al. 2008) and the Sgr dwarf galaxy (McWilliam et al. 2013) fall on
the same locus, despite their low oxygen abundances.  The tight scatter suggests
that both the Cu and O abundances are roughly correct, or at least any error in
one is cancelled by a similar error in the other; this is despite the different
sensitivities of Cu~I and [O~I] lines to stellar atmosphere parameters.

If the apparent large scatter in the [Cu/Fe] ratios of Figure~\ref{fig-bulge-cufe}
is real, this may be due to inhomogeneous chemical evolution, and the fact 
that Cu and O are co-produced in the same mass SNII progenitors, whereas 
Fe is produced in much lower-mass SNII, as well as SNIa.  However, it may be
that the [Cu/Fe] versus [Fe/H] diagram can be explained by normal measurement
errors plus a trend that is slightly more complex than initially expected.  It
is also possible that a small number of foreground disk stars are present, with
lower [Cu/Fe] ratios, and serve to increase the apparent dispersion.

\begin{center}
\begin{figure}
\includegraphics[width=3.0in]{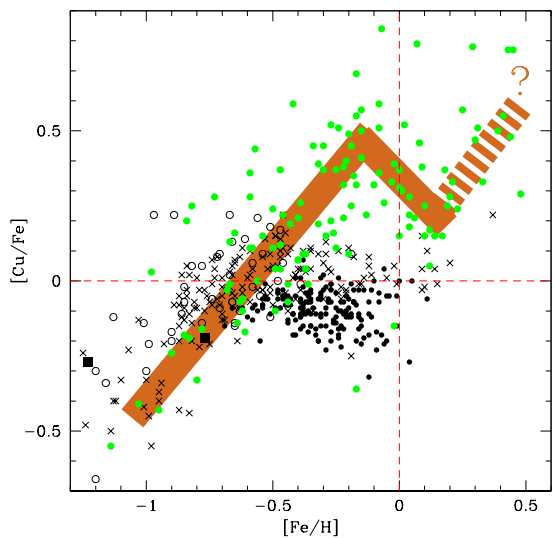}
\caption{ 
A zoomed-in portion of [Cu/Fe] versus [Fe/H] in the bulge, MW disks and halo;
symbols have the same meaning as in Figure~\ref{fig-bulge-cufe}.
The chocolate-colored strip indicates the bulge trend adopted in this work.
The rise in [Cu/Fe] with increasing [Fe/H], above [Fe/H]$\sim$$-$1 dex, 
results from the metallicity-dependent Cu yield from SNII progenitors.
A subsequent decline in bulge [Cu/Fe] above [Fe/H]$\sim$$-$0.2 dex is likely 
due to the delayed addition of Fe from SNIa; similar declines are seen for
the MW thick and thin disks, but at lower [Fe/H] and [Cu/Fe] than the bulge.
The final, uncertain, rise in bulge [Cu/Fe] above solar metallicity (marked
by a dashed chocolate strip), if real, indicates the rising metal-dependent
Cu yield overwhelming the SNIa Fe for super metal-rich stars.  The [Cu/Fe] 
trend seen here, based on the abundances of Johnson et al. (2014), are 
qualitatively consistent with a higher SFR in the bulge than the MW disks, 
as concluded from the bulge [O/Fe] and [Mg/Fe] ratios.  The zig-zag shape
is likely due to metal-dependent Cu yields from massive stars and the
late addition of SNIa iron.
}
\label{fig-bulge-cufezoom}
\end{figure}
\end{center}

Figure~\ref{fig-bulge-cufezoom} shows an expanded portion of the [Cu/Fe] versus
[Fe/H] plot, with a trend for the underlying metallicity-dependent [Cu/Fe] ratios
indicated by a chocolate-colored zig-zag swath.  In this expansion, the same linear,
metal-dependent, increasing [Cu/Fe] trend that is seen in the MW thick disk
can be seen in the bulge; but, the bulge trend continues, linearly, to 
[Cu/Fe]$\sim$$+$0.45 dex by [Fe/H]$\sim$$-$0.2 dex.  Above [Fe/H]=$-$0.2 dex, the
bulge [Cu/Fe] ratio declines with increasing [Fe/H], similar to the slopes
displayed by the thin and thick disks.  

The rise to high [Cu/Fe] in the bulge could be explained by the lack of 
Fe from SNIa, which reduces [Cu/Fe] in the MW disks; instead, the [Cu/Fe] 
trend continues to increase with metallicity, due to the weak s-process 
in massive stars.  The subsequent decline in the [Cu/Fe] ratio in the bulge
can then be due to the addition of SNIa Fe beginning at a higher [Fe/H] in
the bulge than in the disks.  The scatter of points about the chocolate-colored 
swaths in Figure~\ref{fig-bulge-cufezoom} may reasonably be due to measurement
uncertainty, and so there would be no need to appeal to inhomogeneous chemical
enrichment of the bulge.  For this reason, it would be profitable to re-visit
the Cu abundances of bulge stars with improved accuracy.

The dashed chocolate-colored swath in Figure~\ref{fig-bulge-cufezoom}, followed
by a question mark, indicates a further speculation that would explain the
bulge points near [Cu/Fe]~$\sim$$+$0.5 and [Fe/H]$\sim$0.4 dex.  At such
high [Fe/H] it is possible that those points are be due to blended Cu~I lines.
However, an alternative explanation is that after the decline in [Cu/Fe], due
to the late addition of Fe from SNIa, the metal-dependent Cu yield continues
to increase, into the super-metal-rich (SMR) regime, and these higher copper 
yields from SNII progenitors eventually overwhelm the SNIa iron.  In this way, 
the [Cu/Fe] ratio for SMR bulge stars again increases with increasing [Fe/H].
If this is true, then one should expect a similar turn-around in the MW disks.
A few of the most metal-rich disk stars in Reddy et al. (2006) 
trend to higher [Cu/Fe], but the number of points is insufficient for a solid
conclusion; also, these stars may suffer from blends with the Cu~I lines.
Remarkably, the chemical abundance study of stars within 15pc of the sun,
by Allende~Prieto et al. (2004) found a clear trend of rising [Cu/Fe] with
increasing [Fe/H] for thin disk stars above the solar [Fe/H]; below solar 
[Fe/H] the Allende~Prieto et al. (2004) results agree with thin disk [Cu/Fe]
compositions of Reddy et al. (2006).  
Support for this rising [Cu/Fe] trend in thin disk stars above solar [Fe/H] 
is also evident in Figure~3 of the chemical abundance study of Feltzing
\& Gustafsson (1998).  I note that the slope of the SMR [Cu/Fe]
versus [Fe/H] is consistent between these three thin disk studies,
at $\sim$0.1 dex/dex, and also in rough agreement with the slope
of the SMR rise of [Cu/Fe] in the bulge proposed here.

Interestingly, the zig-zag shaped [Cu/Fe] trend with [Fe/H] is
similar to, but larger amplitude than, the [Na/Fe] trend.  Both
are consistent with metal-dependent yields from massive stars and 
the late addition of iron from SNIa.


These conclusions about the bulge [Cu/Fe] trend are
qualitatively consistent with the idea that the bulge SFR was higher than
the MW disks, and the SNIa time-delay scenario, proposed by Tinsley (1979)
and Matteucci \& Brocato (1990), to explain the trend of [O/Fe] in the MW.
A detailed chemical evolution model for [Cu/Fe] needs to be consistent with 
the [O/Fe] trend in the bulge, and should provide constraints on the 
metal-dependent Cu yields and/or the bulge SFR. I note that Ga, Ge and
Se are also expected to be over-produced by the weak s-process, and these
would provide a check on the conclusions drawn from Cu; unfortunately,
these elements are spectroscopically challenging at solar [Fe/H].  Moderate
enhancements of Rb and $^{25}$Mg are also expected from the weak s-process,
and may also form the basis of a consistency check on Cu.  These species
should be measurable through the lines of Rb~I at 7800\AA\ and 7947\AA\
and MgH in the 5100--5150\AA\ region.

I conclude that
the enhanced [Cu/Fe] ratios in bulge stars suggests a greater
role for SNII nucleosynthesis in the bulge than the MW disks, due to a higher
SFR in the bulge.

\subsection{Zinc}

Similar to copper, zinc also thought to be produced by a combination of alpha-rich
freeze-out during core-collapse SNe and the weak s-process in SNII progenitor
stars (see Bisterzo et al. 2004; Pignatari 2008, 2010; Woosley \& Weaver 1995).  However,
the alpha-rich freeze-out is thought to be much more important for zinc, relative to
the weak s-process, compared to the situation for copper (Bisterzo et al. 2004).

The earliest attempted Zn measurements for metal-poor stars includes studies by
Cohen (1978, 1979) and Luck \& Bond (1985).  However, Sneden \& Crocker (1988)
and Sneden, Gratton \& Crocker (1991) were the first to firmly establish
the [Zn/Fe] abundance trend with [Fe/H] in the MW halo and disks.  Modern MW disk/halo
Zn abundance studies, with larger samples and smaller measurement uncertainties,
include the work of Bensby et al. (2003, 2005) and Reddy et al. (2003, 2006).

These studies have employed the 4722 and 4810 Zn~I lines, which are fairly strong, so
better for lower than higher metallicities.  Allende~Prieto et al. (2004) reported 
that the 4722\AA\ line has strongly damped wings, larger than predicted
by the Uns\"old approximation.  Inspection of the Arcturus Atlas (Hinkle et al. 2000)
suggests to me that the line is blended in both wings.  The 6362\AA\ Zn~I line is also
employed, but this line lies within a broad Ca~I auto-ionization feature, which
should be properly included in the analysis; to me, the red wing of this line appears
to be blended with a V~I line in the spectrum of Arcturus.

The above abundance studies indicate [Zn/Fe] roughly flat with [Fe/H] for the 
MW disks, but with a mean [Zn/Fe]$\sim$$+$0.10 to $+$0.15 dex for stars below 
[Fe/H] less than $\sim$$-$0.4 dex.  While the thick disk stars appear 
slightly enhanced compared to the thin disk, this seems to be due to a 
subtle slope of declining [Zn/Fe] with increasing [Fe/H] above [Fe/H]=$-$0.4 dex.  
For stars greater than solar [Fe/H] the results of Bensby et al. (2003, 2005) suggest
a small increase in [Zn/Fe] with [Fe/H], while the Reddy et al. (2003, 2006) results
show a flat trend at [Zn/Fe]=0.0 dex.  Notably, the Reddy et al. (2003, 2006) samples
have many fewer stars with super-solar [Fe/H] than the Bensby et al. (2003, 2005)
studies.  


The study of solar neighborhood of stars within 15pc, by Allende~Prieto
et al. (2004), found strongly increasing [Zn/Fe] for stars above
solar metallicity, up to [Zn/Fe]=$+$0.5 dex for [Fe/H]=$+$0.5 dex,
based on only the 4810\AA\ Zn~I line. While such a strong increase in
[Zn/Fe] is not supported by Bensby et al. (2003, 2005), the fact that both
studies find increasing [Zn/Fe] with increasing [Fe/H] above solar suggests
that there is a real trend.  Interestingly, this would be consistent with
a nucleosynthetic component from the weak s-process, as suggested by
Bisterzo et al. (2004)




To date, only two bulge abundance studies have measured zinc: Bensby et al. (2013)
and Barbuy et al. (2015).  These two studies are in approximate agreement below 
solar [Fe/H], with a $\sim$$+$0.1 dex enhancement in [Zn/Fe].  Above solar metallicity,
and up to [Fe/H]=$+$0.5 dex, the lensed dwarfs of Bensby et al. (2013) show a similar
$\sim$$+$0.1 dex [Zn/Fe] enhancement; however, the RGB stars of Barbuy et al. (2015)
show a strong decline in [Zn/Fe] up to their most metal-rich star.

My interpretation of the Barbuy et al. (2015) result is that the decline in [Zn/Fe]
begins near [Fe/H]=$-$0.2 dex, with [Zn/Fe] starting near $+$0.1 dex, and declining 
to [Zn/Fe]=$-$0.4 dex by [Fe/H]$\sim$$+$0.3 dex.  This indicates a maximal slope in 
the [Zn/Fe] versus [Fe/H] trend, that can only occur by adding Fe but no Zn.  Since
SNIa make iron but effectively no zinc, one way to reproduce the
Barbuy et al. (2015) bulge [Zn/Fe] trend is in the case of iron enrichment 
dominated by SNIa above [Fe/H]$\sim$$-$0.2 dex.

At the moment, we can say that either the bulge [Zn/Fe] trend is identical to the MW disks,
to within the measurement uncertainty, according to Bensby et al. (2003, 2005, 2013),
or that the stars above [Fe/H]=$-$0.2 dex show the maximum possible decline in [Zn/Fe]
with [Fe/H], according to the results of Barbuy et al. (2015).  This latter possibility
requires addition of Fe without Zn, as might be expected from material dominated by SNIa
ejecta; this would have implications for understanding other elements.
The putative paucity of Zn in SMR bulge stars 
would also suggest a sudden absence of the weak s-process contribution from SNII progenitors.


Clearly, further chemical abundance studies are required to resolve the differences
between extant conclusions about the basic trends of [Zn/Fe] in the MW disks and the
bulge.  A new study on [Zn/Fe] in SMR disk dwarf stars would also be profitable.  
Furthermore, it might be helpful to measure [Zn/Fe] for a sample of disk red 
giant stars, in order to check the results from the dwarfs.  For bulge stars, higher
S/N spectra would be valuable.

\section{Neutron-Capture Elements}

All elements beyond the iron-peak are made by neutron-capture, typically
from a combination of both rapid, r-, and/or slow, s-, processes (e.g.
Burbidge et al. 1957).  However, a few rare proton-rich isotopes may result
from a variety of other
mechanisms occurring in special circumstances; for example,
the $\gamma$ process (e.g., Howard, Meyer \& Woosley 1991)
rp-process (e.g., Wallace \& Woosley 1981) and $\nu$p-process
(Fr\"ohlich et al. 2006).


Elements in the s-process peaks (e.g., Sr, Y, Zr, Ba, La, Ce), resulting
from their small neutron-capture cross sections as a consequence of 
nuclear shell structure, are often referred to as ``s-process'' elements, 
even though, in the sun, the r-process contributes 15 to 28\% of these
(Simmerer et al.  2004).  Notably, at low metallicity and in metal-poor
r-process rich stars these same elements are made mostly by the
r-process (e.g., Sneden et al. 1996).

Similarly, elements commonly referred to as ``r-process'',
such as I, Pt, Au, Eu, Tb, Gd, Dy and Ho range from 3 to 18\% 
s-process in the solar composition (e.g., Simmerer et al. 2004).
The only 100\% r-process elements are those beyond Bi/Pb; all of
these are radioactive, although U and Th have very long  half-lives.

\subsection{The Solar Neighborhood}

It is surprising that the abundance trends [X/Fe] of all the 
neutron-capture elements, in the MW thick and thin disks are not yet
fully explored.
Chemical abundance measurements for neutron-capture elements in 
solar neighborhood stars have been investigated for the last 56 years,
beginning with Baschek (1959) and Aller \& Greenstein (1960), with
increasingly sophisticated analysis and ever larger samples of stars.  

For the current discussion of neutron-capture elements in the MW thin
and thick disks it is expedient to rely on a few, relatively recent,
high-precision chemical abundance studies.  In particular, I shall rely on
the chemical abundance studies for hundreds of MW disk dwarf stars,
performed by Edvardsson et al. (1993),
Koch \& Edvardsson (2002), Bensby et al. (2005, 2014) and
Reddy et al. (2003, 2006), Mishenina et al. (2013), and very recently
by Battistini \& Bensby (2016). 
Together, these studies encompass Sr, Y, Zr, Ba, La, Ce, Nd, and Eu;
lines of other neutron-capture elements are not easily detected in
dwarf turnoff stars.

Other useful studies encompassing MW disk stars 
include: RGB stars by Fulbright (2000, 2002) for Y, Zr, Ba and Eu;
Simmerer et al. (2004) for La and Eu in dwarf and RGB stars;
Feltzing \& Gustafsson (1998) for Y, La, Nd (and noisy measurements
of Zr, Mo and Hf) for metal-rich 
dwarf stars above [Fe/H]=$+$0.1 dex, and Koch \& Edvardsson (2002) for Eu
in disk dwarfs.   This list is necessarily incomplete; I apologize to
those excellent researchers whose work I have not mentioned here, but
who's efforts were important for our current understanding. 

%

\begin{center}
\begin{figure}
\includegraphics[width=3.0in]{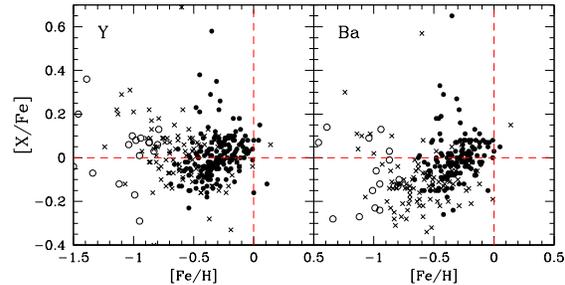}
\caption{ [Y/Fe] and [Ba/Fe] in the MW thin disk (filled circles), thick disk
(crosses) and Halo (open circles) from Reddy et al. (2003, 2006).  The [Y/Fe]
trend is roughly flat, while [Ba/Fe] is deficient by $\sim$0.1 to 0.2 dex
in the thick disk, or metal-poor stars and rises with [Fe/H] in the thin disk.
}
\label{fig-reddy-bay}
\end{figure}
\end{center}

For barium, Edvardsson et al. (1993), Bensby et al. (2005) and
Reddy et al. (2006) found a flat trend of [Ba/Fe] with [Fe/H] in the
thick disk, but slightly deficient, at [Ba/Fe]=$-$0.1 dex over the
range [Fe/H]=$-$1 to $+$0.0 dex; see Figure~\ref{fig-reddy-bay}
for the Reddy et al. (2003, 2006) results.

For the thin disk Edvardsson et al. (1993) found [Ba/Fe] mostly flat, near
the solar ratio; but, the youngest stars showed slightly higher [Ba/Fe]
near [Fe/H]$\sim$$-$0.2 dex and a slight deficiency near [Fe/H]=$+$0.2
dex, suggesting a small downward trend.  The Bensby et al. (2005) results
show some scatter in the thin disk [Ba/Fe] at these metallicities,
roughly consistent with the Edvardsson et al. (1993) results.  On the 
other hand, Reddy et al. (2003) found a flat trend of [Ba/Fe] with [Fe/H] 
at [Ba/Fe]$\sim$0.0 dex for the thin disk at all metallicities, as seen in
Figure~\ref{fig-reddy-bay}.

\begin{center}
\begin{figure}
\includegraphics[width=3.0in]{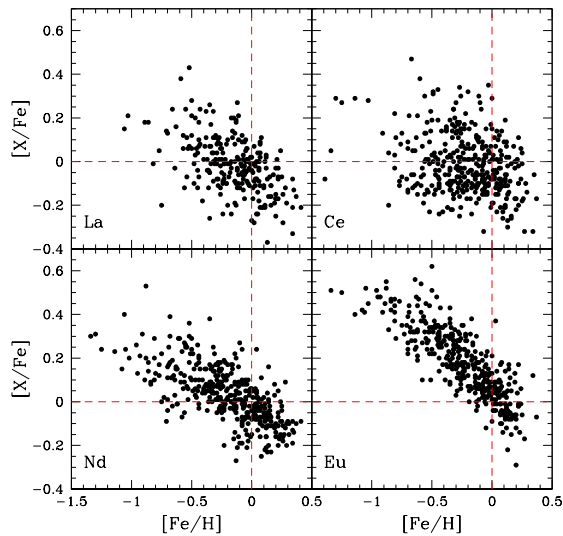}
\caption{Neutron-Capture elements in the MW disks from Battistini \& Bensby (2016).
Shallower slopes suggest larger s-process fractions.  The [Eu/Fe] decline with
[Fe/H] is similar to the $\alpha$-elements and presumably due to the
increasing importance of Fe from SNIa.  Note the average [Eu/Fe]$\sim$0.05 dex at
solar [Fe/H].}
\label{fig-ncapture-bb2015}
\end{figure}
\end{center}

Figure~\ref{fig-reddy-bay} also shows Yttrium with a flat trend, at [Y/Fe]=0.0 
dex, for both the thick and thin MW disks for Reddy et al. (2003, 2006).
This is consistent with the results of Edvardsson et al. (1993) and
Bensby et al. (2005) in dwarf stars, and the giants analyzed
by Fulbright (2002).  Note that, Reddy et al. (2003, 2006) also show a flat
[Ce/Fe] trend at the solar value, similar to the Battistini \& Bensby (2016)
results shown in Figure~\ref{fig-ncapture-bb2015}.

Data from Edvardsson et al. (1993) 
indicate a flat [Zr/Fe] trend with [Fe/H] with a possible enhancement 
of $+$0.10 dex, similar to the giant stars of Fulbright (2002).

However, the 311 stars analyzed by Battistini \& Bensby (2016) show a
clear increasing trend in [Zr/Fe] with decreasing [Fe/H], reaching
[Zr/Fe]$\sim$$+$0.4 dex near [Fe/H]=$-$1.0 dex (see
Figure~\ref{fig-zrfe-diskbulge}).  In this way, the
[Zr/Fe] trend in the MW disk appears to  be similar to the behavior 
of the $\alpha$-elements, suggesting a
strong r-process component for zirconium, or at least significant Zr production
on short time scales.  This is inconsistent with the 96\% Main-s
component assignment for Zr by Bisterzo et al. (2011), the 81\% s-process adopted
by Simmerer et al. (2004), and even the 66\% s-process adopted by 
Travaglio et al. (2004) and Bisterzo et al. (2014).

Lanthanum abundances from Simmerer et al. (2004), and references therein,
show a flat [La/Fe] trend, particularly at thick disk metallicities; a hint
of a downward trend toward solar [Fe/H] is present.  However, this putative
decline in [La/Fe] is not maintained for the super-metal-rich stars of
Feltzing \& Gustafsson (1998), which show [La/Fe]=0.0 dex up to
[Fe/H]$\sim$$+$0.4 dex; thus, it appears that [La/Fe] is roughly flat at the
solar ratio for all stars above [Fe/H]$\sim$$-$1.  However, the results of
Battistini \& Bensby (2016), shown in Figure~\ref{fig-ncapture-bb2015}, suggests
a shallow negative [La/Fe] slope with [Fe/H].  It is not clear whether this
La decline is due to increased Fe from SNIa, metallicity-dependent s-process
yields from AGB stars (e.g., Busso, Gallino \& Wasserburg 1999; Cristallo et al. 2009, 2011),
or from systematic errors in the abundance measurements.

For MW disk dwarf stars,
the studies of Woolf, Tomkin \& Lambert (1995), Koch \& Edvardsson (2002),
Reddy et al. (2003, 2006), Bensby et al. (2005) and Mishenina et al. (2013)
show that the [Eu/Fe] trend with [Fe/H] is much like the $\alpha$-elements: 
enhanced by $\sim$$+$0.4 dex below [Fe/H]$\sim$$-$1, then declining linearly 
with increasing metallicity through the solar composition.

\begin{center}
\begin{figure}
\includegraphics[width=3.0in]{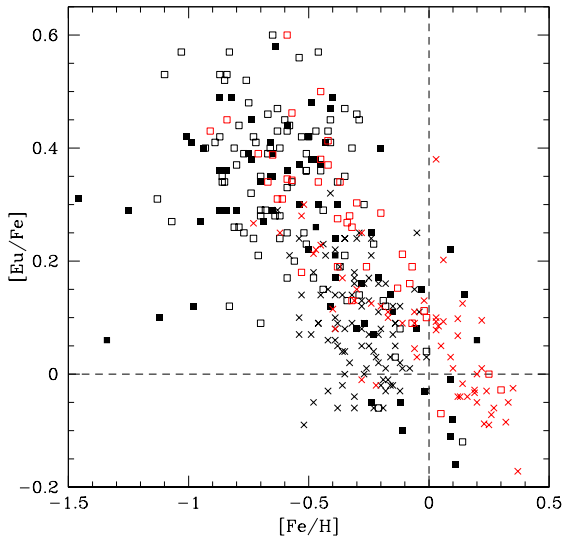}
\caption{[Eu/Fe] versus [Fe/H] in the Milky Way thick and thin disks, reported
by Bensby et al. (2005) and Reddy et al. (2003, 2006).  Open black squares:
thick disk stars from Reddy et al. (2006); filled black squares: disk/halo
stars from Reddy et al. (2006); black crosses: thin disk stars identified in
Reddy et al. (2003); open red squares: thick disk stars from Bensby et al.
(2005); red crosses: thin disk stars from Bensby et al. (2005).
Note that Bensby thin disk stars are higher than Reddy et al. (2003, 2006) by
$\sim$0.1 dex in [Eu/Fe].  }
\label{fig-eufedisk}
\end{figure}
\end{center}

While the Reddy et al. (2003, 2006) results include a number of stars with
low [Eu/Fe] ratios, the mean trend appears to go through [Eu/Fe]$\sim$0.0 dex 
at solar [Fe/H].
On the other hand, the mean Bensby et al. (2005) points intersect
[Fe/H]=0.0 dex at [Eu/Fe]$\sim$$+$0.1 dex (see Figure~\ref{fig-eufedisk});
furthermore, the recent results of Battistini \& Bensby (2016) also show an 
enhanced average [Eu/Fe] ratio, near $\sim$0.05 dex, at solar metallicity.
A small Eu-enhancement is also seen in the results of Misheninia et al. (2013),
indicating the average [Eu/Fe]$\sim$$+$0.03 dex, at solar iron abundance.

If the [Eu/Fe] enhancement seen in Bensby et al. (2005) is real, then the
sun is Eu-deficient relative to other solar-metallicity stars by $\sim$0.1 dex.
However, composition studies of solar twins (e.g. Ramirez et al. 2009)
show very small element abundance scatter, and no obvious zero-point offsets,
in neutron-capture element abundances; although the logic of this argument is
somewhat circular.  

For solar [Fe/H] MW disk stars in Battistini \& Bensby (2016),
the average Sr, La, and Ce lie slightly below the solar [X/Fe] ratio,
while Sm and Eu lie slightly above, and the mean [Zr/Fe] and [Nd/Fe] are
at roughly solar composition. 
The inconsistency across different elements, the scatter between studies,
and the $+$0.03 to $+$0.05 dex level of the putative zero-point offsets
all suggest that they are due to measurement error.
Furthermore, an [Eu/Fe] trend that passes through the solar composition
fits the reported abundances to within the abundance measurement errors.

On the other hand, the existence of a real zero-point abundance offset in
[Eu/Fe] should be considered in the context of the Matteucci \& Brocato (1990) 
chemical evolution scenario, used to explain the trend of [O/Fe] in the MW.  
The $\alpha$-like decline of [Eu/Fe] with [Fe/H] suggests that, like
oxygen, the r-process is associated with short-timescale progenitors.  Presently
favored sites for the r-process include low-mass (8--10 M$_{\odot}$) 
O-Ne-Mg SNII (Wanajo et al. 2003) and merging neutron stars
(Lattimer \& Schramm 1974; Symbalisty \& Schramm 1982), recently resurrected
(e.g., Goriely et al. 2011; Wanajo et al. 2014).

Consistent with this explanation for the $\alpha$-like trend of Eu is 
the decline in [Nd/Fe] with increasing [Fe/H], which shows a shallower
slope than Eu; Nd is thought to be roughly half r- and half s-process,
whereas $\sim$94\% of Eu is attributed to the r-process.

In summary, the current data indicates that the classical s-process elements,
Sr, Y, Ba, La, and Ce, show roughly flat trends with [Fe/H], near the solar
[X/Fe] value, while the r-process element Eu exhibits a steep decline, from
[Eu/Fe]$\sim$$+$0.4 dex down to 0.0 dex at solar [Fe/H], similar to the 
trend seen for the $\alpha$-elements.   

Neutron-capture elements with significant r-process 
contributions show declining [X/Fe] ratios with increasing metallicity, like
the $\alpha$-elements, but the larger the s-process contribution, the shallower
the slope.   The s-process component is the likely reason that Nd
shows a slope declining from [Nd/Fe] of only $+$0.25 dex in the halo, compared
to the 0.4 dex decline for [Eu/Fe].  
The decline in Sm is greater than for Nd, as one might expect from the smaller
s-process contribution to Sm .
However, the samarium trend closely resembles the slope of the Eu and the 
$\alpha$-element trends, suggesting that Sm has a higher r-process fraction
than the value of 66\% to 69\%, suggested by Burris et al. (2000),
Simmerer et al. (2004), and Bisterzo et al. (2014).  

Alternatively, a steeper than expected decline in [Sm/Fe] could be due to
the metallicity-dependent yield of heavy s-process elements, as detailed 
by Busso et al. (1999).  
In this case, the number of iron-peak seed nuclei increase with increasing
metallicity, while the neutron irradiation is roughly constant; thus, at 
higher [Fe/H] each seed nucleus captures fewer neutrons, and the s-process
production of the heaviest nuclei is reduced, while the production of
light s-process elements is increased.

The $\alpha$-like slope of Zr, as measured by Battistini \& Bensby (2016), may
lead one to conclude that it, too, is dominated by the r-process at
[Fe/H]$\sim$$-$1, despite the strong s-process component
in the solar composition.  Alternatively, the strong Zr slope may be due to
other nucleosynthetic processes associated with short timescales, and 
massive stars, such as the alpha-rich freeze-out, or other light element
production process (e.g. Woosley \& Hoffman 1992).

The fact that the classical s-process elements show flat [X/Fe] ratios, even
though the chemical evolution scenario of Matteucci \& Brocato (1990),
indicates that the delayed addition of iron from SNIa cause the [$\alpha$/Fe] 
ratios to decline by $\sim$0.4 dex, shows that the production of the
s-process elements must have increased along with the delayed injection of
SNIa iron.  This is qualitatively consistent with the idea that the main 
component of the s-process is due to relatively low-mass stars, at 
2--3 M$_{\odot}$ (e.g., see summary in Busso et al.  2004), whose lifetimes 
are not too dissimilar to the delayed SNIa progenitors.
Thus, elements produced by the main s-process may use used to
infer long formation timescales.

%
%

\subsection{Neutron-Capture Elements in the Bulge}

While a  number of bulge chemical abundance studies have reported results for
various neutron-capture elements, the situation is not yet in a satisfactory
state, due to measurement errors, number of stars, and number of elements studied;
in particular, key diagnostic abundance ratios remaining poorly constrained.

The neutron-capture element abundance studies considered here include the RGB
stars of MR94 and work performed by Fulbright (unpublished) appearing in
the MFR10 (henceforth MFR10/Fu), the Red Clump giants studied by Johnson et al. (2012)
and van der Swaelmen et al. (2015), and the lensed dwarfs of Bensby et al. (2013).

The main result is that the bulge neutron-capture abundances are roughly
similar to the MW thin and thick disks.  However, results for the [Eu/Fe] 
trend are most consistent with a slight enhancement at solar [Fe/H],
similar to the observed bulge $\alpha$-element enhancements (see
section~3 ).  Furthermore, the [La/Eu] ratios indicate that 
the rise of the s-process component starts at a higher [Fe/H] in the bulge 
than the MW disks.  
In the context of the MB90 chemical evolution scenario, both of these observed 
abundance trends are consistent with the idea that the SFR in the bulge was 
higher than in the MW thin and thick disks.

\subsubsection{The r-process and [Eu/Fe] }

To date, europium is the only nearly pure r-process element whose abundances
have been measured in the bulge; although, bulge abundance measurements for
neodymium (roughly half r-process in the sun) are available.

Very noisy europium abundance measurements were first attempted, for 11
bulge RGB stars, by MR94, who concluded that [Eu/Fe] appeared slightly enhanced,
comparable to the trend with [Fe/H] seen in solar neighborhood disk stars;
but the MR94 Eu measurements were very noisy.

\begin{center}
\begin{figure}
\includegraphics[width=3.0in]{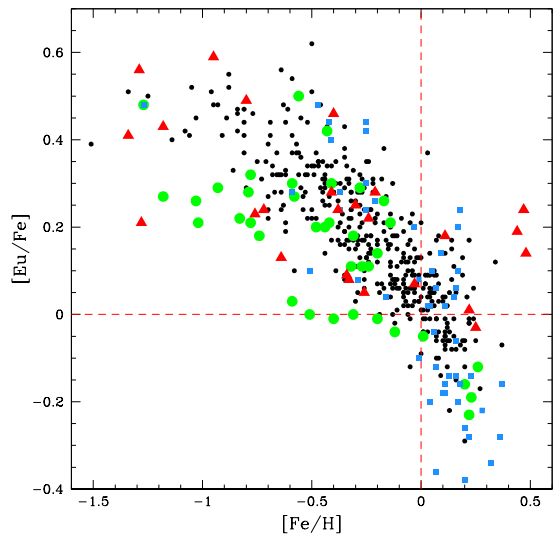}
\caption{[Eu/Fe] versus [Fe/H] in the bulge compared to the MW disk
data of Battistini \& Bensby (2016; filled black circles).  Filled
red triangles: MFR10/Fu, normalized with Fe~II line abundances. Filled
green circles represent [Eu~II/Fe~I] from Johnson et al. (2012), showing
overall deficient [Eu/Fe] at all [Fe/H].
Filled blue squares indicate [Eu~II/Fe~I] from van der Swaelmen et al. 
(2015), with stars mostly above solar [Fe/H].
}
\label{fig-eufebulge}
\end{figure}
\end{center}

MFR10/Fu reported [Eu/Fe] EW abundance measurements for 25 bulge RGB stars
(filled red triangles in Figure~\ref{fig-eufebulge}), where abundances from 
Fe~II lines were employed to ratio with the Eu~II line at 6645\AA .  The 
bulge MFR10/Fu [Eu/Fe] trend with [Fe/H] is identical, within the errors, 
to the $\alpha$-like downward trend with [Fe/H] seen in the MW disk data of 
Bensby et al. (2005).
As previously noted, the [Eu/Fe] versus [Fe/H] trend for MW disk stars, by
Bensby et al. (2005), does not pass through the solar composition, but is
europium-rich, with [Eu/Fe]$\sim$$+$0.1 dex at [Fe/H]=0.0 dex; thus, either
there is a zero-point error in Bensby et al. (2005), or the sun is Eu-deficient.
I note that the three most metal-rich stars in MFR10/Fu, near
[Fe/H]$\sim$$+$0.5 dex (see Figure~\ref{fig-eufebulge}), lie above
the extrapolated bulge trend, possibly signalling unaccounted blending
of the Eu~II line at high metallicity.  

Even excluding the most metal-rich stars, the MFR10/Fu [Eu/Fe] trend (shown
in Figure~\ref{fig-eufebulge}) indicates enhanced [Eu/Fe] near
$\sim$0.1 dex at solar [Fe/H].  In the MFR10/Fu data the bulge reaches
the solar [Eu/Fe] ratio at [Fe/H]=$+$0.2 dex.  While
noisy, and based on relatively few stars, the use of Fe~II lines for the
[Eu/Fe] ratios makes the MFR10/Fu results robust against systematic errors 
in the adopted model atmosphere gravities and [$\alpha$/Fe] ratios.

Europium abundances measured for 39 and 56 mostly RGB stars by
Johnson et al.  (2012) and van der Swaelmen et al. (2015) respectively, 
show considerable scatter in Figure~\ref{fig-eufebulge}.  At least some of
this noise was probably due to the relatively low resolution (R$\sim$25,000)
spectra of the Johnson et al. (2012) Blanco Hydra fiber data;
van der Swaelmen et al. (2015) employed VLT/UVES spectra at
R$\sim$45,000.  The Johnson et al. (2012) S/N was significantly higher
than van der Swaelmen's, such that the overall spectral quality was similar.
By comparison, the spectra used for the MFR10/Fu abundances were based on
Keck/HIRES R$\sim$45,000 to 60,000 spectra and S/N$\sim$45--100 per pixel.
An additional source of noise results from the use of iron
abundances from Fe~I lines, instead of Fe~II lines, to compare with Eu~II
abundances for the [Eu/Fe] ratio in the studies of Johnson et al. (2012)
and van der Swaelmen et al. (2015).  Because [Eu~II/Fe~I] ratios are sensitive 
to the H$^-$ opacity, significant systematic errors in the ratio can result 
from errors in the adopted logg, [Fe/H] and [$\alpha$/Fe] ratios.

The Johnson et al. (2012) [Eu/Fe] ratios show a declining trend
with increasing [Fe/H], but the points lie 0.1 to 0.2 dex below the MW disk
ratio at all [Fe/H].  Notably, points below [Fe/H]$\sim$$-$0.8 dex are roughly 
0.2 dex lower than MFR10/Fu.  If we choose to employ these metal-poor points 
to set and correct a systematic zero-point error, then the more metal-rich 
[Eu/Fe] trend lies above the MW disk trend, roughly consistent with the 
MFR10/Fu results, and indicating an enhanced [Eu/Fe] ratio at solar [Fe/H].  

%

The van der Swaelmen et al. (2015) [Eu/Fe] ratios also show an $\alpha$-like 
decline with [Fe/H], similar to the MW disk, but higher than 
Johnson et al. (2012) with a large scatter above solar [Fe/H].  For the few
points at low metallicity, the [Eu~II/Fe~I] values show no offset relative 
to MFR10/Fu or the MW disk results, and therefore there is no case for an 
overall zero-point shift.  The van der Swaelmen et al. (2015) sample is heavily
weighted to super-solar [Fe/H]; for these metal-rich stars, the range of
[Eu/Fe] at roughly fixed [Fe/H] is $\sim$0.8 dex, suggesting a 1$\sigma$
dispersion of $\sim$0.2 dex.

While all studies of [Eu/Fe] in the bulge show an $\alpha$-like decline with 
[Fe/H], the detailed study-to-study differences makes it difficult to 
determine whether the
bulge [Eu/Fe] is slightly enhanced relative to the MW disks, as one would
expect based on the slightly enhanced bulge [O/Fe] and [Mg/Fe] ratios.
Comparison with the MW disk is complicated by the reported enhanced
disk [Eu/Fe] ratios at solar [Fe/H]; this is most likely due to zero-point 
errors in the disk star measurements, but could be due to a europium 
deficiency in the sun.

Taking these issues into consideration, I think that MFR10/Fu provide the 
most robust estimate of the bulge [Eu/Fe] trend with [Fe/H], due to the 
higher resolution spectra employed and the use of Fe~II lines in the normalization.

I assume that the variance in reported [Eu/Fe] values, for MW disk stars
at solar-metallicity, indicate the presence of small zero-point
measurement errors, rather than a genuine deficiency
of europium in the Sun.  Thus, the bulge [Eu/Fe] data are consistent with 
a small enhancement in [Eu/Fe], roughly $+$0.1 dex at solar [Fe/H], relative 
to the MW disk trend.  This is in qualitative agreement with expectations 
from the measured enhancements of [O/Fe] and [Mg/Fe] in the bulge, 
suggesting that the SFR was higher than the MW thin and thick disks.  

The scatter in reported bulge [Eu/Fe] abundance ratios leaves open the
possibility that the [Eu/Fe] trend with [Fe/H] might not be consistent with
the measured [O/Fe] and [Mg/Fe] enhancements.  In that case, it would be
necessary to add complexity to the chemical evolution scenario of MB90.
 
Clearly, more Eu abundance work is required for both disk and bulge stars,
with the goal of reducing the measurement uncertainties.  This may be
achieved by the acquisition of higher S/N, higher resolution, spectra
of Red Clump bulge stars, and including additional Eu~II and Fe~II lines, 
for line-by-line differential, profile-matching, spectrum synthesis 
abundance analyses.

While europium is the most pure r-process element measured in bulge stars, 
at 94\% r-process in the solar composition (Bisterzo et al. 2014),
neodymium has a significant solar r-process component, at 43\%. 

The trend of [Nd/Fe] with [Fe/H] in the bulge, as reported by Johnson
et al. (2012) and van der Swaelmen et al. (2015) shows  a strong downward
sloping trend, suggesting a dominant r-process component for Nd.  The
overall trend from both studies is lower than the MW disk by 0.15 to 0.20 dex.
This deficiency might be real, say due to a low s-process component for Nd,
or could be due to systematic error, resulting from an improper model
atmosphere gravity and the fact that Nd~II line abundances were ratioed
with Fe~I line abundances, instead of Fe~II.   However, since the bulge Y
and Ba abundance trends for lensed dwarf stars (see section~3) are
similar to the MW disks, and because of similar deficiencies in the [Eu/Fe] 
trend for these studies, we assume that the low bulge [Nd/Fe] ratios from
Johnson et al. (2012) and van der Swaelemen et al. (2015) are likely due
to systematic error, perhaps to the adopted model atmosphere parameters.  
The very large scatter in measured bulge Nd abundances shows that more work 
on this element is required, employing superior spectra and
more robust analysis techniques.


\subsubsection{An R-Process Rich Bulge Star}

Johnson et al. (2012) found [Eu/Fe]=$+$0.93 dex for their most metal-poor bulge 
star, at [Fe/H]=$-$1.54 dex.  A follow-up study of this star, by
Johnson et al. (2013), employed a MIKE-Magellan spectrum with higher S/N
spectra and twice the resolving power, at R=40,000.  The new abundance
analysis included
Fe~II lines, which enabled a gravity-insensitive [Eu/Fe] ratio to be measured;
the improved spectrum yielded [Fe/H]=$-$1.67 and [Eu/Fe]=$+$0.99 dex.  Thus,
[Eu/Fe] for this star is $\sim$0.6 dex higher than the typical for the MW halo.
Furthermore, Johnson et al.  (2013) found that the heavy-element abundance
distribution in this star was well fit by the solar r-process pattern.  Thus,
this star appears to be a mild version of the r-process rich stars in the MW 
halo (e.g. Sneden et al. 1994, 1996), but with [Fe/H] more than 1.0 dex higher than
the halo r-process rich stars.

The explanation for the halo r-process rich stars is that the neutron-capture 
elements come from a single r-process event and that such events are rare,
at most 1/40 of the total SN population (e.g.  McWilliam et al. 1995).
If Johnson's bulge r-process rich star resulted from the same mechanism, 
then at the metallicity of the halo r-process rich stars, the bulge star would
have [Eu/Fe]=$+$2.0 dex, roughly 0.4 dex higher than the halo r-process stars.

Did the bulge r-process star material result from a single unusually strong
r-process event, later diluted by normal material up to [Fe/H]=$-$1.67 dex?
Alternatively, could the r-process dispersion seen in the MW halo be shifted
to higher [Fe/H] in the bulge, perhaps as a result of a faster chemical
enrichment timescale for iron, or dilution with less pristine gas in the bulge?

If the bulge r-process dispersion is shifted to higher [Fe/H] than the halo,
one expects the slightly more metal-poor bulge stars to show a large
dispersion and steeply declining [Eu/Fe] ratios.  In this regard, the study
of Koch et al. (2015) found the [Eu/Fe] of their 4 bulge stars with
[Fe/H]$\sim$$-$2 dex to be near 0.0 dex, roughly 0.4 dex lower than halo
stars at the same metallicity.  Thus, it is possible that the dispersion
in the [r-process/Fe] trend is shifted to higher [Fe/H] in the bulge than
the MW halo.

The relatively high [Fe/H] of Johnson's bulge r-process rich star is similar
to three r-process rich stars in the Ursa Minor dwarf galaxy 
(Sadakane et al. 2004; Cohen \& Huang 2010), with [Fe/H] between $-$1.7 and $-$1.5
dex, and [Eu/Fe]$\sim$$+$0.9 dex.  This raises the possibility that Johnson's bulge
r-process star was captured from a dwarf galaxy.

\subsubsection{The s-/r- process ratio and the rise of the s-process}

\begin{center}
\begin{figure}
\includegraphics[width=3.0in]{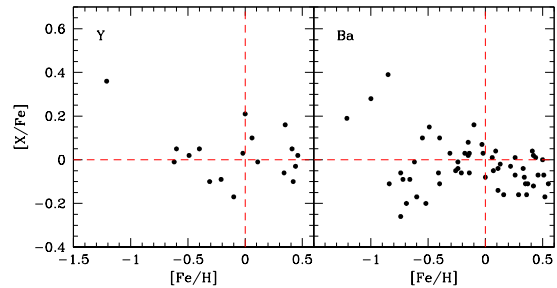}
\caption{Abundance [X/Fe] trends, as a function of [Fe/H], for classical first
and second s-process peak elements (Y and Ba respectively), for lensed
dwarf stars measured by Bensby et al. (2013).  The slight deficiency and
flat trend is similar to the MW disks (see Figure~\ref{fig-reddy-bay}).
}
\label{fig-baybulge}
\end{figure}
\end{center}

In Figure~\ref{fig-baybulge}, the classical first and second
s-process elements Y and Ba, measured in lensed bulge 
dwarf stars by Bensby et al. (2013), show roughly flat trends, and a
slight Ba deficiency;  this is close to the behavior of these two
elements in the MW disk (appearing in Figure~\ref{fig-reddy-bay}),
although the comparisons more closely resemble the thick disk trends.

Regarding lanthanum, the bulge results from MFR10/Fu, Johnson et al. (2012),
and van der Swaelmen et al. (2015) all show a subtle decline in [La/Fe] with
increasing [Fe/H].  It is notable that the Johnson et al. (2012) [La/Fe] 
points lie $\sim$0.15 dex lower than the MFR10/Fu values, and both trends 
are lower than the MW disk results of Battistini \& Bensby (2016).
On the other hand, the MFR10/Fu [La/Fe] trend with [Fe/H] is
practically identical to that found by Simmerer et al. (2004) for the MW 
disks and halo.  Thus, similar to the situation for Eu, the comparison of the
bulge [La/Fe] results with the MW disks is not fully constrained, partly 
because the MW disk trends have not yet converged at the 0.1 dex level.

One potential cause for the variance in reported MW disk [La/Fe] trends may
be related to whether the target stars are red giants or dwarfs.  
Difficulties for the comparison of the disk and bulge [La/Fe] trends include:
the possibility of systematic errors, such as the absence of Fe~II line 
abundances for [X/Fe] ratios in the Johnson et al.  (2012) and van der
Swaelmen et al. (2015) studies; difficulties with accounting
for blends and continuum, and the small line depths, that characterize the
lower-resolution studies of Johnson et al. (2012) and van der Swaelmen et al. 
(2015); and, possible unaccounted blending affecting the EW method for the 
most metal-rich stars studied by MFR10/Fu.

At this point, the question of whether the abundance of classical s-process
elements, compared to iron, is higher, lower, or the same in the bulge as
the MW disk depends entirely on which studies are chosen for the comparison.

Regardless of zero-point uncertainties, [Nd/Fe] and [La/Fe] are lower
at high metallicity in both the bulge and the MW disks and in both 
these Galactic components the slope is steeper for Nd than La.

This metal-dependence may simply be due to the predicted decline in s-process
yields from AGB stars with increasing metallicity (e.g. Cristallo et al. 2009,
2011), or may be due to the decline in available neutrons per iron-peak
seed nucleus: as the number of seed nuclei increase with increasing [Fe/H]
(the [hs/ls] effect as described by Busso et al. 1999); or the decline may
simply be due to an increasing yield of Fe from SNIa as time progressed.
It is likely that all three mechanisms were at work.

\begin{center}
\begin{figure}
\includegraphics[width=3.0in]{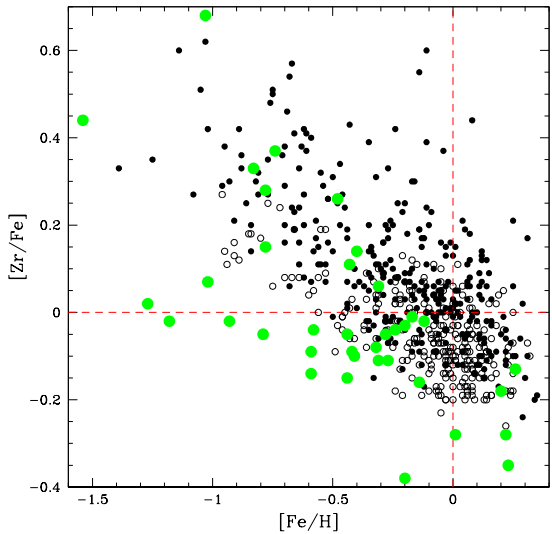}
\caption{[Zr/Fe] versus [Fe/H] in bulge Red Clump stars, from Johnson et al. (2012; 
filled green circles), compared to the Milky Way thick and thin disk
F and G dwarf stars (Battistini \& Bensby 2016; filled black circles).  
While the deficiency in the measured bulge [Zr/Fe] ratios may be
real and require an explanation in nucleosynthesis, differential non-LTE
over-ionization of Zr~I in giant relative to dwarf stars may be
responsible (e.g., Brown et al. 1983).  }
\label{fig-zrfe-diskbulge}
\end{figure}
\end{center}

Regarding the first s-process peak element zirconium,
Figure~\ref{fig-zrfe-diskbulge} shows the bulge results from Johnson
et al. (2012) indicating that the bulge shares the same strong
decline in [Zr/Fe] with [Fe/H] seen in the MW disks by Battistini \& Bensby
(2016); however, the bulge [Zr/Fe] ratios, derived from Red Clump stars, are
shifted lower than the disk values, by 0.1--0.2 dex, at all metallicities.  
Since these [Zr/Fe] ratios were derived from Zr~I and Fe~I lines, they are 
not very sensitive to model atmosphere logg or [$\alpha$/Fe] values, or 
other H$^{-}$ opacity effects.  While it is possible that the putative bulge
Zr deficiency has a nucleosynthetic origin, I note that non-LTE 
over-ionization of Zr~I in MW disk cool RGB stars results in LTE abundance
deficiencies of up to $\sim$0.4 dex, compared to abundances derived from
Zr~II lines (Brown et al.  1983).  For the time being, it is best to assume
that the apparent relative [Zr/Fe] deficiency in the bulge is simply due to
non-LTE effects on Zr~I lines.

Since 2--3 M$_{\odot}$ stars are thought to be responsible for the main
s-process, and the r-process is associated with SNII events (either directly
or from merging neutron stars) with short progenitor lifetimes, the
transition from r-process to s-process, effectively the rise of the s-process,
provides a probe for the rate of chemical enrichment.  Thus, it is possible
to gain
insight into the formation timescale and SFR in the bulge from the relative
proportions of r- and s-process elements.

Ideally, the s- to r-process ratio is best measured from pure s- and pure
r-process elements.  Common choices for nearly pure
s- and r- elements in dwarf stars are Ba and Eu; both elements are represented
by ionized lines, which are relatively robust against non-LTE effects.  In
dwarf stars the Ba~II lines are unsaturated but relatively strong and easy to
detect, while the Eu~II lines are significantly weaker, but measurable with
sufficiently high S/N spectra.  In RGB and Red Clump giant stars, ionized
lines are much stronger than in dwarf stars, resulting in badly saturated
Ba~II lines near solar metallicity.  As a result, the s-process is better
measured using La~II lines (e.g., McWilliam 1997) which are typically
readily detected but unsaturated in solar metallicity red giant stars. 
Notably, La~II lines have very strong hyperfine splitting, resulting in 
significant de-saturation to very high equivalent width and, therefore,
more precise abundance measurement.

Only the bulge abundance studies of MFR10/Fu, Johnson et al. (2012), and van
der Swaelemen et al. (2015) permit measurement of s-/r-process ratios from
[La/Eu] abundances.  Fortunately, the [La/Eu] ratio is robust against
potential systematic errors due to the use of inappropriate model atmosphere
gravities or [$\alpha$/Fe], because the ratio is computed using
ionized lines.  

\begin{center}
\begin{figure}
\includegraphics[width=3.0in]{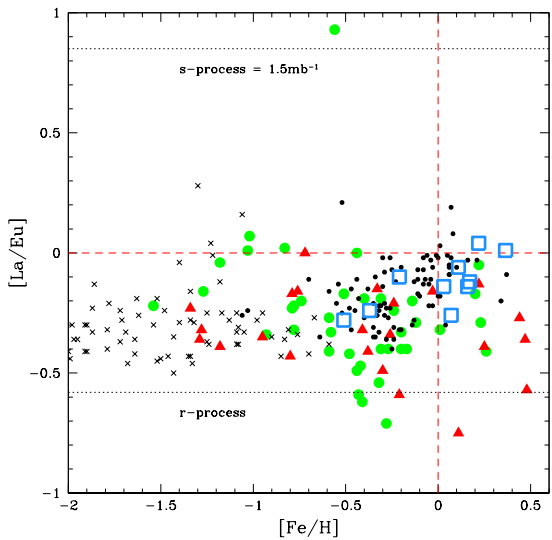}
\caption{
A comparison of [La/Eu] in the MW halo (crosses from Simmerer et al. 2004)
and disk (filled black circles from Battistini \& Bensby 2016) with bulge 
points from MFR10/Fu (red filled triangles), Johnson et al. (2012; filled green
circles) and medianed van der Swaelmen et al. (2015; open blue squares).
The lower dotted line shows the solar-system r-process value from Simmerer et 
al. (2004) and the upper dotted line shows a strong (neutron exposure
of 1.5mb$^{-1}$) s-process ratio, from Malaney (1987).
All three bulge studies indicate a larger r-process fraction (smaller 
s-process) than the MW disk, consistent with a higher SFR in the bulge.
The medianed van der Swaelmen et al. (2015) results, exhibiting the smallest
r-process enhancement of the three, shows [La/Eu] shifted to $\sim$$+$0.2
dex higher [Fe/H] than the MW disk trend (with a range from $+$0.1 to
$+$0.3 dex).
%
}
\label{fig-laeubulge}
\end{figure}
\end{center}

Figure~\ref{fig-laeubulge} shows that the [La/Eu] ratios
from all three bulge studies are systematically lower than the MW disk
trend.  
The Johnson et al. 
(2012) and MFR10/Fu [La/Eu] results are consistent with halo-like s-/r-process 
ratios over the entire metallicity range.  However, as noted earlier the
MFR10/Fu points with [Fe/H] near $+$0.5 dex have suspiciously high [Eu/Fe], 
which may indicate a blend with the single Eu~II line used in their analysis; 
thus, it is probably better to give the points near [Fe/H]=$+$0.5 dex 
lower weight.  Certainly, more detailed spectrum synthesis of the
Eu~II line at 6645\AA\ and the use of other Eu~II lines would be helpful
for understanding the europium content of the most metal-rich bulge stars.
If the [Fe/H]$\sim$$+$0.5 dex points are ignored, it is possible to
interpret both the Johnson et al. (2012) and the MFR10/Fu bulge points as
consistent, with a slowly rising [La/Eu] ratio for stars above the solar
[Fe/H], although the bulge [La/Eu] ratio still lies well below the disk trend.
Extrapolating these remaining bulge points suggests a possible intercept 
with the solar [La/Eu] ratio near [Fe/H]=$+$0.5 dex.

Individual [La/Eu] ratios from van der Swaelmen et al. (2015) show
significant scatter, roughly 1.0 dex near [Fe/H]=$+$0.1 dex; although
the bulk of points are fairly consistent.  Because of this scatter, 
each of the van der Swaelmen et al. (2015) points in
Figure~\ref{fig-laeubulge}, indicated by large open blue boxes, represents
the median of  5 points, consecutive in [Fe/H].  The trend of these
medianed points suggest that the bulge reaches solar [La/Eu] near 
[Fe/H]=$+$0.2 dex, but this intercept may reasonably occur between
[Fe/H] $+$0.1 and $+$0.3 dex.

The low [La/Eu] ratios from all three bulge studies indicates that ejecta
from 2--3 M$_{\odot}$ AGB stars, that produce the main s-process, were less
significant in the bulge than the disk.  This could be explained by a
shorter (i.e., faster) chemical enrichment timescale in the bulge,
compared to the MW disk, or equivalently, that the bulge had a higher SFR
than the disk, so that higher [Fe/H] was reached before the AGB stars had
time to pollute the ISM.  Other possible explanations include a heavier
IMF in the bulge and/or a smaller rate of pristine gas infall into the bulge
than the disk.

In summary, all three studies reached the conclusion that the [La/Eu] ratios
in bulge stars contain less s-process material, or equivalently, a higher
r-process fraction, than MW disk stars.
Figure~\ref{fig-laeubulge} suggests that the bulge
[La/Eu] versus [Fe/H] trend may be shifted to higher [Fe/H], by 0.2 to
0.5 dex, relative to the MW disks, consistent with a SFR 1.6 to 3 times
higher in the bulge.

The neutron capture element ratios in the bulge show that it is
chemically distinct from the MW halo, thick and thin disk stars examined
in the solar vicinity.

%
%


\subsubsection{SPINSTARS?}

I should mention a red herring that was initially exciting: early results for
the bulge globular cluster NGC~6522, based on relatively low-dispersion
GIRAFFE spectra (Barbuy et al. 2009; Chiappini et al. 2011) found evidence
for strong [Y/Fe] and [Sr/Fe] abundances, near $+$1.0 dex; in particular,
the [Y/Ba] ratios were so strong that they were inconsistent with main
s-process AGB nucleosynthesis.
It was suggested that s-processing in rapidly rotating, metal-poor, massive
stars, known as spinstars, may reproduce the measured abundances.  In such
objects $^{12}$C is transported by rotational mixing from the He-burning core
into hydrogen-rich layers, undergoes proton burning, and later
returns to the He-core.  By this mechanism primary $^{22}$Ne is ultimately 
produced, followed by s-processing via the $^{22}$Ne($\alpha$,n)$^{25}$Mg chain.
The most abundant predicted elements from this s-processing are Sr, Y and Zr
(e.g. Pignatari et al. 2008).  

More recent work on this globular cluster, by Barbuy et al. (2014), with 
significantly higher resolving-power spectra and much greater wavelength 
coverage than the original study, did not support their earlier findings.  
Indeed, the heavy element abundances in Barbuy et al. (2014) fit nicely
with the trends now established for the bulge field stars.
Thus, at present, there is no strong evidence for spinstar nucleosynthesis
in the bulge.

\section{Summary and Discussion}

\subsection{Metallicity Distribution Function}

Measurements of the bulge metallicity distribution function (MDF), indicated by [Fe/H], 
show a convergence for RC stars and lensed dwarf stars.  As measured by RC
giants, at a latitude of {\it b}=$-$4$^{\circ}$, the results of Hill et al.
(2011) and Gonzalez et al. (2015) have an average and median [Fe/H] of $+$0.06
dex and $+$0.15 dex, respectively.  The difference between these two measures
results from a skew in the iron distribution function (IDF).  These [Fe/H]
values for the bulge exceed 
the solar neighborhood, mostly thin disk, stars within 15pc from the study 
of Allende Prieto et al. (2004), where the mean and median [Fe/H] values are 
both $-$0.13 dex.  Thus, the {\it b}=$-$4$^{\circ}$ bulge is more metal-rich than
the solar neighborhood thin disk, on average, by almost 0.2 dex, while the peak 
of the two IDFs differ by nearly 0.3 dex.  A higher T$_{\rm eff}$ scale
for dwarf stars, by Casagrande et al. (2010), would increase the Solar
Neighborhood IDF by $\sim$0.10 dex, and give a mean [Fe/H] close to
solar composition, similar to the Stromgren photometry survey of 
Casagrande et al. (2011).  Including these adjustments, the bulge RC IDF is
$\sim$0.10 dex higher than the Solar Neighborhood thin
disk.

The metallicity distribution function measured by Mg, whose production is
overwhelmingly dominated by SNII progenitors, shows a difference between
solar neighborhood and bulge at {\it b}=$-$4$^{\circ}$ of $\sim$0.33 dex, 
for the same studies as the [Fe/H] comparison.  These comparisons show that
the yield of Fe and Mg are higher in the bulge than the Solar Neighborhood
thin disk.  This difference between bulge and Solar Neighborhood average
[Mg/H] distributions is much larger than could result from the potential
systematic temperature uncertainties.  Clearly, a composition study of
RC giant stars in the Solar Neighborhood, using an identical T${\rm eff}$
scale to the bulge RC would be useful for such MDF comparisons.

Given the average thick disk [Fe/H], near $-$0.7 dex (e.g., Gilmore et al. 1995),
the average bulge [Fe/H], at {\it b}=$-$4$^{\circ}$, is about 0.7 dex 
more metal-rich; clearly, the two populations are not very similar 
in iron abundance (nor overall metallicity).

Recalling that, in chemical evolution, the yield is simply the ratio of mass of
metals produced to mass locked-up in low mass stars, possible explanations for 
relatively higher Fe and Mg yields in the bulge include (but are not limited to):
extensive mass-loss from the MW thin disk, or greater retention of SN ejecta in 
the bulge; a bulge IMF strongly under-representing low-mass stars; or 
inward, radial, flows within the MW thin disk feeding metals into the bulge.

The bulge vertical metallicity gradient, found by Terndrup (1988) and
Minniti et al. (1995) and confirmed by numerous others, was explained by 
Hill et al (2011) and later Ness et al. (2013) as due to the super-position
of [Fe/H] sub-populations, whose proportions change with Galactic latitude.
While each sub-population is seen at different latitudes, the higher [Fe/H]
sub-populations occur closer to the Galactic plane.  Hill et al. (2011)
proposed two populations, at [Fe/H] centered at $-$0.30 dex and $+$0.32 dex, 
both [Fe/H] distribution functions well fit by the Simple model, concordant
with Rich (1990), indicating no G-dwarf Problem, unlike the MW disk, 
and therefore no evidence of prolonged infall into the bulge, either 
vertically or radially.  Thus, the relatively high mean metallicity of the
bulge is unlikely to be due to radial inflow.

The bi-model [Fe/H] distribution was explained by Hill et al. (2011) as
indicating the presence of an old metal-poor, spheroidal bulge component
combined with a more metal-rich population at lower latitudes, formed on
much longer timescale, and whose evolution was driven by a bar.

Of the three most notable sub-populations found by Ness et al. (2013), 
near [Fe/H] $+$0.11, $-$0.28, and $-$0.70 dex, the [Fe/H]$\sim$$-$0.70 
dex is associated with the thick disk.  The populations at [Fe/H]$\sim$$+$0.11
and $-$0.28 dex are attributed to instability-driven bar/bulge formation
from the thin disk, similar to the kinematically cold stars of the thin
disk today.  They claimed no obvious evidence for a classical bulge 
component.

Claims of bulge sub-populations evoke questions, such as: do the
different mean [Fe/H] values indicate different effective yields,
suggesting that the sub-populations were made in different environments?  
If so, what environmental parameters were responsible for the yield 
differences?  Did retention of SN ejecta vary over the sub-populations?
Or are the IDF peaks simply due to temporary peaks in the SFR during an
overall evolution?  Perhaps the sub-populations reflect the time 
evolution of yields: the [Fe/H]=$-$0.7 dex peak due to SNII only; the 
$-$0.25 dex peak from yields of SNII plus prompt SNIa, and the $+$0.11 
dex population from SNII, prompt SNIa, and delayed SNIa.

Such questions motivate the need for more precise chemical abundance
measurements of bulge stars, similar to the disk study of Bensby et al.
(2014), showing small-amplitude, but distinct, different chemical paths
for the thick and thin disks.  Perhaps future bulge abundance plots,
similar to Figure~\ref{fig-b14sicati90}, will show distinct loci in [$\alpha$/Fe]
for the [Fe/H] sub-populations.

Presently, the continuity of the [$\alpha$/Fe] trends suggests a chemical
connection between the [Fe/H] sub-populations of Hill et al. (2011)
and Ness et al. (2013).

The scale heights, kinematics, bimodal MDF and [$\alpha$/Fe] ratios of
the bulge MDF sub-populations in Hill et al. (2011) are reminiscent of
the local thin and thick disks, but at higher average [Fe/H]; the 
Ness et al. (2013) MDF peaks are roughly similar, but with a
contamination by local thick disk metallicities near [Fe/H]=$-$0.7 dex.

\subsection{[$\alpha$/Fe]}

For low [Fe/H] bulge stars, below $\sim$$-$0.5 dex, the bulge [O/Fe], [Mg/Fe]
and [$<$SiCaTi$>$/Fe] trends with [Fe/H] appear very close to the MW thick
disk, although [Mg/Fe], and possibly [$<$SiCaTi$>$/Fe], seem to be slightly
larger in the bulge.  Both the bulge
and the thick disk have enhanced [$\alpha$/Fe] ratios
compared to the thin disk.  

The knee in the bulge [$\alpha$/Fe] trends occur near [Fe/H]=$-$0.6 $\pm$0.1 dex,
similar to the MW thick disk.  The bulge [$\alpha$/Fe] ratios, defined by bulge giant
stars, continues a linear decline and passes through solar [Fe/H] with 
[$\alpha$/Fe]$\sim$$+$0.15 dex, and finally
reaches solar [$\alpha$/Fe] ratios at [Fe/H] near $+$0.2 to $+$0.3 dex.

There is evidence in the thick disk [$\alpha$/Fe] results of Bensby et al.
(2014) for exactly
the same behavior as the bulge, if we accept that a number of supposed thin disk
stars with high [$\alpha$/Fe] are really thick disk, and that most of the putative
thick disk stars near solar [Fe/H] are actually misidentified thin disk
objects.  The issue is shown in Figure~\ref{fig-b14sicati90} and described in 
section~2.
Perhaps kinematic mixing of the thick and thin disk may 
have been involved in potential mis-identification.

On the other hand, the actual location of the kinematically identified thick disk
points, with [Fe/H] slightly below solar, lies closer to the thin disk,
as if the thick disk [$\alpha$/Fe] ratios suddenly declined and merged
with the thin disk trend.

If this latter description is correct, the solar neighborhood thick disk
[$\alpha$/Fe] ratios lie well below the bulge ratios above about 
[Fe/H]=$-$0.2 dex.  To understand what the alpha-rich thin disk stars 
are and which of the above best describes the [$\alpha$/Fe] ratios of 
solar metallicity thick disk requires further investigation.  Again, we
are confronted by our imperfect understanding of the MW disk composition!

The comparison of bulge and thick disk compositions is also complicated
by the fact that most of the bulge studies have examined red giants, 
whereas the thick and thin disk chemical composition trends have been 
defined, almost entirely, from dwarf stars.  Only a handful of thick 
disk red giant stars were employed in the comparison of
Alves Brito et al. (2010) and Gonzalez et al.  (2011); and, only four 
near solar metallicity.  Accurate chemical abundance studies for
hundreds of thick disk red giants would be helpful for the comparison.

In this regard, abundance differences between disk dwarfs and a handful
of, mostly thin disk, red giant standards in FMR07 indicated corrections
for Si~I and Al~I of nearly 0.1 dex.  It is possible
that these giant-dwarf differences are due to non-LTE effects in either
the red giants or the dwarfs; although, abundance differences with
standard 1D-LTE results may result from 3D hydrodynamic atmospheres and
even non-plane parallel geometry (in giants).  For this reason, extensive
non-LTE corrections for many lines, of many elements, in a variety of stellar
atmospheres, would be very useful.  See Bergemann \& Nordlander (2014),
and references therein, for a review of the current status of non-LTE
corrections.

Given these difficulties with dwarf-giant comparisons, and the, hopefully,
more robust results of  Alves Brito et al. (2010), Gonzalez et al. (2011)
and Bensby et al. (2013), it is probably best to assume that the bulge 
[$\alpha$/Fe] trends are the same as the MW thick disk.  
However, the putative local thick disk stars with [Fe/H] slightly below 
the solar value have low [$\alpha$/Fe] ratios characteristic of the
thin disk.  Until these stars can be robustly excluded from the
thick disk, there is a non-negligible probability that the Solar
Neighborhood thick and thin disk [$\alpha$/Fe] trends merge near 
[Fe/H]$\sim$0.0 dex.


If the thick disk and bulge [$\alpha$/Fe] trends are the same, then
both the IMF and SFR of the bulge and thick disk are similar.
At the very least, the bulge SFR was higher than for the thin disk.

The observation that the bulge [$\alpha$/Fe] trends reach the solar 
value near [Fe/H]=$+$0.3 dex indicates that the bulge SFR was crudely
twice that of the thin disk and, if all else is assumed equal, the
formation timescale
roughly half that of the thin disk; however, a more robust estimate 
would require a detailed chemical evolution model.  


Besides the well-known alpha-elements O, Mg, Si, Ca and Ti, the bulge
shows an alpha-like trend for [Al/Fe] with [Fe/H].  This is expected,
since, apart from relatively minor proton-burning re-arrangements,
Al production occurs in post carbon-burning hydrostatic phases of 
massive stars that end as SNII events.  This picture is corroborated
by the Al deficiencies in dwarf galaxies (e.g., McWilliam et al 2013),
that also show deficiencies of other hydrostatic alpha elements (O, Mg).
Indeed, the comparison of [Al/Fe] in bulge, MW disks, and Sgr dwarf galaxy
by FMR07 shows startling differences.
However, nucleosynthesis predictions (e.g., Nomoto et al. 2006, WW95)
suggest that the Al yields also depend on progenitor metallicity.
If there is a mild increase in Al yields with SNII metallicity we
would expect to see a slightly lower amplitude of the alpha-like
trend.

While the SNIa time-delay scenario can explain the similarity of
declines in [X/Fe] for most of the alpha-elements by the simple addition
of Fe from SNIa events, oxygen shows a steeper decline than the
other alphas, starting from a higher [O/Fe] plateau in metal-poor
stars.  This suggests that there must be a decline in the yield of
oxygen with increasing [Fe/H].  This effect is also revealed in
the trends of [O/Mg] and [C/O] with [Fe/H] in the bulge and MW disks.
Chemical evolution models (McWilliam et al. 2008; Cescutti et al. 2009)
for SNII element yields without stellar wind mass-loss 
fail to reproduce the observed [O/Mg] and [C/O] ratios, whereas
models including stellar winds match the shape of the observed 
abundance trends.  However, estimated stellar wind mass-loss rates
have declined in recent years, resulting in some uncertainty.  These
issues are connected with the production of WR stars and the importance
of episodic mass-loss in massive stars.

An important observation is that the [$\alpha$/Fe] trends with [Fe/H]
are the same for all locations within the bulge, as established by
Johnson et al. (2013).  This has recently been confirmed for Galactic 
latitudes, {\it b}, of 0$^{\circ}$, $-$1$^{\circ}$ and $-$2$^{\circ}$, 
by Ryde et al. (2016), based on near-IR spectroscopy of bulge M giants,
who's alpha-element trends were identical to those of G15 
and Hill et al. (2011), for {\it b}=$-$4$^{\circ}$ RC giants.


It is remarkable that the [$\alpha$/Fe] trends in the bulge and thick disk are
so similar, suggesting a similar SFR and formation timescale to solar [Fe/H],
despite the vastly different overall metallicities.  Perhaps, the difference
in metallicity simply reflects different efficiencies in retaining SN ejecta.


An interesting characteristic of the bulge alpha-elements, first noticed
by FMR07, is that the trends of [O/Fe], [Mg/Fe], and [$<$SiCaTi$>$/Fe] show
tight relations below [Fe/H]$\sim$$-$1 dex, quite unlike the halo,
for example seen in Figure~\ref{fig-b14sicati90}.  In this regard
the bulge also resembles the thick disk: the large dispersion in halo [$\alpha$/Fe]
ratios below [Fe/H]$\sim$$-$1 is due to an inhomogeneous halo, probably resulting
from accretion of dwarf galaxies with a variety of star formation histories.
On the other hand, both the MW thick disk and bulge appear to have been much
more homogeneous.

\subsection{Notable Iron-Peak Elements}

The trend of [Cu/Fe] with [Fe/H] in the bulge, only measured by J14, 
is very different than the MW thin and thick disks. At low metallicity,
the slope of increasing [Cu/Fe] with [Fe/H] continues to higher [Cu/Fe]
than the MW disks, near [Cu/Fe] of $+$0.4 to $+$0.5 dex; whereas, the MW
thick and thin disks flatten-out near [Fe/H]$\sim$$-$0.7 dex, presumably
due to the addition of SNIa Fe.
The bulge [Cu/Fe] trend declines from $\sim$$+$0.4 dex, near [Fe/H]=$-$0.2
dex, down to [Cu/Fe]=$+$0.2 dex, as if the addition of SNIa Fe occurred
much later in the bulge than the MW disks.  Finally, near [Fe/H]=$\sim$$+$0.2
dex, the [Cu/Fe] trend starts to increase again.
This is similar to the zig-zag shape of the very small amplitude trend
     of [Na/Fe] with [Fe/H].

If the Cu measurements of J14 are correct, these results are qualitatively
consistent with the time-delay scenario of Tinsley (1979) and MB90,
for Cu produced by massive stars with a metal-dependent yield, as expected
for the weak s-process in massive stars (e.g., Pignatari et al. 2010); but,
the delayed SNIa Fe began at a higher [Fe/H] in the bulge,
suggesting involvement of massive star ejecta to higher [Fe/H],
probably due to a higher SFR.  In fact, the unusual bulge [Cu/Fe] trend, 
if real, is consistent with the SNIa time-delay scenario.

While this suggests that the bulge chemistry, and therefore evolution,
was very different than the MW thick disk,  it may still be that the 
bulge [Cu/Fe] trend follows the trajectory that the thick disk would
have produced if it had not run out of gas.

Interestingly, the bulge and thick disk [Cu/Fe] trends differ 
at [Fe/H]$\sim$$-$0.7 dex, right where the thick disk MDF starts
to decline, presumably due to gas-loss.  
This may relate to the low [$\alpha$/Fe] in putative thick disk stars 
just below solar metallicity.  It may indicate that the MW thick disk 
really did decline in [$\alpha$/Fe] after the MDF peak, whereas the 
bulge had plenty of gas to maintain its [$\alpha$/Fe] trend and produce
plentiful amounts Cu.

Clearly, it is important to check the [Cu/Fe] results of J14, for
both bulge dwarfs and giants; in particular, more non-LTE
calculations are desperately needed.

Notably, zinc is expected to share some similarities in its nucleosynthetic
origin (alpha-rich freeze-out plus weak s-process) as copper, and so would
provide a useful comparison.  The two studies of Zn in the bulge show
different measured trends: one identical to the MW disks, and the other
showing a decline in [Zn/Fe] above [Fe/H]$\sim$$-$0.2 dex, similar to
[Cu/Fe] but at a lower enhancement level.  It would be useful to determine
which of the two measured trends for bulge stars is correct.

The [Mn/Fe] trend with [Fe/H] in the bulge, MW thin and thick disks are
identical, to within measurement error.  This is contrary to the
expectation that Mn is over-produced in SNIa with the rising [Mn/Fe]
due to the increasing SNIa/SNII ratio with [Fe/H].  Since the bulge
[$\alpha$/Fe] ratios at solar [Fe/H] are near $+$0.15 dex, less SNIa
material and therefore, less, Mn from SNIa is expected; a low [Mn/Fe]
ratio is predicted for high [$\alpha$/Fe], especially compared to the
thin disk.
If the LTE measurements are correct, then Mn is most likely 
produced proportional to metallicity, by both SNIa and SNII events.

However, present day non-LTE corrections considerably flatten the steeply
increasing [Mn/Fe] trend with [Fe/H] seen in LTE.  If correct, then the
Mn abundances provide very little constraints on its production in SNIa
and SNII events.

\subsection{Neutron Capture Elements}

Due to the alpha-like trend of the r-process element europium, the
[Eu/Fe] trend with [Fe/H] offers the possibility the compare the
SFR in the bulge and disk.  Unfortunately, the disk trends are
not well defined, or at least not completely converged; thus, how
the bulge compares depends upon which MW comparison study is employed.
Notwithstanding, if the Battistini \& Bensby (2016) MW disk data are 
employed, then the bulge [Eu/Fe] follows the disk, but without sufficient 
precision to distinguish between thick and thin components.

A probe into the chemical evolution of the bulge can be obtained
from the rise of the s-process, which is normally driven by low-mass
(2--3 M$_{\odot}$) AGB stars.  The [La/Eu] ratio measures the 
s-process/r-process ratio and only increases when the low-mass AGB stars 
produce s-process (associated with He-shell burning episodes).  Although 
the data possess measurement scatter that is larger than one would like,
it appears that the bulge [La/Eu] ratio begins to increase at a higher 
[Fe/H] than the MW disks, indicating a higher SFR in the bulge.  However, these 
measurements need to be greatly improved upon.


\section{Chemical Evolution of a Secular Bulge}

Evidence in recent years, including the presence of a bar, a boxy/peanut shape, the
X-shape bulge, and disk kinematics (e.g. see Wegg \& Gerhard 2013)
has led to the conclusion that the bulge was built by the growth of a stellar bar, 
through accretion of stars from the inner Galactic disk, and was 
subsequently subject to buckling processes that thickened the bar into the bulge
morphology seen today.

Any bulge evolution model should produce these kinematic and morphological features,
as well as match the observed vertical metallicity gradient, chemical composition trends,
and the multi-, or bi-modal MDF. 
Here, I provide a sketch of this evolution and how the chemical composition trends 
might have occurred (but, see Di~Matteo 2016 for a more detailed discussion of bulge
chemodynamics).

First, it is notable that the metal-rich bulge sub-population, which is concentrated toward
the Galactic plane, has a mean [Fe/H] in excess of the Solar Neighborhood thin disk, with
[$\alpha$/Fe] ratios indicating the presence of nucleosynthesis products from SNIa.  Thus,
this bulge sub-population shares kinematic and chemical characteristics with the local thin
disk, albeit at higher [Fe/H].  

On the other hand, the [Fe/H]$\sim$$-$0.3 dex bulge sub-population shows a relatively 
large vertical scale height and [$\alpha$/Fe] ratios indicating higher SFR and more
rapid formation than the local thin disk, but similar or slightly faster than the local
thick disk; these chemical and kinematic characteristics suggests that this population
is similar to the local thick disk, but at significantly higher [Fe/H].

If these assumptions are correct, both the thick disk and thin disk are 
represented in the bulge, but at significantly higher [Fe/H] than at the solar 
circle (by about 0.3 dex), which suggests radial [Fe/H] gradients for both thin and 
thick disks, near $-$0.04 to $-$0.05 dex per kpc between the bulge and solar circle.
These metallicity gradients might be due to more efficient retention of SN ejecta
in the inner galaxy .

High mean bulge MDFs could also be accomplished by a reduction of the mass locked-up 
in low-mass, unevolved, stars.  While this is a modification of the IMF, the bulge 
[$\alpha$/Fe] ratios require that the relative proportions of SNII and SNIa be similar
to the thick disk, so there can be no large modification of the IMF slope of massive stars,
only a reduced number of unevolved stars.

Although radial inflow of gas could increase the overall MDF of the bulge, this would
also result in a bulge G-dwarf Problem (a deficiency of metal-poor stars) and slightly
lower [$\alpha$/Fe] ratios compared to the solar circle, but these are not seen.

Since the relatively low [Fe/H] of inner thick disk MDF indicates that it did not reach 
the yield, chemical evolution terminated before complete consumption of the gas.  Thus, 
this gas must have gone somewhere else, or have been lost from the Galaxy altogether.  
On the other hand, the inner thin disk gas, represented by the [Fe/H]=$+$0.3 dex
sub-population, must have come from somewhere.

It seems likely, and logical, that the thick disk gas ultimately settled into the thin
disk, both at the solar circle and inner disk region, due to molecular cloud collisions.  
Stars that formed before the gas settled retained the kinematic signature of the gas 
at the time of their formation, in particular with disk kinematics and large vertical 
scale height.

However, at the solar circle
the thick disk gas stopped chemical enrichment at a lower [Fe/H] than for the inner thick 
disk.  The, presumably, higher gas densities in the inner disk region would result
in higher SFR and shorter formation timescale compared to the solar neighborhood thick 
disk, as suggested by the slightly higher [$\alpha$/Fe], lower [La/Eu], and higher [Cu/Fe]
ratios observed in the bulge.

The inner thin disk would then have formed out of the settling inner thick disk gas, 
resulting in a continuity of the chemical composition trends, as observed.  Finally, 
instabilities in the thin disk (e.g., as outlined by Athanassoula \& Misiriotis 2002;
Wegg \& Gerhard 2013) led to a stellar bar and its thickening, by buckling processes,
into the boxy/peanut/X-shape morphology with disk-like kinematics seen today.  The
inner thick and thin disk stars entrained into the X-shape morphology by the bar
retained vertical scale heights similar to their original values, resulting in the 
vertical [Fe/H] gradient and bimodal MDF.

The scenario sketched above suggests a number of consistency checks and questions.
For example, if radial gas inflow were responsible for higher [Fe/H] in the inner regions, 
lower [$\alpha$/Fe] ratios might be expected, but the opposite is observed.
The metal-rich bulge sub-population should be slightly younger than the
metal-poor population, with a longer formation timescale, consistent with the observed
low [$\alpha$/Fe] ratios and high [La/Eu] ratios.  But, what was the actual formation
timescale of the inner disk?  It might be possible to probe this question using chemical 
abundance patterns for s-process elements and connecting those to the mass of the s-process
sites.  For example, most s-process elements in the solar neighborhood thin disk
are thought to be produced by 2--3 M$_{\odot}$ AGB stars with neutrons provided by
the $^{13}$C($\alpha$,$n$)$^{16}$O source.  However, the s-process abundance pattern
produced by intermediate mass AGB stars is expected to be somewhat different 
(e.g. Busso et al. 2004); in particular, the
$^{22}$Ne($\alpha$,$n$)$^{25}$Mg neutron source and high neutron densities dominates
for 5--8 M$_{\odot}$ AGB stars, and should result in relative enhancements of $^{25}$Mg,
$^{26}$Mg, $^{96}$Zr and $^{87}$Rb.  These isotopes can be measured in RGB stars,
from MgH, ZrO and Rb~I lines, and might provide a constraint on the inner disk formation 
timescale.  

It would be interesting to measure the [$\alpha$/Fe] ratios in the bulge to much higher 
precision.  High-precision chemical abundances of local thin and thick disk stars by
Reddy et al. (2006) show considerable overlap in [Fe/H], but the two populations are
separated in [$\alpha$/Fe] by 0.15 to 0.2 dex.  This is clearly related to the evolution
of the local thin and thick disks; therefore, it would be useful to know whether the 
bulge/inner thin and thick disk populations also show similar [Fe/H] overlap and small 
separations in [$\alpha$/Fe].

More precise abundance measurements would also permit a better evaluation of the relative
SFR for the inner thick disk/bulge compared to the solar neighborhood thick disk, based
on the [$\alpha$/Fe] and [La/Eu] trends with [Fe/H].  Indeed, a systematic study of the 
detailed elemental abundances in the thick disk between the solar circle and bulge region
would provide a useful test of the above scenario: one expects a general increase in 
[Fe/H] with decreasing Galactocentric radius, and subtle increases in the SFR, reflected
in [$\alpha$/Fe], [Cu/Fe] and [La/Eu] abundance trends.


\section{Conclusions}

The bulge shows a vertical [Fe/H] gradient, at $\sim$0.5 dex/kpc, with
more metal-rich stars concentrated toward the plane. 
The average and median [Fe/H] in Baade's Window bulge field, at 
{\it b}=$-$3.9$^{\circ}$ is $+$0.06 dex and $+$0.15 dex respectively.

Hill et al. (2011) identified two sub-populations, centered at [Fe/H]
of $-$0.30 dex and $+$0.32 dex, at {\it b}=$-$3.9$^{\circ}$, while
Ness et al. (2013) suggest three sub-populations in the main bulge MDF, 
with [Fe/H] in their {\it b}=$-$5$^{\circ}$ field of $+$0.12 dex,
$-$0.26 dex and $-$0.66 dex.  Both studies suggest that the vertical
gradient is due to changing proportions of these sub-populations.

The higher mean and median [Fe/H] values for the bulge, compared to the
local thin and thick disks, indicates a higher yield for the bulge.  This 
could easily be due to more efficient retention of SN ejecta in the bulge, 
especially SNIa.  Other possible explanations include: 1. an IMF deficient
in the lowest mass stars in the bulge, so locking-up less gas, or 2. radial
gas inflow into bulge from the inner disk, although this
should result in a deficit of metal-poor stars, which is not seen.

If these bulge sub-populations originated from inner thick and thin disk
stars entrained into a secular bulge through bar formation, then the high
mean [Fe/H] values of the bulge sub-populations suggest radial [Fe/H] gradients
from bulge to solar neighborhood of $-$0.04 to $-$0.05 dex/kpc for both thin
and thick disks.

The [$\alpha$/Fe] ratios in bulge, below [Fe/H]$\sim$$-$0.5 dex, are much like
the local thick disk trends, suggesting similar IMF and SFR; but, 
the thick disk is metal deficient, compared to the bulge,
by more than 0.7 dex.  Possible slight enhancement of the bulge [Mg/Fe] and
[$<$SiCaTi$>$/Fe]compared
to the thick disk may indicate SFR differences, but could be due to measurement
errors, or systematic uncertainties
in the comparison of abundances for bulge giants with thick disk dwarfs.

Above [Fe/H]=$-$0.5 dex, the kinematically identified thick disk stars merge into
the thin disk [$\alpha$/Fe] trends by solar [Fe/H], whereas the bulge [$\alpha$/Fe]
is enhanced compared to the thin disk, by $\sim$$+$0.15 dex, indicating a higher
SFR in the bulge than thin disk, at least.  On the other hand,
a small number of kinematically identified local thin disk stars seems to extend 
the slope established by metal-poor thick disk stars, to solar [Fe/H] and beyond.
The status of these high-$\alpha$ thin disk stars should be investigated further.

It is remarkable that the [$\alpha$/Fe] trends of the thick disk and bulge are
so close, suggesting similar SFR, even though their metallicities differ
enormously.

%
%

The [La/Eu] ratio, indicating the onset of the s-process and presence
of ejecta from 2--3 M$_{\odot}$ AGB stars, is consistent with a slightly
higher SFR and shorter formation timescale for the bulge, compared to the
MW disks.

The [Cu/Fe] trend, measured by only one study, J14, shows a stunningly different
trend than thick disk or thin disk.  However, the  zig-zag [Cu/Fe] trend is
qualitatively consistent with the combination of a high bulge SFR and 
metal-dependent Cu yields from massive stars (as expected) in the presence of
the SNIa time delay scenario.  
Curiously, the [Na/Fe] ratios in the bulge, while close to the solar ratio,
shows a small amplitude zig-zag trend, similar to [Cu/Fe], suggesting the presence 
of metal-dependent Na yields from massive stars.

The trend of LTE [Mn/Fe] ratios are the same in the bulge as the MW thick 
and thin disks, to within measurement uncertainty.   This is contrary to 
the expectation that SNIa over-produce Mn.  Since the ratio of SNIa/SNII material
is lower in the bulge than the thin disk, as evidenced from [$\alpha$/Fe] ratios, 
lower [Mn/Fe] are be expected in the bulge, but are not seen.  Predicted non-LTE 
corrections to the LTE Mn abundances suggest that the trend of this element could 
be seriously affected by non-LTE effects.

\section{Recommendations for Future Studies} 
  
The remarkable bulge [Cu/Fe] trends found by J14 need to be checked.  The
information may already be present in the lensed dwarf spectra of B13.
Since Cu~I lines are strong, and possibly vulnerable to saturation effects
in K giants, more robust abundances may be obtained from the warmer RC
giants and lensed dwarf stars.  However, strong hfs effects in Cu~I lines
are helpful for de-saturating the lines and increasing abundance sensitivity.

High precision abundances are required for bulge alpha-elements.  The bulge
MDF sub-populations in Ness et al. (2013)  and Hill et al. (2011) should be
carefully studied, in order to see whether there is a continuous chemical
composition trend, or discrete [$\alpha$/Fe] trends for each sub-population,
similar to the [$\alpha$/Fe] differences between local thick and thin disks
(e.g. Figure~\ref{fig-b14sicati90}).

Neutron-capture elements, especially Eu, La, should be measured to better
precision than heretofore obtained, in order to further identify the [Fe/H]
of the nucleosynthetic onset of low-mass AGB stars, for comparison with
the MW disks.

Other elements of interest include Rb, which is thought to be made in
s-processing driven by the $^{22}$Ne($\alpha$,n)$^{25}$Mg neutron
source, prevalent in intermediate mass AGB and massive stars.  
Increased Mg isotopic ratios (25/24 and 16/24) reveal the same 
neutron source in similar stars; thus, isotopic Mg measurements, based
on the MgH lines near 5130\AA\ would be of interest.  

For all the above abundance goals it would be best to obtain high resolving
power spectra (R=40,000 to 60,000) in order to separate-out blends, to 
detect continuum, and for increased line depths and abundance sensitivity.
However, for Mg isotopes high S/N, $\sim$100 and high resolving power (R$\sim$80,000
to 120,000) are best.

Warmer RC giants offer the possibility for superior abundance measurements
than the more luminous K giants, due to reduced line blending, greater continuum
and frequently less saturated lines.  However, higher S/N is
also required.  One possibility is to look at metal-normal and metal-rich
bulge stars at higher latitudes (e.g. {\it b}=$-$8$^{\circ}$), where the
extinction is much reduced.
This may be effective, as it appears that the [$\alpha$/Fe] versus [Fe/H] trends are
identical over the whole bulge, as if kinematic heating simply brought some of
the metal-rich populations into higher latitude orbits.

Temperatures should be based on the excitation of Fe~I lines, and measured from
line-by-line differential analysis relative to solar neighborhood standards, like
Arcturus (e.g. FMR07, Hill et al. 2011).

An impediment to our understanding is the ability to compare the composition of 
the same types of stars in the bulge and thick and thin disks.  Therefore,
we need to robustly determine the abundance effects involved with dwarf and giant
star comparisons; in this regard, a survey of MW thin and thick disk giants, 
using the identical abundance analysis methods to the bulge stars would be helpful.
The $\alpha$-rich thin disk stars near solar [Fe/H] in the Bensby et al. (2014)
study should be examined more closely to verify their validity.

A study of the radial [Fe/H] and composition gradients in the thick disk would
provide a useful consistency check on the bulge formation scenario: as Galactocentric
radius decreases, gradual increases in the mean [Fe/H], higher
[$\alpha$/Fe] and lower [La/Eu] ratios are expected, consistent with higher SFR and 
shorter formation timescales toward the bulge.  

Finally, we need non-LTE corrections for Cu, alphas, and all other elements of interest
in the bulge.  These non-LTE corrections should be verified empirically.  For example,
by studying the composition of dwarf and giant stars in clusters of different 
metallicity, similar to work done by Korn et al. (2007) for NGC6397.


\end{document}